\documentclass[aps,onecolumn,reprint,superscriptaddress,showpacs,nofootinbib]{revtex4}
\pdfoutput=1

\usepackage{graphicx}
\usepackage{subfigure}
\usepackage{filecontents}
\usepackage{amsmath}
\usepackage{amssymb}
\usepackage{bm}
\usepackage{color}

\usepackage{latexsym}
\usepackage{dcolumn}
\usepackage{amsxtra}

\def\e{\mathop{\rm \mbox{{\Large e}}}\nolimits}
\newcommand{\bn}[1]{\mbox{\boldmath $#1$}}

\newcommand{\be}{\begin{equation}}
\newcommand{\ee}{\end{equation}}
\newcommand{\bc}{\begin{center}}
\newcommand{\ec}{\end{center}}
\newcommand{\bea}{\begin{eqnarray}}
\newcommand{\beann}{\begin{eqnarray}}
\newcommand{\eeann}{\end{eqnarray}}
\newcommand{\eea}{\end{eqnarray}}
\newcommand{\ba}{\begin{array}}
\newcommand{\ea}{\end{array}}
\newcommand{\nn}{\nonumber}

\def\a{\mbox{\it{\Large $a$}}}

\def\1{_{1}}
\def\2{_{2}}

\def\t

{\mbox{q}_{\mbox{\small t}}}

\def\a{\mbox{\it{\large $a$}}}

\def\e{\mathop{\rm \mbox{{\Large e}}}\nolimits}

\def\a{\mbox{\it{\Large $a$}}}

\def\1{_{1}}
\def\2{_{2}}

\def\t{\mbox{q}_{\mbox{\small t}}}

\def\a{\mbox{\it{\large $a$}}}

\def\ni{\noindent}
\def\mb{\mbox}
\def\bm{\boldmath}
\def\e{\mathop{\rm \mbox{{\Large e}}}\nolimits}
\def\ae{\mathrm{\mb{\Large \textbf{e}}}}

\def\p{^{\;\prime}}
\def\t{^{\;\dag}}

\def\a{\mb{\it{\Large $a$}}}
\def\aa{\mathrm{\mb{\Large \textbf{a}}}}
\def\asa{\mathrm{\mb{\Large a}}}

\def\Jsf{\mathrm{\mb{\large \textbf{\textsf{J}}}}}

\def\Csf{\mathrm{\mb{\large \textbf{\textsf{C}}}}}

\def\Ssf{\mathrm{\mb{\large \textbf{\textsf{S}}}}}

\newtheorem{thm}{Theorem}[section]
\newtheorem{defn}{Definition}[section]

\newtheorem{propo}[thm]{Proposition}

\newtheorem{casos}[thm]{Cases}

\begin{document}

\title{Surveying the Multicomponent Scattering Matrix: Unitarity and Symmetries}

\author{L. Diago-Cisneros}
\email[E-mail and orcid:$\;$]{ldiago@fisica.uh.cu; leovildo.diago@ibero.mx \& http/orcid.org/0000-0001-9409-1545/}
\affiliation{Facultad de F\'{\i}sica. Universidad de La Habana, La Habana, Cuba.}
\affiliation{Departamento de F\'{\i}sica y Matem\'{a}ticas, Universidad Iberoamericana,  Ciudad de M\'exico.}

\author{J. J. Flores-Godoy}
\email[E-mail and orcid:$\;$]{jose.flores@ucu.edu.uy \& http/orcid.org/0000-0003-0569-0162/}
\affiliation{Departamento de F\'{\i}sica y Matem\'{a}ticas, Universidad Iberoamericana, Ciudad de M\'exico,  M\'exico.}

\author{G. Fern\'{a}ndez-Anaya}
\email[E-mail and orcid:$\;$]{guillermo.fenandezo@ibero.mx \& http/orcid.org/0000-0002-2396-5046/}
\affiliation{Departamento de F\'{\i}sica y Matem\'{a}ticas, Universidad Iberoamericana, Ciudad de M\'exico,  M\'exico.}

\date{\today}

\begin{abstract}
   Multicomponent-multiband fluxes of spim-charge carriers, whose components propagate mixed and synchronously, with \emph{a priori} nonzero incoming amplitudes, do not obey the standard unitarity condition on the scattering matrix for an arbitrary basis set. For such cases, we have derived a robust theoretical procedure, which is fundamental in quantum-transport problems for unitarity preservation and we have named after \emph{structured unitarity condition}. Our approach deals with $(N \times N)$ interacting components (for $N \geq 2$), within the envelope function approximation (EFA), and yet the standard unitary properties of the ($N = 1$) scattering matrix are recovered. Rather arbitrary conditions to the basis-set and/or to the output scattering coefficients, are not longer required, if the \emph{eigen}-functions are orthonormalized in both the configuration and the spinorial spaces. We expect the present model to be workable, for different kind of multiband-multicomponent physical systems described by Hermitian Hamiltonians within the EFA, with small transformations if any. We foretell the interplay for the state-vector transfer matrix, together with the large values of its condition number, as a novel complementary tools for a more accurate definition of the threshold for tunnelling channels in a scattering experiment.
\end{abstract}

\maketitle


\section{Introductory outline}
 \label{Cap:Tunel::Intro}
\begin{figure}[ht!]
\centering
  \includegraphics[angle=0,width=0.8\linewidth]{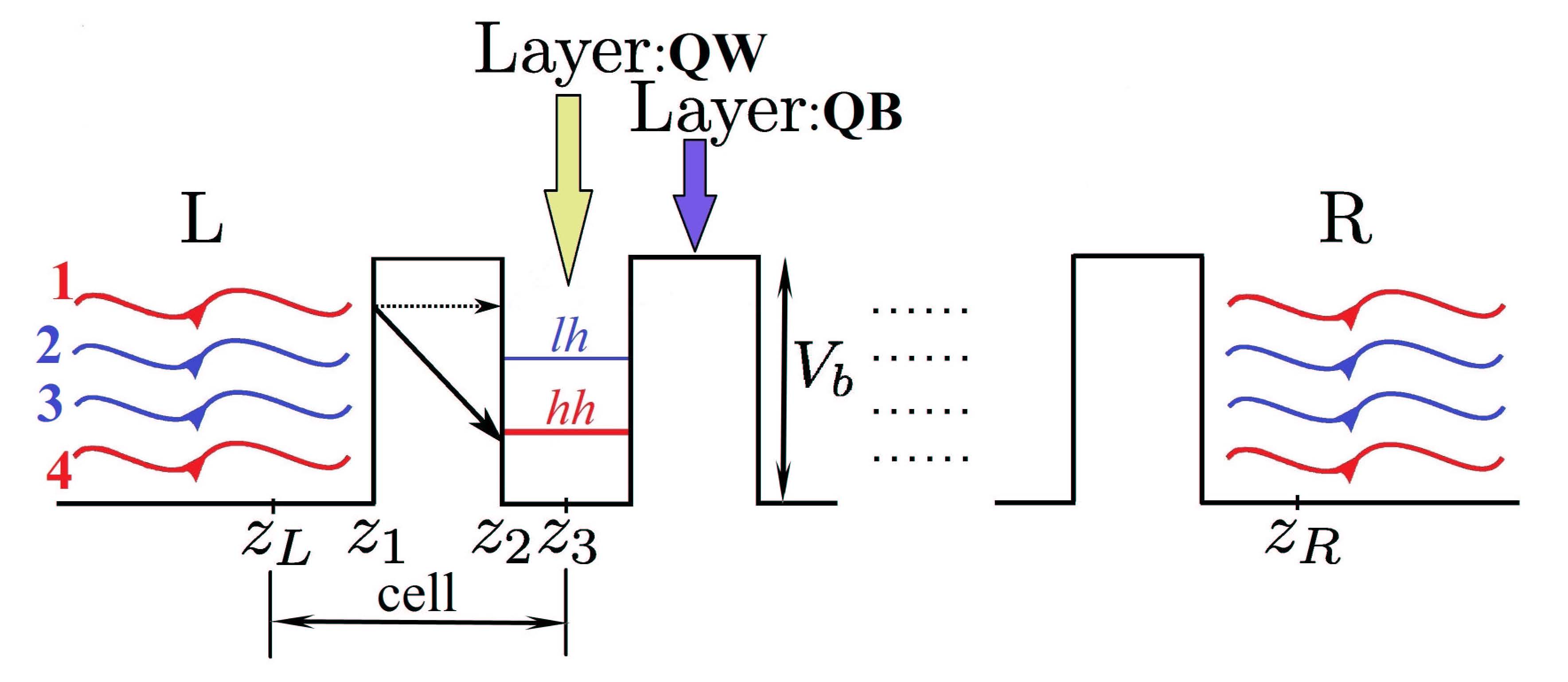}
    \caption{\label{Fig:BVprofile} (Color online). Schematic representation of quantum transport of heavy holes ($hh$) and light holes ($lh$) in a superlattice of $GaAs$- electrode ($L$) /$(AlAs/GaAs)^n$/ $GaAs$- electrode($R$), with no external fields and strains. QW(QB) stands for quantum well(barrier), respectively.}
\end{figure}

The unitarity and symmetries properties in the multicomponent-multiband scattering theory (MMST) is a subtle problem, with several difficulties to overcome. We have developed in a fairly general fashion, an analysis of the unitarity and several analytic symmetry properties of the MMST, mainly by means  of the scattering matrix (SM) workbench. Though undeniably not exhaustively detailed in every mathematical entity, we thought the present theoretical modelling as a useful workbench to deal with $N$-component synchronous mixed-particle quantum transport. Moreover, instead of completely rigorous mathematical formalisms, we choose less abstract --as posible--, practical tools to deal with unitarity preservation and symmetries in multicomponent-multiband systems. The focus has been put in problems well described by a matrix system of second-order differential equations, with first-derivative terms (responsible for the coupled interplay) included. Provided a consistent use of the present orthonormalization procedure, no flux conservation (FC) mismatches should arise. In this study, an exercise is devoted to the quantum transport of holes in $Q2D$ multiband-multichannel physical systems, within the framework of our theoretical procedure. The numerical simulations were based on the $2$-bands Kohn-L\"uttinger model Hamiltonian, which only consideres the two highest in energy sub-bands of the valence band (VB). It is important to stress that, most of the properties, definitions and propositions that have been presented, are valid for any physical layered-model, as the one sketched in Fig.\ref{Fig:BVprofile}.

The quantum transport of electrons and holes in semiconductor heterostructures, are important subjects on Solid State Physics. In comparison to electrons in the conduction band (CB), the case of VB holes have been less studied due to mathematical difficulties of the models. Nevertheless, when both charge carriers are involved, as in opto-electronic devices, the response time threshold would be determined by holes due to its bigger effective mass. Additionally, in experiments with $GaAs-AlAs$ superlattices, when the VB is in resonance and the CB is not, the tunneling of holes occurs more rapidly than the tunneling of electrons regardless the effective masses \cite{Schneider85}. The actual models of single-component fluxes \cite{Wess89,Erdogan93,Kumar97,Morifuji95,Sanchez95} are not sufficient to describe the quantum transport of mixed multi-component fluxes, due to the lack of enough physical information about the dispersion processes. We present an alternative approach, in which all the propagating modes are taking into account collectively and simultaneously\footnote{From now on, as \underline{simultaneous} we will understand that the $N$-component coupled modes propagating throughout a system represented in Fig.\ref{Fig:BVprofile}, coexist. They have been assumed with nonzero initial amplitudes and then, they are simultaneously accesible for all energies of the incident flux. The synchronization of our approximation must not be confused with temporal simultaneity of events, because we are dealing with a strictly stationary problem.}. Then, the multi-component and multi-channel synchronous transmission of amplitudes, can be described without arbitrary assumptions. In the present modelling, both the formalism of the transfer matrix (TM) and the $N$-component SM ($N \geq 2$) are combined, and we have called it the multi-component scattering approach (MSA)\cite{Diago02,Diago06}. Many physical phenomena, can be understood as scattering problems and thus, they are susceptible to be studied within the framework of the SM, which relates the incoming flux with the outgoing one. It is well known, that the SM is unitary within the single-band effective mass approximation (EMA). Nevertheless, when the problem need to be described by a matrix differential system like (\ref{Eqmaestra}), then the fulfillment of this crucial property is not a simple task. As we will see later, the properties of the basis set of expanded linear-independent solutions (LI) of the physical system, play an important role to achieve the unitarity condition on the SM. In the specialized literature for multi-band problems \cite{Wess89,Erdogan93,Broido85,Kumar97,Morifuji95,Klimeck01}, it is standard to impose the orthonormalization in the configuration space, complementing in some cases with other numerical conditions. Though successful for several practical situations, that treatment could be insufficient whenever the mixing and simultaneously propagating carriers are involved. This for example is the case of heavy holes ($hh$) and light holes ($lh$), with different total angular momentum projection, traversing throughout a layered heterostructure [see Fig.\ref{Fig:BVprofile}] with finite in-plane energy. The first mark of this relevant problem was given at 1995 by S\'{a}nchez and Proetto \cite{Sanchez95}, who revisited the form of the unitary for the SM, in the particular case of the ($2 \times 2$) Kohn-L\"{u}ttinger (KL) model. Let us consider a problem described by a system of two or more lineal ordinary second-order coupled differential equations. The eigenvalue equation of that problem, for a multi-component system with translational symmetry in the [$x,y$] plane perpendicular to axis $z$ [see Fig.\ref{Fig:BVprofile}], can be written in the matrix form as \cite{RPA04}
\begin{eqnarray}
 \label{Eqmaestra}
  \frac{d}{dz} \left[ \bn{B}(z) \cdot \frac{d\bn{F}(z)}{dz} + \bn{P}(z) \cdot
  \bn{F}(z)\right] + \bn{Y}(z) \cdot \frac{d\bn{F}(z)}{dz} + \bn{W}(z) \cdot \bn{F}(z)
  & = & \bn{O}_{\mb{\tiny N}}\,, \nonumber \\
\end{eqnarray}
where $\bn{B}(z), \bn{P}(z), \bn{Y}(z)$ and  $\bn{W}(z)$ fulfil
\begin{eqnarray}
 \label{CoefMat-Prop(i)}
    \bn{B}(z)\t & = & \bn{B}(z) \\
    \bn{Y}(z) & = & -\bn{P}(z) \\
    \bn{P}(z)\t & = & \pm \bn{P}(z) \\
    \bn{W}(z)\t & = & \bn{W}(z) = \bn{V}(z) - E\bn{I}_{\mb{\tiny N}}
 \label{CoefMat-Prop(f)}
\end{eqnarray}
\ni and all matrices are $(N \times N)$. Hereinafter $\bn{O}_{\mb{\tiny N}}/\bn{I}_{\mb{\tiny N}}$ stands for the $N$ order null/identity matrix, respectively. We represent by $\bn {F}(z)$ the {\em{field}} under study (for example: the envelope function for $hh$ or $lh$). As $\bn {F}(z)$ has $N$-components, we refereed to it as a {\em{super-vector}} that belongs to the functional vector space of the problem. This is completely analogous --although not in the same sense--, when dealing with the position $\vec{r}$, or the velocity $\vec{v}$ vectors. The later are characterized by the way they change under an orthogonal transformation of coordinates in ordinary $3D$ space. When examining (\ref{Eqmaestra}), there highlights the linear form associated to this system, here refereed as \cite{RPA04}.
\begin{eqnarray}
 \label{FormaLineal}
  \bn {A}(z) & = & \bn{B}(z) \cdot \frac{d\bn{F}(z)}{dz} + \bn{P}(z) \cdot \bn{F}(z)\,.
\end{eqnarray}
If $\bn{B}(z)$, $\bn{P}(z)$, $\bn{Y}(z)$ and $\bn{W}(z)$ have the properties required by (\ref{CoefMat-Prop(i)})-(\ref{CoefMat-Prop(f)}), the {\em adjoint} operator (Hermitian conjugated ) has the same rule to operate than the original operator. Its property of Hermiticity, {\em formal} or not\footnote{See a detailed analysis of that subject on page $99$ of the reference \cite{RPA04}.}, will depend on the boundary conditions that fulfill the operator and its {\em adjoint}. Mathematically speaking, this linear form play a relevant role in the continuity of $\bn {F}(z)$ for all $z$ as we will see in the Subsec. \ref{Cap:SM::Converge:::Fun-Ortonor} and is the cornerstone in the \emph{Surface Green Function Method}\cite{FGM92}.


\section{Flux Tunneling}
 \label{Cap:SM::Tun-Flujo}

Lets turn now to the central point \textit{i.e.}, the unitarity property within the MMST in the fashion of the SM. The procedure starts from the known expression of the flux density \cite{Malik99,RPA01}, which in accordance with the equation of motion (\ref{Eqmaestra}), reads
\begin{equation}
 \label{flujo1}
  j(z) = -i\left[\bn{A}(z)\t \cdot \bn{F}(z)-\bn{F}(z)\t \cdot \bn{A}(z)\right],
\end{equation}
\ni and can be conveniently modified with (\ref{Supervec-ffl}). Then, $ \forall z$ we can write
\begin{eqnarray}
 \label{flujo2}
  j & = & - i \; \bn{\Omega}\t \cdot \Jsf
  \;\cdot \bn{\Omega}\,, \\
 \nonumber
 \label{Jota}
 \Jsf & = & \left\Vert
                \begin{array}{cc}
                  \bn{O}_{\mb{\tiny N}} & - \bn{I}_{\mb{\tiny N}} \\
                  \bn{I}_{\mb{\tiny N}} &  \bn{O}_{\mb{\tiny N}}
                \end{array}
              \right\Vert\,.
\end{eqnarray}
It is important to emphasize, that in the $(2 \times 2)$ KL model, where $\bn{P}(z)$ is anti-Hermitian, expression (\ref{flujo1}) is reduced to \cite{Diago05}
\begin{equation}
 \label{flujo:KL(2x2)}
  j(z) = 2 \Im m \left[ \bn{F}(z)\hspace{0.3mm}'^{\dag}\bn{B}(z)\bn{F}(z) \right] -
       2\bn{F}(z)^{\dag}\bn{P}(z)^{\dag}\bn{F}(z)\,,
\end{equation}
\ni meanwhile in another interesting case like the $1D$ Schr\"{o}dinger equation, with $N = 1$ and for multi-channel $3D$ cases \cite{Borland61,Mello88,Pereyra98}, it reduces to the widely known expression
\begin{equation}
 \label{flujo:1D-Sch}
 j(z) = \bn{F}(z)^{\dag}\bn{F}(z)' - \bn{F}(z)\hspace{0.3mm}'^{\dag}\bn{F}(z)\,.
\end{equation}

\begin{figure}[ht!]
\centering
    \includegraphics[width=0.5\linewidth]{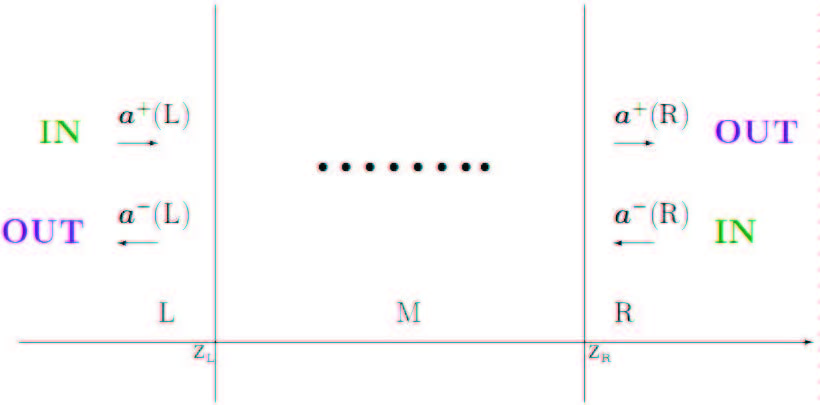}
    \caption{\label{Fig:PotDis1} General scheme of a scattering process in a typical layered system. Between layers ${\rm L}$ and ${\rm R}$ could be a single interface --\textit{i.e.}, ${\rm L}$ and ${\rm R}$ directly matched--, or any intermediate structure \cite{RPA04}.}
\end{figure}
Relations (\ref{flujo1}) and (\ref{flujo2}), are valid for layered systems with $N$-coupled components described by the EFA model. If we now use (\ref{Supervec-ffl2}), the relation (\ref{flujo2}) becomes
\begin{eqnarray}
 \nonumber
  & j = -i\; \Bigl(\bn{Q} \cdot \bn{\a}\Bigr)\t \cdot \Jsf \cdot
                 \Bigl(\bn{Q} \cdot \bn{\a}\Bigr) & \\
 \nonumber
  & \hspace{3mm} = -i \bn{\a}\t \cdot \bn{Q}\t \cdot \Jsf \cdot \;\bn{Q} \cdot \bn{\a}\,.& \\
 \nonumber
   & \mb{We define:}\hspace{2mm}  \bn{X} = \bn{Q}\t \cdot \Jsf\; \cdot \bn{Q}=
       \left\Vert
         \begin{array}{cc}
          \bn{X}_{++} &  \bn{X}_{+-} \\
          \bn{X}_{-+} &  \bn{X}_{--}
         \end{array}
       \right\Vert \,, & \;\; \mb{then,}\\
 \label{flujo3}
 & j =  -i \bn{\a}\t \cdot \bn{X} \cdot \bn{\a}\,. & \\
 \nonumber
 & \hspace{10mm} _{(1 \times 2N)(2N \times 2N)(2N \times 1)} &
\end{eqnarray}

Let\\
$\bn{\a}_{+} \Rightarrow$ the coefficients of the LI solutions that propagates form left to right. \\
$\bn{\a}_{-} \Rightarrow$ the coefficients of the LI solutions that propagates from right to left.  \\
$L \Rightarrow$ be the region at left of the scattering system (barrier). \\
$R \Rightarrow$ be the region at right of the scattering system (barrier).

For regions $L$ and $R$ [see Fig.\ref{Fig:PotDis1}], we can build
$$
 \bn{\a}^{\mb{\tiny L}}= \left\Vert
   \ba{c}
    \bn{\a}^{\mb{\tiny L}}_{+} \\
    \bn{\a}^{\mb{\tiny L}}_{-}
   \ea \right\Vert \;\;\;,\;\;\;
   \bn{\a}^{\mb{\tiny R}}= \left\Vert
   \ba{c}
    \bn{\a}^{\mb{\tiny R}}_{+} \\
    \bn{\a}^{\mb{\tiny R}}_{-}
   \ea \right\Vert.
$$
$\hspace{62mm} _{(2N \times 1)\hspace{26mm}(2N \times 1)}$\\
Then\footnote{To reduce the notation, henceforth we will omit in some cases the symbol ``$\; \bn{\cdot} \;$" in the matrix and vector products.}
\begin{eqnarray}
 \label{flujo3a}
 j^{\,n} = -i\left\{ \left(\bn{\a}^{n}_{+}\right)\t\, \bn{X}^{n}_{++}\,
                 \bn{\a}^{n}_{+}\;
        +\;
       \left(\bn{\a}^{n}_{+}\right)\t\, \bn{X}^{n}_{+-}\,  \bn{\a}^{n}_{-}\;
        +\;
        \left(\bn{\a}^{n}_{-}\right)\t\, \bn{X}^{n}_{-+}\, \bn{\a}^{n}_{+}\;
        + \left(\bn{\a}^{n}_{-}\right)\t\, \bn{X}^{n}_{--}\, \bn{\a}^{n}_{-}\right\},
\end{eqnarray}
$$
 \; \mb{para}\; n = L, R.
$$

\paragraph{\textsf{Flux conservation (FC)}$\,$:} For a scattering experiment, without boundary conditions or restrictions to the incident package, FC implies that:
\begin{defn}
 The number of particles at both sides of the obstacle is the same.
\end{defn}
This is why $j(z)$, evaluated at both sides of the scattered system is the same, whenever an elastic scattering process take place in the system, or rather
\begin{eqnarray}
 \label{CC1}
  \hspace{25mm} j^{\mb{\tiny L}} & = & \;j^{\mb{\tiny R}}  \\
 \nonumber
      & \Downarrow & \\
 \label{CC2}
  -i \Bigl(\bn{\a}^{\mb{\tiny L}}\Bigr)\t \bn{X}^{\mb{\tiny L}}
     \bn{\a}^{\mb{\tiny L}}
       & = &
 -i \Bigl(\bn{\a}^{\mb{\tiny R}}\Bigr)\t \bn{X}^{\mb{\tiny R}}
     \bn{\a}^{\mb{\tiny R}}\,.
\end{eqnarray}
\noindent This expression is a real scalar and will be used below in a similar representation, to extract several properties of the SM within the framework of the MMST.

\section{Structured unitarity of $\bn{S}$}
 \label{Cap:SM:::cuasi-uni}

Lets begin by recalling the standard definition of the SM $\bn{S}$ \cite{RPA04},
\begin{defn}
 \label{SM-def1}
\begin{eqnarray}
\begin{array}{||c||}
  \bn{a}^{-}({\rm L}) \\
  \bn{a}^{+}({\rm R})
 \end{array}_{\hspace*{1mm}\mb{\textit{out}}}
& = &
 \bn{S}(\mb{\textit{output}},\mb{\textit{input}}) \; \cdot \;
 \begin{array}{||c||}
  \bn{a}^+({\rm L}) \\
  \bn{a}^-({\rm R})
 \end{array}_{\hspace*{1mm}\mb{\textit{in}}} \;,
\end{eqnarray}
\end{defn}
\ni where the coefficients have been divided into two groups: those of the waves that travel from left to right and those that travel the other way around [see Fig.\ref{Fig:PotDis1}]. The first ones have been denoted by the supra-index ``+"; meanwhile to the others corresponds the supra-index ``-". Lately we develop a relation between the coefficients of the envelope function $\bn{F}(z)$ from (\ref{Eqmaestra}) --whose basis functions remains so far, free of any special condition--, in regions $L$ and $R$, with the incident and emergent functions of the scattering system under study [see Fig.\ref{Fig:PotDis1}]. Using the formalism of the SM within the MMST, we finally achieve a reliable representation, which contains the envisioned condition corresponding to the unitarity of the SM, in general, for EFA models and particularly for the KL hamiltonian. Now turn to define
\begin{eqnarray}
 \label{In-Out}
 \bn{\mathcal{I}}= \left\Vert
   \ba{c}
    \bn{\a}^{\mb{\tiny L}}_{+} \\
    \bn{\a}^{\mb{\tiny R}}_{-}
   \ea \right\Vert_{in} \;\;\;;\;\;\;
 \bn{\mathcal{O}}= \left\Vert
   \ba{c}
    \bn{\a}^{\mb{\tiny L}}_{-} \\
    \bn{\a}^{\mb{\tiny R}}_{+}
   \ea \right\Vert_{out}, \\
   \nonumber
   \hspace*{5mm}_{(2N \times 1)\hspace{24mm}(2N \times 1)}
\end{eqnarray}
\ni as the amplitude vectors of the incident and emerging propagating modes, respectively, that keep the following relation with $\bn{S}$
\begin{equation}
 \label{SM}
\bn{\mathcal{O}}_{out} =  \bn{S}\bn{\mathcal{I}}_{in}\,.
\end{equation}
We introduce the transformations
\begin{eqnarray}
 \left.{\hspace{1mm}}
  \begin{tabular}{ccc}
        $\bn{\mathcal{I}}_{in}$ & $\Longrightarrow $ & $\bn{\a}^{\mb{\tiny L}}$ \\
        $\bn{\mathcal{O}}_{out}$ & $\Longrightarrow $ & $\bn{\a}^{\mb{\tiny R}}$
  \end{tabular}
 \right\},
\end{eqnarray}
$$
 \left\Vert
   \begin{array}{c}
    \a^{\mb{\tiny L}}_{\mb{\tiny 1+}} \\
    \a^{\mb{\tiny L}}_{\mb{\tiny 2+}} \\
    \vdots \\
    \a^{\mb{\tiny L}}_{\mb{\tiny N+}} \\
    \a^{\mb{\tiny L}}_{\mb{\tiny 1\small -}} \\
    \a^{\mb{\tiny L}}_{\mb{\tiny 2\small -}} \\
    \vdots \\
    \a^{\mb{\tiny L}}_{\mb{\tiny N\small -}}
   \end{array}
  \right\Vert_{L}
   =
 \left\Vert
   \begin{array}{cccccccc}
    1 & 0 & \ldots & 0 & 0 & 0 & \ldots & 0 \\
    0 & 1 & \ldots & 0 & 0 & 0 & \ldots & 0 \\
    \vdots & \vdots & \ddots & \vdots & \vdots & \vdots & \ddots & \vdots \\
    0 & 0 & \ldots & 1 & 0 & 0 & \ldots & 0 \\
    0 & 0 & \ldots & 0 & \frac{\a^{\mb{\tiny L}}_{1-}}{\a^{\mb{\tiny R}}_{1-}} & 0 &
      \ldots & 0 \\
    0 & 0 & \ldots & 0 & 0 & \frac{\a^{\mb{\tiny L}}_{2-}}{\a^{\mb{\tiny R}}_{2-}} &
      \ldots & 0 \\
    \vdots & \vdots & \ddots & \vdots & \vdots & \vdots & \ddots & \vdots \\
    0 & 0 & \ldots & 0 & 0 & 0 & \ldots & \frac{\a^{\mb{\tiny L}}_{N-}}{\a^{\mb{\tiny R}}_{N-}}
   \end{array}
  \right\Vert
  \left\Vert
   \begin{array}{c}
    \a^{\mb{\tiny L}}_{\mb{\tiny 1+}} \\
    \a^{\mb{\tiny L}}_{\mb{\tiny 2+}} \\
    \vdots \\
    \a^{\mb{\tiny L}}_{\mb{\tiny N+}} \\
    \a^{\mb{\tiny R}}_{\mb{\tiny 1\small -}} \\
    \a^{\mb{\tiny R}}_{\mb{\tiny 2\small -}} \\
    \vdots \\
    \a^{\mb{\tiny R}}_{\mb{\tiny N\small -}}
   \end{array}
  \right\Vert_{in},
$$
\ni and hence
$$
 \left\Vert
   \ba{c}
    \a^{\mb{\tiny R}}_{\mb{\tiny 1+}} \\
    \a^{\mb{\tiny R}}_{\mb{\tiny 2+}} \\
    \vdots \\
    \a^{\mb{\tiny R}}_{\mb{\tiny N+}} \\
    \a^{\mb{\tiny R}}_{\mb{\tiny 1\small -}} \\
    \a^{\mb{\tiny R}}_{\mb{\tiny 2\small -}} \\
    \vdots \\
    \a^{\mb{\tiny R}}_{\mb{\tiny N\small -}}
   \ea \right\Vert_{R}
   =
 \left\Vert
   \ba{cccccccc}
    0 & 0 & \ldots & 0 & 1 & 0 & \ldots & 0 \\
    0 & 0 & \ldots & 0 & 0 & 1 & \ldots & 0 \\
    \vdots & \vdots & \ddots & \vdots & \vdots & \vdots & \ddots & \vdots \\
    0 & 0 & \ldots & 0 & 0 & 0 & \ldots & 1 \\
    \frac{\a^{\mb{\tiny R}}_{1-}}{\a^{\mb{\tiny L}}_{1-}} & 0 & \ldots & 0 & 0 & 0 &
     \ldots & 0 \\
    0 & \frac{\a^{\mb{\tiny R}}_{2-}}{\a^{\mb{\tiny L}}_{2-}} & \ldots & 0 & 0 & 0 &
     \ldots & 0 \\
    \vdots & \vdots & \ddots & \vdots & \vdots & \vdots & \ddots & \vdots \\
    0 & 0 & \ldots & \frac{\a^{\mb{\tiny R}}_{N-}}{\a^{\mb{\tiny L}}_{N-}} & 0 & 0 &
     \ldots & 0
   \ea \right\Vert
   \left\Vert
   \ba{c}
    \a^{\mb{\tiny L}}_{\mb{\tiny 1\small -}} \\
    \a^{\mb{\tiny L}}_{\mb{\tiny 2\small -}} \\
    \vdots \\
    \a^{\mb{\tiny L}}_{\mb{\tiny N\small -}} \\
    \a^{\mb{\tiny R}}_{\mb{\tiny 1+}} \\
    \a^{\mb{\tiny R}}_{\mb{\tiny 2+}} \\
    \vdots \\
    \a^{\mb{\tiny R}}_{\mb{\tiny N+}}
   \ea \right\Vert_{out}.
$$
$\mbox{\hspace{35 mm}}_{(2N \times 1)\hspace{32 mm}}(2N \times 2N)\hspace{25 mm}(2N \times 1)$

We take
\begin{eqnarray*}
 \Csf =
  \left\Vert
   \begin{array}{cccc}
    \frac{\a^{\mb{\tiny L}}_{1-}}{\a^{\mb{\tiny R}}_{1-}} & 0 & \ldots & 0  \\
    0 & \frac{\a^{\mb{\tiny L}}_{2-}}{\a^{\mb{\tiny R}}_{2-}} & \ldots & 0 \\
    \vdots & \vdots & \ddots & \vdots \\
    0 & 0 & \ldots & \frac{\a^{\mb{\tiny L}}_{N-}}{\a^{\mb{\tiny R}}_{N-}}
   \end{array}
  \right\Vert, \\
\end{eqnarray*}
$\hspace{59mm} _{(N \times N)}$\\

\ni which we can write
\begin{eqnarray}
 \label{In-L}
  \left\Vert
   \begin{array}{c}
    \bn{\a}^{\mb{\tiny L}}_{+} \\
    \bn{\a}^{\mb{\tiny L}}_{-}
   \end{array} \right\Vert_{L}
    & = &
  \left\Vert
   \begin{array}{cc}
    \bn{I}_{\mb{\tiny N}} & \bn{O}_{\mb{\tiny N}} \\
    \bn{O}_{\mb{\tiny N}} & \Csf
   \end{array}
  \right\Vert
  \left\Vert
    \begin{array}{c}
     \bn{\a}^{\mb{\tiny L}}_{+} \\
     \bn{\a}^{\mb{\tiny R}}_{-}
    \end{array}
  \right\Vert_{in} \\
  \nonumber
  \\
  \label{In-R}
  \left\Vert
   \begin{array}{c}
    \bn{\a}^{\mb{\tiny R}}_{+} \\
    \bn{\a}^{\mb{\tiny R}}_{-}
   \end{array}
  \right\Vert_{R}
    & = &
  \left\Vert
   \begin{array}{cc}
     \bn{O}_{\mb{\tiny N}} & \bn{I}_{\mb{\tiny N}} \\
     \Csf^{-1} & \bn{O}_{\mb{\tiny N}}
   \end{array}
  \right\Vert
   \left\Vert
    \begin{array}{c}
     \bn{\a}^{\mb{\tiny L}}_{-} \\
     \bn{\a}^{\mb{\tiny R}}_{+}
    \end{array}
   \right\Vert_{out}.
\end{eqnarray}

Now, if we define
$$
 \bn{\Pi}=
  \left\Vert
   \begin{array}{cc}
    \bn{I}_{\mb{\tiny N}} & \bn{O}_{\mb{\tiny N}} \\
    \bn{O}_{\mb{\tiny N}} & \Csf
   \end{array}
  \right\Vert,
$$
\ni then it is simple to express
$$
 \bn{\Pi}^{-1}\bn{J}_{x}=
  \left\Vert
   \begin{array}{cc}
    \bn{O}_{\mb{\tiny N}} &  \bn{I}_{\mb{\tiny N}} \\
    \Csf^{-1} & \bn{O}_{\mb{\tiny N}}
   \end{array}
   \right\Vert,
$$
\ni being
$$
 \bn{J}_{x}=
 \left\Vert
   \begin{array}{cc}
    \bn{O}_{\mb{\tiny N}} & \bn{I}_{\mb{\tiny N}} \\
    \bn{I}_{\mb{\tiny N}} & \bn{O}_{\mb{\tiny N}}
  \end{array}
 \right\Vert,
$$
\ni and then we can rewrite (\ref{In-L}) and (\ref{In-R}) as
\begin{eqnarray}
 \left.{\hspace{1mm}}
  \begin{tabular}{ccc}
        $\bn{\a}^{\mb{\tiny L}}$ & $=$ & $\bn{\Pi}\bn{I}_{in}$ \\
        $\bn{\a}^{\mb{\tiny R}}$ & $=$ & $\bn{\Pi}^{-1}\bn{J}_{x}
         \bn{\mathcal{O}}_{out}$
  \end{tabular}
 \right\}.
\end{eqnarray}

Note that it is fulfilled
\begin{equation}
 \label{flux-Jx}
  \bn{J}_{x} = -\Jsf\; \bn{\Sigma}_{z} \,,
\end{equation}
\ni where $\bn{\Sigma}_{z} $ is the generalized Pauli matrix $\bn{\sigma}_{z}$ of $(2N \times 2N)$.

\subsection{EFA general case: $N \geq 2$}
 \label{Cap:SM::PseudoU:::EFA}

To deal with the physical problem posted in the Sec.\ref{Cap:Tunel::Intro}, the system is divided into three regions as $L, M, R$ [see the Figure on \ref{Fig:PotDis1}]. The external regions $L$ and $R$ are supposed to have constant parameters and in these slabs, the states of the system have eigenvalues (energy, \textit{momentum}) which are constants in principle. The region $M$ of the system might be conformed by different layers of different materials or by a single material with $z$-dependent composition. Getting back to the FC condition (\ref{CC1}), it is simple to put
\begin{eqnarray}
 \nonumber
  \Bigl(\bn{\Pi}\bn{\mathcal{I}}_{in}\Bigr)\t \bn{X}^{\mb{\tiny L}}\;
  \bn{\Pi}\; \bn{\mathcal{I}}_{in}
 & = &
  \Bigl(\bn{\Pi}^{-1}\bn{J}_{x}\bn{\mathcal{O}}_{out}\Bigr)\t
   \bn{X}^{\mb{\tiny R}}\;
  \bn{\Pi}^{-1}\; \bn{J}_{x}\bn{\mathcal{O}}_{out}, \\
 \label{CC3}
  \bn{\mathcal{I}}_{in}\t\; \bn{\Pi}\t\; \bn{X}^{\mb{\tiny L}}\; \bn{\Pi}\;
   \bn{\mathcal{I}}_{in}
 & = &
  \bn{\mathcal{O}}_{out}\t\; \bn{J}_{x}\t\; \Bigl(\bn{\Pi}^{-1}\Bigr)\t\;
  \bn{X}^{\mb{\tiny R}}\; \bn{\Pi}^{-1}\; \bn{J}_{x}\;
  \bn{\mathcal{O}}_{out}.
\end{eqnarray}

From (\ref{SM}), it is straightforward to state
\begin{equation}
 \label{SM-def2}
  \bn{\mathcal{O}}_{out}\t =  \bn{\mathcal{I}}_{in}\t \bn{S}\t,
\end{equation}
\ni thereby
\begin{eqnarray}
 \nonumber
  \hspace{10mm}\bn{\mathcal{I}}_{in}\t\; \bn{\Pi}\t\; \bn{X}^{\mb{\tiny L}}\;
  \bn{\Pi}\;\bn{\mathcal{I}}_{in}
 & = &
  \bn{\mathcal{I}}_{in}\t\; \bn{S}\t\; \bn{J}_{x}\t\;
  \Bigl(\bn{\Pi}^{-1}\Bigr)\t\;
  \bn{X}^{\mb{\tiny R}}\; \bn{\Pi}^{-1}\; \bn{J}_{x}\; \bn{S}\;
  \bn{\mathcal{I}}_{in}, \\
 \nonumber
 & \Downarrow & \\
  \hspace{24mm}\bn{\Pi}\t\; \bn{X}^{\mb{\tiny L}}\; \bn{\Pi}
 & = &
 \bn{S}\t\; \bn{J}_{x}\t\;
  \Bigl(\bn{\Pi}^{-1}\Bigr)\t\;
  \bn{X}^{\mb{\tiny R}}\; \bn{\Pi}^{-1}\; \bn{J}_{x}\; \bn{S}\,.
\end{eqnarray}
\ni Let
$$
 \bn{\Pi}_{\mb{\tiny L}} = \bn{\Pi}\t\; \bn{X}^{\mb{\tiny L}}\; \bn{\Pi}
$$
$$
 \bn{\Pi}_{\mb{\tiny R}} = \Bigl(\bn{\Pi}^{-1}\;\bn{J}_{x}\Bigr)\t\;
  \bn{X}^{\mb{\tiny R}}\;\Bigl(\bn{\Pi}^{-1}\; \bn{J}_{x}\Bigr)
   =
 \bn{J}_{x}\;  \Bigl(\bn{\Pi}^{-1}\Bigr)\t\;
  \bn{X}^{\mb{\tiny R}}\; \bn{\Pi}^{-1}\; \bn{J}_{x}\,.
$$
\begin{propo}
 \label{Propo:pseudo-uni}
Thereafter the structured unitarity on the SM $\bn{S}$ within the MMST, for non-unitary fluxes of coupled particles --including those away from the scattering center--, can be represented as
\begin{equation}
 \label{cuasi-uni1}
 \bn{S}\t \bn{\Pi}_{\mb{\tiny R}} \bn{S} = \bn{\Pi}_{\mb{\tiny L}}\,.
\end{equation}
\end{propo}

It is worthy to remark that the crucial proposition (\ref{cuasi-uni1}), is a general property that must fulfil the SM in the framework of the MMST, just under the only condition of equal fluxes (\textit{i.e.} elastic processes). We stress that we are dealing so far with an arbitrary LI basis set.
\begin{defn}
 \label{Def:Ortnonor-Full}
  We will consider that a basis set of linearly independent functions is \underline{completely orthonormalized}, when it is fully orthonormalized in both the configuration and the spinorial spaces simultaneously. Otherwise, the basis set of linearly independent functions will be named arbitrary or \underline{incomplete-orthonormal}.
 \end{defn}

In a sense, the proposition (\ref{cuasi-uni1}) can be thought of as structured unitarity for  $\bn{S}$, following the classification of structured matrices  given by D. Steven MacKey, N. MacKey and Francoise Tisseur \cite{Tisseur03}. The physical meaning of the matrices $\bn{\Pi}_{\mb{\tiny R, L}}$ is given in $\bn{X}^{\mb{\tiny R, L}}$, whose diagonal elements correspond, in general, to the coupling-free quantum transport at the asymptotic regions through the allowed channels of the system. This later behavior, do not exclude the interference due to the scattering potential effects. The off-diagonal elements in this matrix correspond, in general, to the interplay between the incident (emerging) modes at the asymptotic regions, far away from the zone where the scatterer obstacle is located and also have information on the mixing of modes. Later on we will see more details on this fact.


\subsection{EFA particular case: $N = 4$}
\label{Cap:SM::PseudoU:::KL}

In the seminal reference \cite{Sanchez95}: A. D. S\'{a}nchez and C. R. Proetto have analyzed the symmetry properties of the SM, for the scattering problem of pure (mixing-free) states in the VB (heavy holes and light holes). They work with a particular representation of the KL model, where the total angular momentum components and spin of the hole staes, remain embedded in the canonical transformation of the basis\footnote{In this section and in others, where we refer to the report \cite{Sanchez95}, we will use the labelling of the authors. For instance: heavy holes (H) and light holes (L). This allows a simple identification of our results form theirs.}. Defining

\begin{figure}[t]
 \begin{center}
 \includegraphics[width=3.5in,height=3.5in]{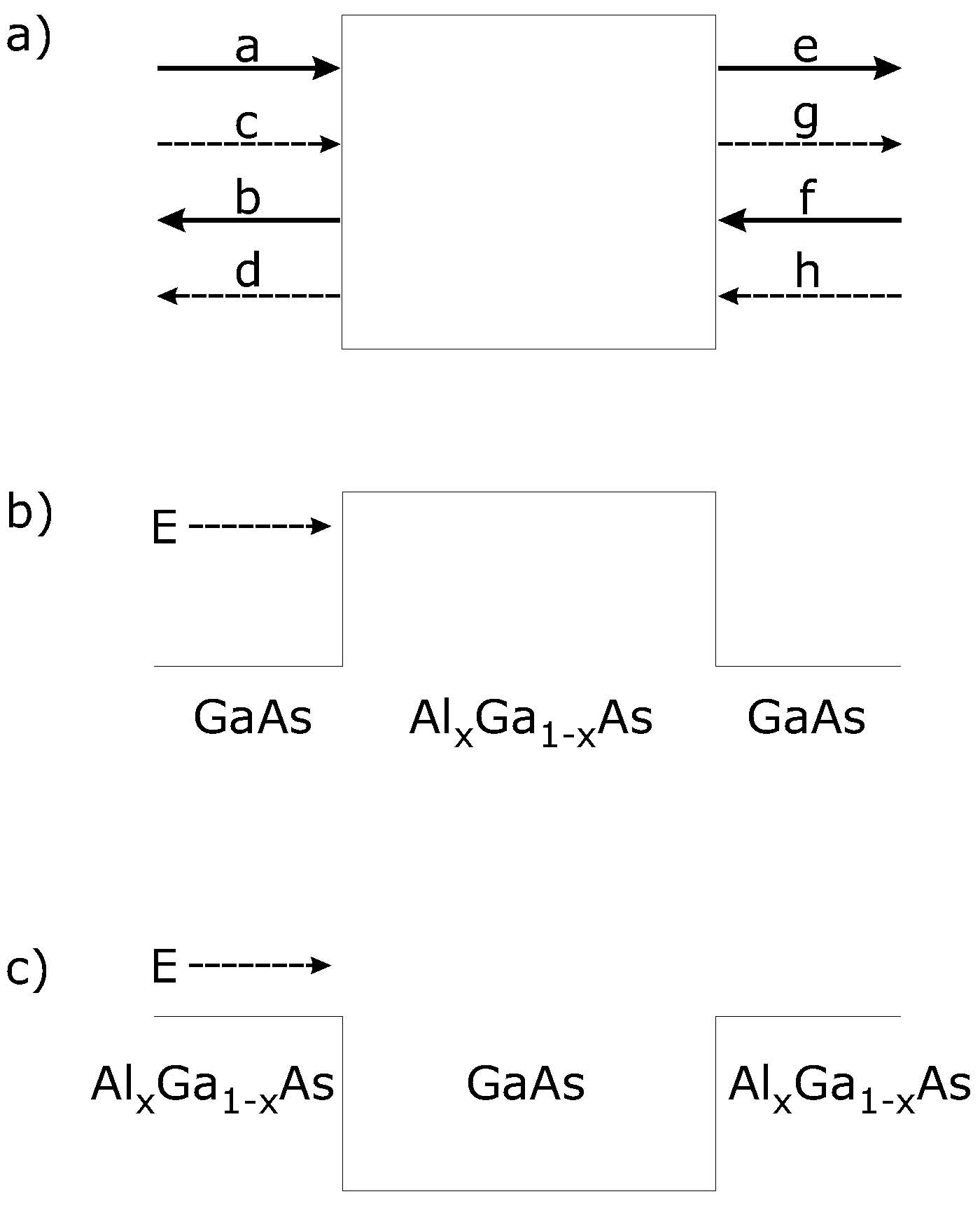}\\
 \caption{\label{Fig:Potdis2} (a) Schematic representation of the quantum scattering problem from the SM viewpoint as considered in reference \cite{Sanchez95}. The arrows stand for incoming (outgoing) probability amplitudes of the propagating modes. (b) Scatterer as a quantum barrier. (c) Scatterer as a quantum well \cite{Sanchez95}}
 \end{center}
\end{figure}

$$
 \mathbf{J}=
  \left(
   \begin{array}{cccc}
    j_{\mb{\tiny H}} & 0 & 0 & 0  \\
    0 & j_{\mb{\tiny L}} & 0 & 0 \\
    0 & 0 & j_{\mb{\tiny H}} & 0 \\
    0 & 0 & 0 & j_{\mb{\tiny L}}
   \end{array}
  \right),\;\;\;
   \mathbf{O} =
  \left(
   \begin{array}{c}
     b \\
     d \\
     e \\
     g
   \end{array}
  \right),\;\;\;
  \mathbf{I} =
  \left(
   \begin{array}{c}
     a \\
     c \\
     f \\
     h
   \end{array}
  \right)\,\,,
$$
\ni where $a-h$ represent the propagating modes amplitudes [see Fig.\ref{Fig:Potdis2}], the condition of FC turns
\begin{equation}
 \label{flujo5}
 \mathbf{O}\t\, \mathbf{J}\, \mathbf{O} = \mathbf{I}\t\, \mathbf{J}\, \mathbf{I}\,.
\end{equation}
\ni Taking into account the condition (\ref{SM}), together with the Hermitian conjugate and substituting into (\ref{flujo5}), S\'{a}nchez and Proetto have obtained as final result \cite{Sanchez95}
\begin{equation}
 \label{cuasi-uni2}
 \bn{\mathcal{S}}\t\, \mathbf{J}\, \bn{\mathcal{S}} = \mathbf{J}\,.
\end{equation}
\ni This is what they denominate pseudo-unitarity and also have been considered as a generalization of the unitary condition for electrons' probability current conservation within the EMA and given by $\bn{S}\t = \bn{S}^{-1}$.


\subsection{Reduction from the structured case to the generalized one.}
\label{Cap:SM::PseudoU:::EFAvsKL}

Given the formulations (\ref{cuasi-uni1}) and (\ref{cuasi-uni2}) presented above, we now will analyze the requirements to transform one into anther. The idea is simple; we are be looking at the circumstances that satisfy:
\begin{eqnarray}
 \label{cuasi-12-NS}
 \left.{\hspace{1mm}}
  \begin{tabular}{c}
        $\bn{\Pi}_{\mb{\tiny R}} = \mathbf{J} $ \\
        $\bn{\Pi}_{\mb{\tiny L}} = \mathbf{J} $
  \end{tabular}
 \right\}\,,
\end{eqnarray}
\ni and then
\begin{equation}
 \label{cuasi-12-N}
 \bn{\Pi}_{\mb{\tiny R}} = \bn{\Pi}_{\mb{\tiny L}}\,.
\end{equation}
Relation (\ref{cuasi-12-NS}) is the \emph{necessary} and \emph{sufficient} condition to perform $  \bn{S}\t \bn{\Pi}_{\mb{\tiny R}} \bn{S} = \bn{\Pi}_{\mb{\tiny L}} \Longrightarrow \bn{S}\t\, \mathbf{J}\, \bn{S} = \mathbf{J}\,.$ Meanwhile, (\ref{cuasi-12-N}) is only a \emph{necessary} condition, and we will discuss it now on. On one hand, we will consider the structure of $\bn{\Pi}_{\mb{\tiny R}}$, $\bn{\Pi}_{\mb{\tiny
L}}$ and the diagonal form of $\Csf$. While on the other hand, we took into account the diagonal character of $\mathbf{J}$. Thereby, these facts lead to derive \begin{eqnarray}
  (\Csf^{*})^{-1}\,\bn{X}^{\mb{\tiny R}}_{--}\,\Csf^{-1}
 = \bn{X}^{\mb{\tiny L}}_{++}\,, \\
 \nonumber
  \Csf^{*}\,\bn{X}^{\mb{\tiny L}}_{--}\,\Csf
 = \bn{X}^{\mb{\tiny R}}_{++}\,.
\end{eqnarray}

If these relations are guaranteed, then the condition (\ref{cuasi-12-N}) can be fulfilled. Furthermore, if one considers that the parameters of equation (\ref{Eqmaestra}) are constant by-layer and following the general form (\ref{flujo3}) for the matrix $\bn{X}$, then it is possible to find the equality of its blocks, which can be cast as
$$
 \bn{X}^{\mb{\tiny R}}_{\jmath} =
 \bn{X}^{\mb{\tiny L}}_{\jmath}\,, \hspace{1cm} \mb{where} \hspace{5mm}
 \jmath = ++,\, +-,\, -+,\, -- \;.
$$
\ni Departing from this, it is not difficult to demonstrate that $\Csf^{-1} = \Csf$, yielding $\Csf = \bn{I}_{\mb{\tiny N}}$, which is an alternate path that leads finally to the \emph{necessary} condition (\ref{cuasi-12-N}). Concrete physically observable (or theoretically predicted) situations for $\Csf$ to becomes the identity matrix within the MMST, are yet to be defined. However, so far the essential point is that the SM in the EFA framework, has a structured unitarity upon an arbitrary LI basis set, which differs from standardized unitarity within EMA problems.

\section{Convergence from EFA to EMA: flux and unitarity}
 \label{Cap:SM:::Converge}

Below, we will look at how the formulations within the MMST for flux equation and the structured unitarity requirement, converge to those of the EMA representation. The clue idea for such transformation, involves mainly working with the character of the $N$-component flux of coupled propagating modes.  There are several ways to deal with that convergence, namely: (i) Start from coupled emerging/incident $N$-component states (EFA framework) and pass to the limit of uncoupled emerging/incident $N$-component states (EMA framework). (ii) Start from coupled emerging/incident $N$-component states build over an incomplete-orthonormalized basis set and perform the complete orthonormalization procedure on the LI \emph{eigen}-functions. Next, we present the convergence criteria for each case.

\subsection{Limit of uncoupled $N$-component flux}
 \label{Cap:SM::Converge:::N-desacopla}

If we pursue this limit further, we will be able to recover the standard unitary condition of $\bn{S}$ (EMA framework). There are three alternatives by which this can be done. Firstly, it is possible to convert the expression of FC in the EFA to that in the EMA. From this, it is straightforward to demonstrate the expected unitarity. Secondly, one can take the structured-unitary condition (\ref{cuasi-uni1}) and derive the standard unitary of $\bn{S}$. As the third and last way, we start from the particular generalized-unitary condition (\ref{cuasi-uni2}) and transfer it into the standardized-unitary of the SM.

\subsubsection{Flux Convergence}
 \label{Cap:SM::Converge:::Psedo-Unitar:Flujos}

To illustrate the first alternative, we will consider that the flux $j_{\mb{\tiny \textbf{EFA}}} = -i \Bigl(\bn{\a}\Bigr)\t\bn{X} \bn{\a},$ corresponds to the general EFA case, as was seen in the Sec. \ref{Cap:SM::Tun-Flujo}. There are not explicit specific requirements to the basis of LI functions. We have to demonstrate that:
\begin{thm}
 \label{Teo:jEFA-jEMA}
\begin{eqnarray}
  \nonumber
  \vartriangle
 \\
 \lim_{modes(c) \rightarrow \, modos(u)}
  \mb{\LARGE$j_{\mb{\tiny \textbf{EFA}}} \equiv j_{\mb{\tiny \textbf{EMA}}}$ }\,,
\end{eqnarray}
\end{thm}
\ni by $\,(c,\,u)\,$ modes we understand $\,(coupled,\,uncoupled)$ modes, respectively. P.A.Mello, P. Pereyra and N. Kumar \cite{Mello88}, have shown that for a physical system described by $N$ uncoupled differential equations --corresponding to the general EMA case--, the probability current density is given by
\begin{equation}
 \label{j-EMA}
  j_{\mb{\tiny \textbf{EMA}}} = \aa\t \bn{\Sigma}_{z} \aa \,,
\end{equation}
\ni where
$$
 \bn{\Sigma}_{z} =
  \left\Vert
   \ba{cc} \bn{I}_{\mb{\tiny N}} & \bn{O}_{\mb{\tiny N}} \\
    \bn{O}_{\mb{\tiny N}} & -\bn{I}_{\mb{\tiny N}}
   \ea \right\Vert\,,
$$
\ni is the generalized Pauli matrix $\sigma_{z}$ and $\aa$ is a matrix with the coefficients of the linear combination of the solutions taken as plane waves. The study is made as if it were a scattering from a certain region $L$ [see Fig.\ref{Fig:PotDis1}], where the interaction of modes is unplugged, to a region $R$ where also the interaction of modes remains unplugged. Considering the modes as independent at the electrodes $L$ and $R$ implies that
\begin{eqnarray*}
  \bn{X} = \bn{Q}\t \bn{J}\;\bn{Q} =
   \left\Vert
    \ba{cc}
     \bn{X}_{++} &  \bn{X}_{+-} \\
     \bn{X}_{-+} &  \bn{X}_{--}
    \ea
   \right\Vert
    =
   \left\Vert
    \ba{ccc}
     \bn{Q}_{22}\bn{Q}_{11}-\bn{Q}_{12}\bn{Q}_{21} &\;\;\;& \bn{O}_{\mb{\tiny N}} \\
     \bn{O}_{\mb{\tiny N}} &\;\;\;& -\Bigl[\bn{Q}_{22}\bn{Q}_{11}-\bn{Q}_{12}\bn{Q}_{21}\Bigr]
    \ea
   \right\Vert \,.
\end{eqnarray*}
\ni Note that the crossed blocks satisfied: $\bn{X}_{+-} = \bn{X}_{-+}= \bn{O}_{\mb{\tiny N}}$ due to the lack of interaction of modes, then the propagating modes in one direction and in the opposite are independent. In general the vectors $\bn{\a} \,$ and $\,\aa \,$ are different, although in this analysis it was assumed that they fulfil $\a_{j} = \alpha_{j} \mathrm{\mb{\Large a}}_{j} \,$ being $\alpha_{j}$ some proportionality coefficients. If the normalization of the LI functions is taken as
  $$
   2\alpha_{j}\,k_{j}|b_j|^2=1 \Longrightarrow |b_j|^2=\frac{1}{2\alpha_{j}\,k_j}; \;
   \forall j=1,...,N\,.
  $$
\ni then, it is possible to get
\begin{equation}
 \bn{Q}_{22}\bn{Q}_{11}-\bn{Q}_{12}
 \bn{Q}_{21}=-i\bn{I}_{\mb{\tiny N}} \;,
\end{equation}
\ni and hence
\begin{equation}
 \label{flujo(EFA-EMA)}
 j_{\mb{\tiny \textbf{EFA}}}\Bigl|_{(d)} \Bigr. = -\aa\t
   \left\{-(i)^2 \left\Vert \ba{cc}
     \bn{I}_{\mb{\tiny N}}  & \bn{O}_{\mb{\tiny N}} \\
     \bn{O}_{\mb{\tiny N}} & -\bn{I}_{\mb{\tiny N}}
   \ea \right\Vert \right\} \aa = -\aa\t \bn{\Sigma}_{z} \aa \,.
\end{equation}
\ni Finally, with the accuracy of a phase, we obtained
\begin{eqnarray}
 \label{EFA-EMA}
 j_{\mb{\tiny \textbf{EFA}}}\Bigl|_{(d)} \Bigr. \equiv j_{\mb{\tiny \textbf{EMA}}}\,,
 \nonumber
 \\
 \hspace{2cm} \blacktriangle
 \end{eqnarray}
\ni when it is considered in the limit, that in the left-hand side member, the modes are uncoupled at the asymptotic regions, which is what we wanted to to demonstrate. Once the demonstration (\ref{Teo:jEFA-jEMA}) have been derived, it is useful to verify one of its main consequences. From (\ref{flujo(EFA-EMA)}), one can get back the standard properties for the unitarity of the SM. To do so we will take, for simplicity, the case $N=1$ and use the convention adopted in Sec. \ref{Cap:SM::Tun-Flujo}. Afterwards the FC (\ref{CC1})-(\ref{CC2}) at the asymptotic regions $L,\,R\;$ are given by
\begin{eqnarray}
 \label{CC5}
  \hspace{1cm}\left\Vert
   \ba{c}
    \asa^{\mb{\tiny L}}_{+} \\
    \asa^{\mb{\tiny L}}_{-}
   \ea \right\Vert\t
    \left\Vert
   \ba{cc}
    1 & 0 \\
    0 & -1
   \ea \right\Vert
   \left\Vert
   \ba{c}
    \asa^{\mb{\tiny L}}_{+} \\
    \asa^{\mb{\tiny L}}_{-}
   \ea \right\Vert
  & = &
  \hspace{3mm}\left\Vert
   \ba{c}
    \asa^{\mb{\tiny R}}_{+} \\
    \asa^{\mb{\tiny R}}_{-}
   \ea \right\Vert\t
    \left\Vert
   \ba{cc}
    1 & 0 \\
    0 & -1
   \ea \right\Vert
   \left\Vert
   \ba{c}
    \asa^{\mb{\tiny R}}_{+} \\
    \asa^{\mb{\tiny R}}_{-}
   \ea \right\Vert
  \\
 \nonumber
  \hspace{1.2cm}\left\Vert
   \ba{cc}
    (\asa^{\mb{\tiny L}}_{+})^{*} \;\; (-\asa^{\mb{\tiny L}}_{-})^{*}
   \ea \right\Vert
    \left\Vert
   \ba{c}
    \asa^{\mb{\tiny L}}_{+} \\
    \asa^{\mb{\tiny L}}_{-}
   \ea \right\Vert
 & = &
   \hspace{3mm}\left\Vert
   \ba{cc}
    (\asa^{\mb{\tiny R}}_{+})^{*} \;\; (-\asa^{\mb{\tiny R}}_{-})^{*}
   \ea \right\Vert
    \left\Vert
   \ba{c}
    \asa^{\mb{\tiny R}}_{+} \\
    \asa^{\mb{\tiny R}}_{-}
   \ea \right\Vert\,,
  \\
 \nonumber
      & \Downarrow &
 \\
 \nonumber
 \hspace{4cm}|\asa^{\mb{\tiny L}}_{+}|^{2} - |\asa^{\mb{\tiny L}}_{-}|^{2}
 & = &
 \hspace{2mm}|\asa^{\mb{\tiny R}}_{+}|^{2} - |\asa^{\mb{\tiny R}}_{-}|^{2}\,,
 \\
 \nonumber
 \mb{when regroup the terms, it is found}
 \\
 \nonumber
 \hspace{4cm}|\asa^{\mb{\tiny L}}_{+}|^{2} + |\asa^{\mb{\tiny R}}_{-}|^{2}
 & = &
  \hspace{2mm}|\asa^{\mb{\tiny L}}_{-}|^{2} + |\asa^{\mb{\tiny R}}_{+}|^{2}\,.
\end{eqnarray}

Using the definitions (\ref{In-Out}), it is possible to rewrite the last expression into a matrix form, that is to say
\begin{equation}
 \label{CC6}
 (\bn{\mathcal{I}}_{in})\t\,\bn{\mathcal{I}}_{in}
   =
 (\bn{\mathcal{O}}_{out})\t\,\bn{\mathcal{O}}_{out},
\end{equation}
\ni now, if we use the formalism of the SM expressed in (\ref{SM}) and its Hermitian conjugated, the right-hand side of the last identity can be written as
$$
 (\bn{\mathcal{I}}_{in})\t\,\bn{\mathcal{I}}_{in}
   =
 (\bn{\mathcal{I}}_{in})\t\,\bn{S}\t\,\bn{S} \,\bn{\mathcal{I}}_{in}\,.
$$
\ni Moreover, it can be finally extracted the familiar unitarity property we are be looking for
\begin{equation}
 \label{Unitar}
 \bn{S}\,\bn{S}\t = \bn{S}\t\,\bn{S}=
  \bn{I}_{\mb{\tiny 2N}}\,.
\end{equation}
\ni From (\ref{Unitar}) is straightforward
$$
 \bn{S}\t = \bn{S}^{-1},
$$
\ni and if the scattering system possesses the time reversal invariance (TRI) symmetry, subsequently it satisfies $\bn{S}^{*} = \bn{S}^{-1}$, leading to
$$
 \bn{S}\t = \bn{S}^{*}.
$$
\ni Yet derived this last, next one find the complex conjugated and we end up reaching another interesting property: the SM is symmetric, which means
\begin{equation}
 \label{Simet}
 \bn{S}^{T} = \bn{S}\,.
\end{equation}

In short words, doing this leads the structured-unitary condition (\ref{cuasi-uni1}) for EFA models (with coupled modes) goes correctly to the familiar properties of unitarity and symmetry valid for EMA models (with uncoupled modes)), \emph{via} the convergence between the corresponding fluxes.

\subsubsection{Reduction of the structured-unitarity: $N \geq 2$}
 \label{Cap:SM::Converge:::Psedo-Unitar:EFA}

Next we try the second alternative posted above. In what follows we show directly how to pass from the structured-unitary condition (\ref{cuasi-uni1}) for the MMST (EFA framework), to the usual property of standardized-unitary condition (\ref{Unitar}). Some algebraic manipulations are required here on the normalization of the involved coefficients. Besides, we modify the matrices $\bn{\Pi}_{\mb{\tiny R}}$ and $\bn{\Pi}_{\mb{\tiny L}}$, by noting that we could then write in the form
$$
 \bn{\Pi}_{\mb{\tiny L}} = \ae_{\mb{\tiny L}}\,\bn{\Pi}\t\; \bn{X}\; \bn{\Pi}
$$
$$
 \bn{\Pi}_{\mb{\tiny R}} = \ae_{\mb{\tiny R}}\,\bn{J}_{x}\;
  \Bigl(\bn{\Pi}^{-1}\Bigr)\t\; \bn{X}\; \bn{\Pi}^{-1}\; \bn{J}_{x}\,
$$
\ni where
$$
 \ae_{\mb{\tiny L}} = \left\Vert
         \ba{cc}
          \bn{I}_{\mb{\tiny N}} & \bn{O}_{\mb{\tiny N}} \\
          \bn{O}_{\mb{\tiny N}} &  \e^{i\theta}\,\bn{I}_{\mb{\tiny N}}
         \ea \right\Vert
 \;\;\;\; \mb{and} \;\;\;\;\;\;
  \ae_{\mb{\tiny R}} = \left\Vert
         \ba{cc}
          \e^{i\theta}\,\bn{I}_{\mb{\tiny N}} & \bn{O}_{\mb{\tiny N}} \\
          \bn{O}_{\mb{\tiny N}} &  \bn{I}_{\mb{\tiny N}}
         \ea \right\Vert\,,
$$
\ni since the phase factor satisfies
\begin{eqnarray}
 \label{fixfac}
 \theta =
 \left\{
  \begin{tabular}{cc}
        $ 0 \;;$ & for \textit{coupled modes} \\
        $ \pi \;; $ & for \textit{uncoupled modes}
  \end{tabular}
\right. \,.
\end{eqnarray}

When one uncouples the propagating modes at regions $L$ and $R$ [see Fig.\ref{Fig:PotDis1}], it was demonstrated in the Subsec.\ref{Cap:SM::Converge:::N-desacopla}, that $\bn{X}= -i\,\bn{\Sigma}_{z}$. If make no difference what propagation direction we choose for the state vectors, the coefficients for the LI solutions of (\ref{Eqmaestra}) become complex at the asymptotic zones. If the coefficients of $\;\Csf\;$ are selected as
  $$
   \frac{|\a^{\mb{\tiny L}}_{j-}|^{2}}{|\a^{\mb{\tiny R}}_{j-}|^{2}}=1; \;\; \forall\, j=
   1,\ldots,N \Longrightarrow |\a^{\mb{\tiny L}}_{j-}|^{2}=|\a^{\mb{\tiny R}}_{j-}|^{2}\,,
  $$
\ni after that, we can express
$$
 \nonumber
 \bn{\Pi}_{\mb{\tiny L}} \Bigl|_{(u)} \Bigr.
  =
  \;\bn{\Pi}_{\mb{\tiny R}} \Bigl|_{(u)} \Bigr.
  =
 -i\,\left\Vert
    \ba{cc}
     \bn{I}_{\mb{\tiny N}} & \bn{O}_{\mb{\tiny N}} \\
     \bn{O}_{\mb{\tiny N}} & \bn{I}_{\mb{\tiny N}}
    \ea \right\Vert
  =
 -i\,\bn{I}_{\mb{\tiny 2N}}\,,
$$
\ni thus
$$
   \bn{S}\t \bn{\Pi}_{\mb{\tiny R}} \Bigl|_{(u)} \Bigr. \bn{S} =
   \bn{\Pi}_{\mb{\tiny L}} \Bigl|_{(u)} \Bigr.
   =
 -i\,\bn{S}\t\,\bn{I}_{\mb{\tiny 2N}}\bn{S}
   =
 -i\,\bn{I}_{\mb{\tiny 2N}}\,,
 $$
\ni which yields the expected traditional unitary property, \emph{i.e.}
$$
 \nonumber
  \bn{S}\t \, \bn{S} = \bn{I}_{\mb{\tiny 2N}}.
$$
\ni Despite this development correspond strictly to the EFA model with $N \geq 2$ components, is completely analogous to the case of $N$-component mixing-free flux within the EMA theory.

\subsubsection{Reduction of the generalized-unitarity: $N = 4$}
 \label{Cap:SM::Converge:::Psedo-Unitar:KL}

For completeness, we describe below the third alternative posted in Subsec \ref{Cap:SM::Converge:::N-desacopla}, which offers a complementary route to recover the standard unitarity condition on $\bn{S}$. For this case, it is important to keep in mind that the analysis in the KL model, is made in the scheme of pure heavy- and light-hole states, proposed in the reference \cite{Sanchez95}. The following analysis addresses the same situation, but more appropriated conditions are imposed. For the sake of focus to what is essential, we assume that $k_{x} = k_{y}=0$. Under these conditions, the $L$ and $H$ states\footnote{We recall to the readers, to follow the labelling of the authors as indicated in the footnote of the Subsec. \ref{Cap:SM::PseudoU:::KL}.} described by (\ref{Eqmaestra}) are uncoupled. Thus, the resulting Hamiltonian has solely diagonal terms with kinetic energy like that of the electron, but with effective masses in the form $m_{\mb{\tiny H}}= m_{0}/(\gamma_{1}-2\gamma_{2})\;\;$ and $\;\; m_{\mb{\tiny L}}= m_{0}/(\gamma_{1}+2\gamma_{2})$. This physical scenario of $L$ and $H$ modes, segregated into two ($N = 1$) independent systems, reliable agrees with the description of the EMA and will be applied as starting platform for the convergence criterium we are be searching for. The potential in which the mixing-free $H$ and $L$ states are scattered, is that of a single quantum barrier (QB) or a simple quantum well (QW) and is given schematically in the figures \ref{Fig:Potdis2}(b) and \ref{Fig:Potdis2}(c), respectively. Consequently, the FC is now separately treated by two independent identities
\begin{eqnarray}
 \label{CC7}
 \left.{\hspace{1mm}}
  \begin{tabular}{ccc}
        $|a|^{2}j_{\mb{\tiny H}}+|f|^{2}j_{\mb{\tiny H}} $ & $=$ &
         $|b|^{2}j_{\mb{\tiny H}}+|e|^{2}j_{\mb{\tiny H}}$ \\
        $|c|^{2}j_{\mb{\tiny L}}+|h|^{2}j_{\mb{\tiny L}} $ & $=$ &
         $|d|^{2}j_{\mb{\tiny L}}+|g|^{2}j_{\mb{\tiny L}}$
  \end{tabular}
 \right\}\,.
\end{eqnarray}

Under the circumstances imposed to the $H$ and $L$, the SM is reduced due to the fact that the crossed probabilities\footnote{A crossed path is represented by a sloping solid line at the layer QB in Fig.\ref{Fig:BVprofile}.} for reflection and transmission amplitudes are forbidden, that is: $r_{\mb{\tiny HL}} = r_{\mb{\tiny LH}} = r_{\mb{\tiny HL}}\p = r_{\mb{\tiny LH}}\p = t_{\mb{\tiny HL}} = t_{\mb{\tiny LH}} = t_{\mb{\tiny HL}}\p = t_{\mb{\tiny LH}}\p = 0$. After some transformations, this lead us to
\begin{eqnarray*}
 \left.{\hspace{1mm}}
  \begin{tabular}{c}
        $ j_{\mb{\tiny H}}\,\bn{S}_{\mb{\tiny H}}\t\,\bn{I}_{\mb{\tiny 2N}}\,
         \bn{S}_{\mb{\tiny H}} = j_{\mb{\tiny H}}\,\bn{I}_{\mb{\tiny 2N}}$ \\
        $ j_{\mb{\tiny L}}\,\bn{S}_{\mb{\tiny L}}\t\,\bn{I}_{\mb{\tiny 2N}}\,
         \bn{S}_{\mb{\tiny L}} = j_{\mb{\tiny L}}\,\bn{I}_{\mb{\tiny 2N}}$
  \end{tabular}
 \right\}\,,
\end{eqnarray*}
\ni and additionally we have:
\begin{equation}
 \label{Unitar:KL}
 \left.{\hspace{1mm}}
  \begin{tabular}{c}
        $ \bn{S}_{\mb{\tiny H}}\t\,\bn{S}_{\mb{\tiny H}} = \bn{I}_{\mb{\tiny 2N}}$ \\
        $ \bn{S}_{\mb{\tiny L}}\t\,\bn{S}_{\mb{\tiny L}} = \bn{I}_{\mb{\tiny 2N}}$
  \end{tabular}
 \right\}\,,
\end{equation}
\ni being this what we had to demonstrate. Let us consider the scattering from a QW of a III-V semiconducting material $A_{3}B_{5}$ [see Fig.\ref{Fig:Potdis2}(c)], between semi-infinite layers of a ternary-alloy composite of variable molar composition. Doing this, one can obtain the composite concentration that preserves the unitarity (\ref{Unitar}) of the SM, when $L$ and $H$ independent states interact with the QW. Shortly will be clear that, an isomorphic problem, where the $L$ and $H$ are scattered by a QB of identical structural characteristic [see Fig.\ref{Fig:Potdis2}(b))], it is not possible to be worked out. The semi-empiric L\"{u}ttinger parameters, depending on the concentration $x$, are given by a linear recurrence, which is
$$
 \gamma_{i}(x)= (1-x)\gamma_{ie} + x\gamma_{iw}\,,\;\; \mb{con}\;\;  i=1,2,3 \,,
$$
\ni here $e/w$ represents $electrode/well$, respectively. If now one writes $\gamma_{1}(x)\;\; \mb{y} \;\, \gamma_{2}(x)\,$  and later on substitutes in the correspondent FC condition, subsequently solving --at the electrodes--, the equation
\begin{equation}
 \label{fixflux}
 j_{\mb{\tiny H}}(x)\Bigl|_{k_x=k_y=0} \Bigr. =
  j_{\mb{\tiny L}}(x)\Bigl|_{k_x = k_y=0} \Bigr.
 \,,
\end{equation}
\ni for the variable $x$, then one ends up getting
\begin{equation}
 \label{fixcox}
 x = \frac{2\,\gamma_{2e}\,\widetilde{\Delta k} - \gamma_{1e}\,\Delta k}
           {2\widetilde{\Delta k}(2\,\gamma_{2e}-\gamma_{2w}) +
            \Delta k(\gamma_{1w}-\gamma_{1e})}\,,
 \hspace{0.1cm} \mb{been

 } \hspace{0.1cm} \Delta k= k_{\mb{\tiny H}}-k_{\mb{\tiny L}}\,;
 \hspace{0.1cm} \mb{y} \hspace{0.1cm} \widetilde{\Delta k}= k_{\mb{\tiny H}}+
  k_{\mb{\tiny L}}\,.
\end{equation}
\ni When selecting concentrations from (\ref{fixcox}), it is satisfied (\ref{fixflux}), so it is likely to write
$$
  j(x)_{\mb{\tiny H}}\,\Bigl( \bn{S}\t\,\bn{I}_{\mb{\tiny 2N}}\, \bn{S}
   \Bigr) \Bigl|_{k_x = k_y=0} \Bigr. =
   j(x)_{\mb{\tiny H}}\, \bn{I}_{\mb{\tiny 2N}} \Bigl|_{k_x=k_y=0}
   \Bigr. \,,
$$
\ni and then to finally achieve the standard unitarity of the SM
$$
 \bn{S}\t\,\bn{S} = \bn{I}_{\mb{\tiny 2N}}\,.
$$
\ni Unfortunately one can not complete the analogy, which means that the identity (\ref{fixflux}) turns into a nonsense if the scattering system is a QB of ternary-alloy, embedded by semi-infinite layers of some $A_{3}B_{5}$ material. The point is: in those layers, it can not be found such variable-concentration composite.

To this end we further saw, that several alternatives could be put into direct correspondence with different criteria upon unitary condition, though note that all paths converge to the same standard result $\bn{S}\t\,\bn{S} = \bn{I}_{\mb{\tiny 2N}}$, whatever representation of unitarity is used at the beginning.

\subsection{Completely Orthonormalized Basis}
\label{Cap:SM::Converge:::Fun-Ortonor}

We have already mentioned that coupled incident/emergent modes described within the MMST (EFA framework), demand specific orthonormalization requirements, which we earlier showed in the definition \ref{Def:Ortnonor-Full}. The LI solutions are supposed to be, \emph{a priori}, orthonormalized to the Dirac's $\delta$ in the coordinates space. Let us start by determining certain orthonormality conditions in the spinorial space of functions, which are not unique as we will see later. If we denominate by $\mb{\bm $f$}_{j}(z)$ the $(N \times 1)$ super-vectors that form an orthonormal basis set, to represent one state of the system described \emph{via} (\ref{Eqmaestra}), we may write
\begin{equation}
 \label{eq2}
  \mb{\bm $F$}(z) = \sum_{j = 1}^{2N} a_{j}\mb{\bm $f$}_{j}(z)\,.
\end{equation}
\ni Particularly at regions $L$ and $R$ [see Fig.\ref{Fig:PotDis1}] of by-layer constant parameters, the $\mb{\bm $f$}_j(z)$ can be taken as
\begin{equation}
 \label{eq3}
 \mb{\bm $f$}_j(z) = \mb{\bm $\Gamma$}_j\e^{iq_{j}z},
\end{equation}
\ni where the vectors $\mb{\bm $\Gamma$}_j$ are certain $(N\times 1)$ spinors. Being independent of the spatial coordinates, the $q_{j}$ are the $2N$ eigenvalues which correspond to them as solution of the quadratic eigenvalue problem (QEP) \cite{Diago06,Diago19} associated to (\ref{Eqmaestra}). If we now substitute (\ref{eq3}) in (\ref{Eqmaestra}) we have
\begin{equation}
 \label{QEP1}
 -q_{j}^{2} \,\mb{\bm $B$} \cdot \mb{\bm $\Gamma$}_{j} + \imath q_{j}\left(\mb{\bm $P$} +
 \mb{\bm $Y$}\right) \cdot \mb{\bm $\Gamma$}_{j} + \mb{\bm $W$} \cdot \mb{\bm $\Gamma$}_{j} =
 \bn{O}_{\mb{\tiny N}}\,,
\end{equation}
\ni so that, for instance represents a typical QEP \cite{Tisseur01,Diago11,Diago19}.  If ${\bf P}$ is Hermitian (formal Hermiticity), there is no a coupling term for first derivative states of the field $\bn{F}(z)$. This is not valid for an anti-Hermitian matrix ($\bn{P} = -\bn{P}\t$). This is precisely the case of interest, since the presence of coupling among modes due to the existence of the linear term in $q_{j}$. Such cases are possible for the KL system, the Kane model and others. If we also use the property $\bn{Y} = -\bn{P}\t$ then we have
\begin{equation}
 \label{eq5}
 -q_{j}^{2} \, \mb{\bm $B$} \cdot \mb{\bm $\Gamma$}_{j} + 2\imath q_{j}\, \mb{\bm $P$} \cdot \mb{\bm $\Gamma$}_{j} +
 \mb{\bm $W$} \cdot \mb{\bm $\Gamma$}_{j} =
 \bn{O}_{\mb{\tiny N}}.
\end{equation}
If we apply to (\ref{eq5}) the operation of Hermitian conjugation and use the properties (\ref{CoefMat-Prop(i)})-(\ref{CoefMat-Prop(f)}), we will get
\begin{equation}
 \label{eq6}
 \bn{\Gamma}_{j}\t \cdot \left[-(q_{j}^{2})^{*}\mb{\bm $B$} + 2\imath(q_{j})^{*}\, \mb{\bm $P$} +
 \mb{\bm $W$} \right] = \bn{O}_{\mb{\tiny N}}\,.
\end{equation}
\ni Next we multiply (\ref{eq5}) --on the left-hand side--, by $\bn{\Gamma}_{k}\t$, which yields
\begin{equation}
 \label{eq7}
  -q_{j}^{2} \,\bn{\Gamma}_{k}\t \cdot \mb{\bm $B$} \cdot \mb{\bm $\Gamma$}_{j} +
  2\imath q_{j}\, \bn{\Gamma}_{k}\t \cdot \mb{\bm $P$} \cdot \mb{\bm $\Gamma$}_{j} +
  \bn{\Gamma}_{k}\t \cdot \mb{\bm $W$} \cdot \mb{\bm $\Gamma$}_{j} =
  \bn{O}_{\mb{\tiny N}}.
\end{equation}
\ni Now we write (\ref{eq6}) for $\bn{\Gamma}_{k}$ and multiply by $\bn{\Gamma}_{j}$ on the right-hand side to get
\begin{equation}
 \label{eq8}
  -(q_{k}^{2})^{*}\, \bn{\Gamma}_{k}\t \cdot \mb{\bm $B$} \cdot \bn{\Gamma}_{j} +
  2\imath q_{k}^{*}\, \bn{\Gamma}_{k}\t \cdot \mb{\bm $P$} \cdot \bn{\Gamma}_{j} +
  \bn{\Gamma}_{k}\t \cdot \mb{\bm $W$} \cdot \bn{\Gamma}_{j} = \bn{O}_{\mb{\tiny N}}\,,
\end{equation}
\ni subtracting (\ref{eq7}) and (\ref{eq8}) results in
\begin{eqnarray}
 \label{eq8(a)}
  \left(-q_{j}^{2} + (q_{k}^{2})^{*}\right) \,\bn{\Gamma}_{k}\t \cdot \mb{\bm $B$} \cdot \mb{\bm $\Gamma$}_{j}
  + \imath 2\left(q_{j} - q_{k}^{*}\right)\, \bn{\Gamma}_{k}\t \cdot \mb{\bm $P$} \cdot \mb{\bm $\Gamma$}_{j} +
  \bn{\Gamma}_{k}\t \cdot \mb{\bm $W$} \cdot \mb{\bm $\Gamma$}_{j}
  & = &  \bn{O}_{\mb{\tiny N}}\,,  \nn \\
  \bn{\Gamma}_{k}\t \cdot \mb{\LARGE {[}} \left\{(q_{k}^{2})^{*} - q_{j}^{2} \right\} \mb{\bm $B$} -
  2\imath \left(q_{k}^{*} - q_{j} \right) \mb{\bm $P$} \mb{\LARGE {]}} \cdot \bn{\Gamma}_{j}
  & = &  \bn{O}_{\mb{\tiny N}}\,. \nn \\
\end{eqnarray}
\ni If now we factorize this expression, then we obtain
\begin{equation}
 \label{eq9}
  \bn{\Gamma}_{k}\t \cdot \mb{\LARGE {[}} \left\{q_{k}^{*} + q_{j} \right\} \mb{\bm $B$} -
  2\imath \mb{\bm $P$} \mb{\LARGE {]}} \cdot \bn{\Gamma}_{j}
   = \bn{O}_{\mb{\tiny N}}\,.
\end{equation}
\ni If in (\ref{eq9}) we consider $q_{i}$, with $i = k, j\,$ as real and assuming
$(q_{k} \neq q_{j})$, after that is obtained the following expression
\begin{equation}
  \label{eq10}
  \bn{\Gamma}_{k}\t \cdot \mb{\LARGE {[}} \left\{q_{k} + q_{j} \right\} \mb{\bm $B$} -
  2\imath \mb{\bm $P$} \mb{\LARGE {]}} \cdot \bn{\Gamma}_{j}
  =  \bn{O}_{\mb{\tiny N}}\,.
\end{equation}

The orthogonality conditions (\ref{eq9}) and (\ref{eq10}), suggest the following normalization criteria
\begin{eqnarray}
 \label{Ortonor1}
  \bn{\Gamma}_{k}\t \cdot \mb{\LARGE {[}} \left( q_{k}^{*} + q_{j} \right) \mb{\bm $B$} -
  2\imath \mb{\bm $P$} \mb{\LARGE {]}} \cdot \bn{\Gamma}_{j}
  & = &  \delta_{kj} \,, \\
  \label{Ortonor2}
  \bn{\Gamma}_{k}\t \cdot \mb{\LARGE {[}} \left\{q_{k} + q_{j} \right\} \mb{\bm $B$} -
  2\imath \mb{\bm $P$} \mb{\LARGE {]}} \cdot \bn{\Gamma}_{j}
  =  \delta_{kj} \,,
\end{eqnarray}
\ni for $q_{i}$, with $i= k, j\,$ complex and real, respectively. From the reference \cite{Tisseur01}, we linearized the QEP to its forms (\ref{QEP1}) or (\ref{QEP2}) and finally we get an associated standard eigenvalue problem (SEP), with the same eigenvalues that the QEP. Doing this linearization procedure, leads the expected conditions to be imposed to the eigenvectors of (\ref{QEP1}), to build a \underline{completely orthonormalized basis} as described in definition \ref{Def:Ortnonor-Full}, that is to say
\begin{defn}
 \label{Orto-QEP(gen)}
 \begin{subequations}
  \begin{equation}
   \label{Orto-QEP(1)}
    \bn{\Gamma}_{k}^{\,\dag}\mb{\large{[}}q_{j}\bn{I}{\mb{\tiny N}} -
    q^{*}_{k}\bn{K} + q^{*}_{k}q_{j}\bn{\mathbb{C}}\mb{\large{]}}\bn{\Gamma}_{j}
    =\bn{\Gamma}_{k}^{\,\dag}\bn{L}^{kj}\bn{\Gamma}_{j}= q_{k}\delta_{kj}.
   \end{equation}
   \begin{equation}
    \label{Orto-QEP(2)}
     \bn{\Gamma}_{k}^{\,\dag}\mb{\large{[}}\bn{I}{\mb{\tiny N}} +
     q^{*}_{k}q_{j}\bn{\mathbb{M}}\mb{\large{]}}\bn{\Gamma}_{j}
     = \bn{\Gamma}_{k}^{\,\dag}\bn{D}^{kj}\bn{\Gamma}_{j} = \delta_{kj}.
    \end{equation}
 \end{subequations}
\end{defn}

Once one have achieved the definition \ref{Orto-QEP(gen)} in the spinorial space, we further could have equally chosen either (\ref{Orto-QEP(1)}) or (\ref{Orto-QEP(2)}), since they are equivalent. Sometimes the physical problem under investigation, could determine which one is more convenient \cite{Diago06}. Next we revisit the convergence to the standard unitarity of the SM, but upon the platform of a completely orthonormalized basis. We assume a physical region, with by-layer constant parameters and coefficients, so it could be described by $2N$ plane waves with energy $E$, then $N$ of these waves travel to the right and $N$ travel to the left [see Fig.\ref{Fig:Potdis2}(a)]. We will take the expressions (\ref{eq2}) and (\ref{FormaLineal}) for the envelope function, and for the linear form $\bn{A}(z)$ associated to the operator in (\ref{Eqmaestra}), respectively. Afterwards, we substitute them in (\ref{flujo1}). However, before that it is necessary to express
\begin{eqnarray}
 \label{FormaLineal2}
  \bn{A} & = & \imath \bn{B} \cdot \sum_{k = 1}^{2N}a_{k}q_{k}\bn{f}_{k} +
  \bn{P} \cdot \sum_{k = 1}^{2N}a_{k}\bn{f}_{k}\,, \\
 \label{FormaLineal3}
  \bn{A}\t & = & -\imath\sum_{k = 1}^{2N}a_{k}^{*}q_{k}^{*}\bn{f}_{k}\t \cdot \bn{B} -
  \sum_{k = 1}^{2N}a_{k}^{*}\bn{f}_{k}\t \cdot \bn{P}\,,
\end{eqnarray}
\ni in terms of (\ref{eq2}). Here, the properties (\ref{CoefMat-Prop(i)})-(\ref{CoefMat-Prop(f)}) have been taken into account, and also we took $\bn{P}$ and its anti-Hermitian because of the reasons explained above. By substituting (\ref{eq2}), (\ref{FormaLineal2}) and (\ref{FormaLineal3}) in (\ref{flujo1}), we can get
\begin{eqnarray}
 \label{eq11}
 j & = & -i \left[-i\sum_{k = 1}^{2N}a_{k}^{*}\bn{f}_{k}\t q_{k}^{*} \bn{B} -
  \sum_{k = 1}^{2N}a_{k}^{*}{f}_{k}\t \bn{P} \right] \bn{F} +
  i \bn{F}\t \left[ i \bn{B} \sum_{k = 1}^{2N}a_{k}q_{k}\bn{f}_{k} +
  \bn{P}  \sum_{k,j = 1}^{2N}a_{k}\bn{f}_{k} \right] \nn , \\
 j & = & -\sum_{k,j = 1}^{2N}a_{k}^{*}a_{j}\bn{f}_{k}\t q_{k}^{*} \bn{B} \bn{f}_{j} +
  i \sum_{k,j = 1}^{2N}a_{k}^{*}a_{j}\bn{f}_{k}\t \bn{P}\bn{f}_{j} -
  \sum_{k,j = 1}^{2N}a_{j}^{*}a_{k}\bn{f}_{j}\t \bn{B}q_{k} \bn{f}_{k} +
  i \sum_{k,j = 1}^{2N}a_{j}^{*}a_{k}\bn{f}_{j}\t \bn{P} \bn{f}_{k}, \nn \\
  & & \nn \\
  && \mb{changing conveniently the order of the subscripts and regrouping we get} \nn
  \\
   j & = & -\sum_{k,j = 1}^{2N}a_{k}^{*}a_{j}\bn{f}_{k}\t \left( q_{k}^{*} + q_{j} \right)
   \bn{B}\bn{f}_{j} + 2i\sum_{k,j = 1}^{2N}a_{k}^{*}a_{j}\bn{f}_{k}\t \bn{P}
   \bn{f}_{j}\,, \nn \\
  & & \nn \\
  & & \mb{doing the contracted product, the preceding expression can be written down in the form} \nn \\
   j & = & -\sum_{k,j = 1}^{2N}a_{k}^{*}a_{j}\bn{f}_{k}\t \mb{\Large {[}} \left( q_{k}^{*} + q_{j} \right)
   \bn{B} - 2i\bn{P} \mb{\Large {]}} \bn{f}_{j}\,.
\end{eqnarray}

If now we take the condition (\ref{Ortonor1}) or (\ref{Ortonor2}), in accordance to the physical case, we may write for the probability current density the expression
\begin{equation}
 \label{flujo6}
  j = -\sum_{k,j = 1}^{2N}a_{k}^{*}a_{j}\delta_{kj}\,.
\end{equation}
Now we make a brief digression, to analyze some particularities of the starting differential system (\ref{Eqmaestra}). Substituting (\ref{eq2}) into (\ref{Eqmaestra}), one has a QEP --which is analogous to (\ref{QEP1})--, and has the form
\begin{equation}
 \label{QEP2}
 \sum_{j = 1}^{2N}a_{j} \left[ -q_{j}^{2}\bn{B} + 2\imath q_{j}\bn{P} + \bn{W} \right] \bn{f}_{j} =
 \bn{O}_{\mb{\tiny N}}\,.
\end{equation}
\ni Next, we make a similar procedure to that from (\ref{eq5}) to (\ref{eq8(a)}), thus we obtain
\begin{eqnarray}
 \label{eq13}
 \sum_{k,j = 1}^{2N}a_{k}^{*}a_{j}\bn{f}_{k}\t \left[ \left\{(q_{k}^{2})^{*} - q_{j}^{2}\right\}\bn{B} -
  2\imath \left( q_{k}^{*} - q_{j} \right) \bn{P} \right] \bn{f}_{j}
  & = & \bn{O}_{\mb{\tiny N}}\,, \nn \\
  & & \nn \\
  \mb{and after factorizing it can be readily get} & & \nn \\
  \sum_{k,j = 1}^{2N}a_{k}^{*}a_{j}\left( q_{k}^{*} - q_{j} \right)\bn{f}_{k}\t
  \mb{\Large {[}} \left\{ q_{k}^{*} + q_{j} \right\}\bn{B} - 2\imath\bn{P} \mb{\Large {]}} \bn{f}_{j}
  & = & \bn{O}_{\mb{\tiny N}}\,, \nn \\
  & & \nn \\
  \mb{from where by using (\ref{Ortonor1}), we have obtained} & & \nn \\
  \sum_{k,j = 1}^{2N}a_{k}^{*}a_{j}\left( q_{k}^{*} - q_{j} \right) \delta_{kj}
 & = & \bn{O}_{\mb{\tiny N}}\,.
 \end{eqnarray}
\ni  Let us see which are the implications in (\ref{eq13}), regarding the eigenvalues $q_{i}$, with $i =
k, j$
\begin{casos}
 \label{Casos-QEP}
 \underline{Implications in (\ref{eq13}) attending to the eigenvalues $q_{i}$, with $i = k, j$}
  \begin{itemize}
    \item For $k \neq j \; \Longrightarrow  \delta_{kj} = 0 \;
      \Rightarrow$ the equation is always satisfied.
    \item For $k = j$
     \begin{itemize}
      \item If $q_{i} \in $ Reals $\; \Longrightarrow (q_{k}-q_{k}) = 0\;
       \Rightarrow$ the equation is always satisfied.
      \item If $q_{i} \in $ Complex, the eigenvalues arise in conjugated pairs
       $(q_{i}^{*}, q_{i})$. As $(q_{i}^{*} - q_{i}) = -2i\, \Im[q_{i}]$, then in the sum
       (\ref{eq13}) will appear pairs such that:
       $$
        -2\imath |a_{n}|^{2}\,\Im[q_{n}] - 2\imath|a_{m}|^{2}\,\Im[q_{m}] = 0 \,,
       $$
       \ni which is the same as to say
       \begin{eqnarray*}
        -2\imath|a_{n}|^{2}\,\Im[q_{n}] + 2\imath |a_{m}|^{2}\,\Im [q_{n}] & = & 0 \, \\
        2\imath \,\Im[q_{n}] \{|a_{m}|^{2} - |a_{n}|^{2}\} & = & 0 \,. \\
        \Downarrow
       \end{eqnarray*}
       \ni To satisfy the preceding expression, it is sufficient to entail that the coefficients of the eigenvalues, which are conjugated by pairs, be $a_{m} \equiv a_{n}$.
     \end{itemize}
  \end{itemize}
\end{casos}
After this unavoidable parenthesis in the development, we come back to the formula (\ref{flujo6}), that we now rewrite for $k = j$, due to they are finite terms, being their non-zero character of special interest as have been commented earlier, then
\begin{equation}
 \label{flujo7}
  j = -\sum_{k,j = 1}^{2N}a_{k}^{*}a_{k}\delta_{kj} = -\bn{\a}\t \cdot \bn{\a} \,.
\end{equation}
\ni If one considers the form to chose the coefficients (as was earlier shown), it is then possible to transform the given expression in (\ref{flujo7}), into that of the propagating modes representation [see reference \cite{Diago06} and references therein]
\begin{eqnarray*}
  j & = & -\left(
          \ba{cc}
           \aa_{1}\t & \aa_{2}\t \\
          \ea
         \right)
         \cdot
         \left(
          \ba{c}
           \aa_{1} \\
           -\aa_{2}
          \ea
         \right)\,,
\end{eqnarray*}
\ni whose widely-accepted form is given by \cite{Mello88}
\begin{eqnarray*}
  j & = & \aa\t \cdot \bn{\Sigma}_{z} \cdot \aa \,.
\end{eqnarray*}
\ni Attending to what was demonstrated in the Subsec. \ref{Cap:SM::Converge:::Psedo-Unitar:Flujos}, it is straightforward that from the preceding expression one can readily obtained (\ref{Unitar}) and (\ref{Simet}), in other words
\begin{eqnarray*}
  \bn{S}\t & = & \bn{S}^{-1} \,, \\
  \bn{S}^{T} & = & \bn{S}\,,
\end{eqnarray*}
\ni which are the properties of unitarity and symmetric character, respectively, what we wanted to get. Being aware that the choice of the basis vectors set, is not unique, we underline the fundamental importance of the completely orthonormalized basis, for unitarity preservation of mixed-particle fluxes in the MMST. As a bonus, it also provides a direct route to recover the convergence to the standardized unitarity of the EMA framework, as had been demonstrated right above.

\section{Symmetry Relations}
 \label{Cap::SM:::Simetry}

 In this section, the purpose is to obtain the symmetry relations with regard of $\bn{S}$ blocks. In the specialized literature these relations are usually  derived from (\ref{cuasi-uni1}), or from the conditions imposed by means of the TRI symmetry and the spatial inversion invariance (SII) over $\bn{S}$ \cite{Pereyra98}. Owing to brevity, we drop a thorough analysis of the discrete symmetries for the MMST, since a detailed description on that subject --though within the viewpoint of the TM formalism--, was reported by Diago \emph{et al.} elsewhere \cite{Diago05}. Hence, what appears nextly, following our target in the present work, is the direct analysis of the consequences of (\ref{cuasi-uni1}) and (\ref{cuasi-uni2}), over the coefficient matrices of quantum transmission and reflection. From the obtained relations, it should be feasible , in principle, to calculate any of the magnitudes relevant for the quantum transport within the MMST, by means of the others one. The main utility of these expressions --and probably the most important one--, is to extract quantities that could be cast, in terms of physically meaningful objects, or rather, terms that yield a reliable interpretation of physical observables of any concrete quantum problem. In that concern, one need to solve (\ref{cuasi-uni1}), that is
$$
  \left\Vert
   \ba{cc}
    \bn{r} & \bn{t}\p   \\
    \bn{t} & \bn{r}\p
   \ea
  \right\Vert \t \,
   \left\Vert
    \ba{cc}
     \bn{\Pi}_{\mb{\tiny R}}^{\mb{\tiny 11}} & \bn{\Pi}_{\mb{\tiny R}}^{\mb{\tiny 12}} \\
     \bn{\Pi}_{\mb{\tiny R}}^{\mb{\tiny 21}} & \bn{\Pi}_{\mb{\tiny R}}^{\mb{\tiny 22}}
    \ea
   \right\Vert
   \left\Vert
    \ba{cc}
     \bn{r} & \bn{t}\p   \\
     \bn{t} & \bn{r}\p
    \ea
   \right\Vert
  =
   \left\Vert
    \ba{cc}
     \bn{\Pi}_{\mb{\tiny L}}^{\mb{\tiny 11}} & \bn{\Pi}_{\mb{\tiny L}}^{\mb{\tiny 12}} \\
     \bn{\Pi}_{\mb{\tiny L}}^{\mb{\tiny 21}} & \bn{\Pi}_{\mb{\tiny L}}^{\mb{\tiny 22}}
    \ea
   \right\Vert \,,
$$
$$
  \left\Vert
    \ba{cc}
     \bn{r}\t & \bn{t}\t   \\
     (\bn{t}\p)\t & (\bn{r}\p)\t
    \ea
   \right\Vert \,
   \left\Vert
    \ba{cc}
     (\Csf^{*})^{-1}\,\bn{X}_{--}\,\Csf^{-1} &
     (\Csf^{*})^{-1}\,\bn{X}_{-+} \\
     \bn{X}_{+-}\,\Csf^{-1} & \bn{X}_{++}
    \ea
   \right\Vert
   \left\Vert
    \ba{cc}
     \bn{r} & \bn{t}\p   \\
     \bn{t} & \bn{r}\p
    \ea
   \right\Vert
  =
$$
$$
  = \left\Vert
    \ba{cc}
      \bn{X}_{++} & \bn{X}_{+-}\,\Csf \\
      \Csf^{*}\,\bn{X}_{-+} &
      \Csf^{*}\,\bn{X}_{--}\,\Csf
    \ea
   \right\Vert \,,
$$
\ni becoming the last, into the following non-trivial symmetry relations:
\begin{subequations}
 \begin{equation}
  \label{sim(r-t)a}
   \bn{r}\t\,\bn{\Pi}_{\mb{\tiny R}}^{\mb{\tiny 11}}\,\bn{r} +
   \bn{t}\t\,\bn{\Pi}_{\mb{\tiny R}}^{\mb{\tiny 21}}\,\bn{r} +
   \bn{r}\t\,\bn{\Pi}_{\mb{\tiny R}}^{\mb{\tiny 12}}\,\bn{t} +
   \bn{t}\t\,\bn{\Pi}_{\mb{\tiny R}}^{\mb{\tiny 22}}\,\bn{t}  =
   \bn{\Pi}_{\mb{\tiny L}}^{\mb{\tiny 11}}
 \end{equation}
 \begin{equation}
 \label{sim(r-t)b}
  \bn{r}\t\,\bn{\Pi}_{\mb{\tiny R}}^{\mb{\tiny 11}}\,\bn{t}\p +
  \bn{t}\t\,\bn{\Pi}_{\mb{\tiny R}}^{\mb{\tiny 21}}\,\bn{t}\p +
  \bn{r}\t\,\bn{\Pi}_{\mb{\tiny R}}^{\mb{\tiny 12}}\,\bn{r}\p +
  \bn{t}\t\,\bn{\Pi}_{\mb{\tiny R}}^{\mb{\tiny 22}}\,\bn{r}\p =
  \bn{\Pi}_{\mb{\tiny L}}^{\mb{\tiny 12}}
 \end{equation}
 \begin{equation}
 \label{sim(r-t)c}
  (\bn{t}\p)\t\,\bn{\Pi}_{\mb{\tiny R}}^{\mb{\tiny 11}}\,\bn{r} +
  (\bn{r}\p)\t\,\bn{\Pi}_{\mb{\tiny R}}^{\mb{\tiny 21}}\,\bn{r} +
  (\bn{t}\p)\t\,\bn{\Pi}_{\mb{\tiny R}}^{\mb{\tiny 12}}\,\bn{t} +
  (\bn{r}\p)\t\,\bn{\Pi}_{\mb{\tiny R}}^{\mb{\tiny 22}}\,\bn{t} =
  \bn{\Pi}_{\mb{\tiny L}}^{\mb{\tiny 21}}
 \end{equation}
 \begin{equation}
 \label{sim(r-t)d}
 (\bn{t}\p)\t\,\bn{\Pi}_{\mb{\tiny R}}^{\mb{\tiny 11}}\,\bn{t}\p +
  (\bn{r}\p)\t\,\bn{\Pi}_{\mb{\tiny R}}^{\mb{\tiny 21}}\,\bn{t}\p +
  (\bn{t}\p)\t\,\bn{\Pi}_{\mb{\tiny R}}^{\mb{\tiny 12}}\,\bn{r}\p +
  (\bn{r}\p)\t\,\bn{\Pi}_{\mb{\tiny R}}^{\mb{\tiny 22}}\,\bn{r}\p =
  \bn{\Pi}_{\mb{\tiny L}}^{\mb{\tiny 22}}\,.
 \end{equation}
\end{subequations}
Only if the variables $\bn{t},\bn{r},\bn{t}\p\, \mb{and}\; \bn{r}\p$ were Hermitian, this system could be mathematically well defined and one can obtain expressions for each one, in terms of the others. Even so, it is not evident that these expressions could be useful, in contrast with what happens when one develops (\ref{cuasi-uni2}). In that case, one obtains simple relations with physical useful meaning, due to the diagonal form of $\mathbf{J}$.
If for instance, in (\ref{sim(r-t)a}) we demand the boundary condition to obey incidence only from the left of the scattering system ( that is to say, from $-\infty$), implies to make $\a^{\mb{\tiny R}}_{j-}=0; \, \forall j=1,\ldots,N$, then $\bn{\Pi}_{\mb{\tiny R}}^{\mb{\tiny 11}}= \bn{\Pi}_{\mb{\tiny R}}^{\mb{\tiny 12}}= \bn{\Pi}_{\mb{\tiny R}}^{\mb{\tiny 21}}=0$, and in (\ref{sim(r-t)a}) we have that $\;\bn{t}\t\,\bn{\Pi}_{\mb{\tiny R}}^{\mb{\tiny 22}}\,\bn{t}=\bn{\Pi}_{\mb{\tiny L}}^{\mb{\tiny 11}},\,$ which more explicitly means that
\begin{equation}
 \label{sim(r-t)a2}
 \bn{t}\t\,\bn{X}_{++}\,\bn{t}=\bn{X}_{++}\,.
\end{equation}

 Symmetry requirements as that of (\ref{sim(r-t)a2}), is a key tool for an intermediate control of the numerical quotation correctness, if dealing with quantum transport phenomena in the framework of the MMST \cite{Diago06}. Regretfully, in (\ref{sim(r-t)b})-(\ref{sim(r-t)d}), under the above imposed conditions, appears in the right-hand side member of the equation, an indetermination of division by zero. For the sake of completeness, we present without derivation, the major symmetry requirements within the TM formalism, which are readily derived from TM's definitions and symmetries \cite{Diago06}. They represent a very useful alternate way to preserve the FC, o rather, the unitarity of the SM, so we call them as \emph{filters}. More to the point, what is provided as a bonus, is a reduction of the computational effort, avoiding as well undesirable numerical artifacts.

\begin{propo}
 \label{Propo:Filtros}
  \underline{Filters for layered $Q2D$ systems for the MMST:}
   \begin{enumerate}
    \item[(a)] Determinant of the FTM\footnote{FTM stands for full transfer matrix [see \ref{Sec:AppB}]} $\bn{M}_{fd}(z,z_{0})$ must be equal to $1$.
    \item[(b)] Determinant of the  STM\footnote{STM stands for state-vectors transfer matrix [see \ref{Sec:AppB}]} $\bn{M}_{sv}(z,z_{0})$ must be equal to $1$.
    \item[(c)] Theorem of Liuoville
          $\mb{\Large {\rm (}}$
            $Tr \mb{\large {\rm [}} \mathcal{P}_{\omega}(z) \mb{\large {\rm ]}} = 0$
          $\mb{\Large {\rm )}}$.
    \item[(d)] FC general principle, for example:
             $\bn{\Sigma}_{z}= \bn{M}_{sv}^{\dag}(z,z_{0})\bn{\Sigma}_{z}\bn{M}_{sv}(z,z_{0})$.
    \item[(e)] TRI symmetry, for example:
             $\bn{M}_{fd}(z,z_{0}) = \bn{\Sigma}\bn{M}_{fd}^{\ast}(z,z_{0})\bn{\Sigma}^{-1}$.
    \item[(f)] SII symmetry, for example:
             $\bn{T}(z,z_{0}) = \left(\Ssf_{\mb{\tiny I}}\right)_{\mb{\tiny T}}^{-1}\bn{T}(-z,-z_{0})
             \left(\Ssf_{\mb{\tiny I}}\right)_{\mb{\tiny T}}$.
    \item[(g)] Charge conservation law
         $\mb{\Large {\rm (}}\bn{\cal I}_{\mb{\tiny FLUX}} + {\bn{\cal R}}_{\mb{\tiny FLUX}}=
         {\bn{\cal T}}_{\mb{\tiny FLUX}} \mb{\Large {\rm )}}$.
    \item[(h)] Hermiticity of the matrices $\bn{\cal I}_{\mb{\tiny FLUX}}$, ${\bn{\cal R}}_{\mb{\tiny FLUX}}$
          and ${\bn{\cal T}}_{\mb{\tiny FLUX}}$.
    \item[(i)] Commutation rules
           $\mb{\Large {\rm (}}$
            $\mb{\large {\rm [}}\bn{\cal I}_{\mb{\tiny FLUX}}^{-1},
            \bn{\cal R}_{\mb{\tiny FLUX}} \mb{\large {\rm ]}} = \bn{O}_{\mb{\tiny
            N}}\,;\;\;$
            $\mb{\large {\rm [}}\bn{\cal I}_{\mb{\tiny FLUX}}^{-1},
            \bn{\cal T}_{\mb{\tiny FLUX}} \mb{\large {\rm ]}} = \bn{O}_{\mb{\tiny N}}$
           $\mb{\Large {\rm )}}$.
   \end{enumerate}
\end{propo}

\begin{figure}[ht!]
\centering
  \includegraphics[width=0.5\linewidth]{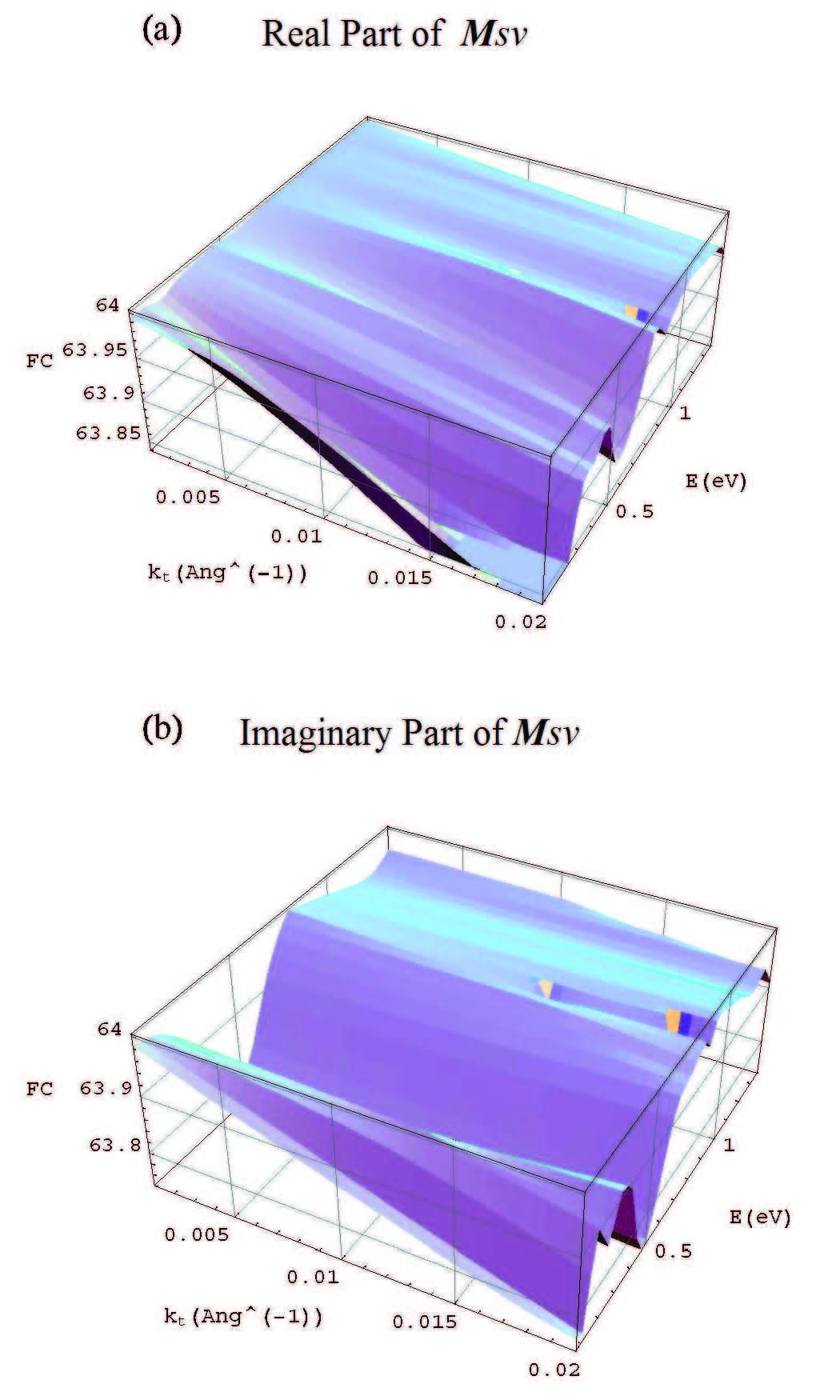}
 \caption{\label{Fig:FC(4x4)Msv} (Color online). Verification of the FC general principle,
  associated to the STM $\bn{M}_{sv}(z,z_{0})$, for a layered heterostructure of a single $AlAs$-QB of thickness $10\,\AA$, embedded in semi-infinite $GaAs$. We have taken for the in-plane (transversal) quasi-vector $[\,10^{-4} < \kappa_{\mb{\tiny T}} < 0.02\,]\,$ $\AA^{-1}\,$, while the incident energy is in the rank $[\,10^{-10} < E < 0.1\,V_{b}\,]\,$eV. The VB´s {\em band-offset} was fixed as $V_{b} = 0.5\,$eV.}
\end{figure}

All these filters have been used in former reports, indeed: (i) they were numerically evaluated for consistency \cite{Diago05}; (ii) they have been quoted to work out expected values for quantum transport entities within the MMST \cite{Diago06}, and more recently, some of them were successfully invoked in related problems \cite{Diago19}. As punchline, we next illustrate in Fig.\ref{Fig:FC(4x4)Msv} --in accordance with the target of the present study--, the FC general principle item of the proposition \ref{Propo:Filtros}, which is an analogous of the above derived requirement (\ref{sim(r-t)a2}).

The numerical evaluation of most of the filters in the proposition \ref{Propo:Filtros}, is far from a trivial task. We have worked this out by means of a mathematical \emph{trick}. Note that the majority of them, have the following generic form
\begin{equation}
 \label{Generic}
  \bn{Z}_1 = \bn{Z}_2 \,,
\end{equation}
\ni being $\bn{Z}_{1,2}$ certain $(2N \times 2N)$ matrices. To get a numeric result, feasible for graphical simulation, we proceed as follow: If (\ref{Generic}) is fulfilled, then $\bn{Z}_{1} - \bn{Z}_2 = \bn{Z}_3\;,$ where $\bn{Z}_{3} =
\bn{O}_{\mb{\tiny 2N}}$. Now, we take the absolute value of $\bn{Z}_3$ and after that, we subtract it from certain matrix
$\bn{Z}_{ini}$, whose elements have the form $(Z_{ini})_{\imath \jmath} = 1;\, \forall \, \imath,
\jmath$. Subsequently, we should obtain
$$
 \bn{Z}_{ini} - \left\|\bn{Z}_{3} \right\| = \bn{Z}_{fin}\,.
$$
\ni Rigorously, the matrix $\bn{Z}_{fin}$, most be equal to the matrix $\bn{Z}_{ini}$, if $\bn{Z}_{3}$ is the null matrix. The later means, that the corresponding symmetry or general principle, fulfills. Finally, we sum all the matrix elements of $\bn{Z}_{fin}$, element by element. It is straightforward that, for a $2N$ order matrix, where each element is the number $1$ --which is the case of the matrix $\bn{Z}_{fin}$--, this sum equals $4N^{2}$. For the case of the KL model Hamiltonian with $N = 4$, the sum of interest results in $64$ [see the vertical axis in Fig.\ref{Fig:FC(4x4)Msv}]. The last, is the number that one has to calculate, for the rank of chosen physical parameters. Worth noting, that matices can have complex-number entries, in such cases, one has to verify both, the real and the imaginary parts of the envisioned expression of the proposition \ref{Propo:Filtros}

\section{Tunneling amplitudes}
 \label{Cap::SM:::Tun-Amplit}

We have already commented that the MSA comprises, in a common base, two approaches of the TM formalism, potentiating the advantages of each technique. As the scattering is the central point of our approximation, now we initiate its study, and we underline that in the specialized literature, there exist different views of the SM. Next we insert an analysis about that, pretending to unify criteria about the transmission and reflection matrices for different ways to define the SM [see Tab.\ref{Tab:SM-verso}]. Not always when it says: Scattering Matrix it has in mind the same object, although they are quite similar. The purpose of this section, is precisely to remark the relation between different objects, that are connected to the SM and to show the similarities and differences they have. In the specialized literature, the SM is defined to connect different objects associated to the incident flux of particles (or cuasi-particles) with their similar of the emerging flux after the ``collision" with the scattered system [see Tab.\ref{Tab:SM-verso}]. To review this diversity we will describe a generalization of (\ref{In-Out}).
\begin{eqnarray}
 \label{In-Out3}
 \bn{\mathcal{I}}=
 \left\Vert
   \ba{c}
    \bn{\mathcal{A}}^{\mb{\tiny L}}_{+} \\
    \bn{\mathcal{A}}^{\mb{\tiny R}}_{-}
   \ea
 \right\Vert_{in} \;\;\;;\;\;\;
 \bn{\mathcal{O}}=
 \left\Vert
   \ba{c}
    \bn{\mathcal{A}}^{\mb{\tiny L}}_{-} \\
    \bn{\mathcal{A}}^{\mb{\tiny R}}_{+}
   \ea
 \right\Vert_{out}, \\
 \nonumber
 \hspace{11mm} _{(2N \times 1)\hspace{20mm}(2N \times 1)}
\end{eqnarray}
\ni and in each case, the elements of $\bn{\mathcal{A}}\,$ will represent one of the objects of Tab.\ref{Tab:SM-verso}.
\begin{table}[h!]
 \caption{\label{Tab:SM-verso} Different approaches to the SM in the literature.}
 \begin{center}
  \begin{tabular}{||l|c|c|c||} \hline\hline
    Case & \textbf{Object that connects} &  Symbol  & Basic Reference \\
   \hline
     (i) & Coefficients  & $\a_{j}$ & Mello, P. Pereyra y N. Kumar \cite{Mello88} \\
         & (of traveling waves)                                   &          & A. D. S\'{a}nchez y C. R. Proetto \cite{Sanchez95} \\
   \hline
    (ii) & Wave Vectors    & $\a_{j}\bn{f}_{j}(z)$ & P. Pereyra \cite{Pereyra98} \\
         & (propagating waves)     &            & \\
   \hline
   (iii) & State vectors   & $\a_{j}\bn{\varphi}_{j}(z)$ &
       L. Diago, P. Pereyra,   \\
         &  \mb{\Large {(}}$z$-part of $\bn{f}_{j}(z)\mb{\Large {)}}$  &                      &
         H. Coppola, R. P\'{e}rez \cite{Diago02} \\
   \hline
   (iv)  &  Wave function   &  $\bn{F}_{j}(t=\pm\infty)$  & A. S. Davydov \cite{Davydov65} \\
         & (time dependent)   &              & \\
   \hline\hline
  \end{tabular}
 \end{center}
\end{table}

The cases to be unified here are (i) and (iii), because they are directly related to our approach of the MSA. Nevertheless, you can notice the proximity between (ii) and (iii): In the KL model, for example, the difference between them is that  $\bn{\varphi}_{j}$ excludes the $(4 \times 1)$ spinors, whose orthonormalization goes from a QEP. The reason to unify cases (i) and (iii), is the relation they have with our MSA modelling. Initially, we express the matrices of transmission and reflection amplitudes, in each one of the mentioned cases. After the corresponding algebraic transformations, we grouped and obtained the contracted products for the quantities of interest, namely
\begin{equation}
 \label{coef-Disp3-4}
 \left.{\hspace{1mm}}
  \begin{tabular}{ccc}
    $ \bn{t} $
    & = & $ (\mb{\large \bn{\varphi}}^{\mb{\tiny R}}_{\mb{\tiny +}})^{-1}\,
    \left\{\bn{\alpha}-\bn{\beta}\,\bn{\delta}^{-1}\,\bn{\gamma}\right\}\,
    \mb{\large \bn{\varphi}}^{\mb{\tiny L}}_{\mb{\tiny +}} $ \\
    $ \bn{r} $
    & = & $ -(\mb{\large \bn{\varphi}}^{\mb{\tiny L}}_{\mb{\small -}})^{-1}\,
     \left\{\bn{\delta}^{-1}\,\bn{\gamma}\right\}\,
     \mb{\large \bn{\varphi}}^{\mb{\tiny L}}_{\mb{\tiny +}} $ \\
    $ \bn{t}\p $
    & = & $ (\mb{\large \bn{\varphi}}^{\mb{\tiny L}}_{\mb{\small -}})^{-1}\,
     \left\{\bn{\delta}^{-1}\right\}\,
     \mb{\large \bn{\varphi}}^{\mb{\tiny R}}_{\mb{\small -}} $ \\
    $ \bn{r}\p $
    & = & $ (\mb{\large \bn{\varphi}}^{\mb{\tiny R}}_{\mb{\tiny +}})^{-1}\,
     \left\{\bn{\beta}\,\bn{\delta}^{-1}\right\}\,
     \mb{\large \bn{\varphi}}^{\mb{\tiny R}}_{\mb{\small -}} $
  \end{tabular}
 \right\} \,.
\end{equation}

This analysis is independent of the type of unitary condition the SM satisfies (standard, pseudo-unitarity or structured). This system is expressed  in terms of the $(N \times N)$ matrices $\bn{\alpha}, \bn{\beta}, \bn{\gamma}, \bn{\delta}$; which represent the $[11]$, $[12]$, $[21]$ and $[22]$ blocks of the matrix $\bn{M}_{sv}(z_{\mb{\tiny R}},z_{\mb{\tiny L}})$, respectively. The difference between the cases (i) and (iii) is on phase factors that do not contribute to the expected transmission and reflection coefficients. This is independent of the fact, that the materials of electrodes are the same or different. The following example illustrate these considerations. Relations (\ref{coef-Disp3-4}) will be evaluated for the $(4 \times 4)$ KL model. Let us consider the problem of simultaneous scattering of $hh$ and $lh$ in a simple cell [see Fig.\ref{Fig:BVprofile}], which in this case is taken as a QB between layers of identical material with no external field. The exponentials that are ordered by $hh_{\mb{\tiny +3/2}}$, $lh_{\mb{\tiny -1/2}}$, $lh_{\mb{\tiny +1/2}}$, $hh_{\mb{\tiny -3/2}}$ \cite{Broido85,Diago02}, will be expressed on non-dimensional magnitudes $q = k_{z}a_{\mb{\tiny o}}\,$ and $\,\xi = z/a_{\mb{\tiny o}}$, where $a_{\mb{\tiny o}}$ is the Bohr radius. Then
$$
 \mb{\large \bn{\varphi}}^{\mb{\tiny L}}_{\mb{\tiny +}} =
   \left\Vert
   \ba{cccc}
    \e^{iq_{hh}\mb{\tiny $\xi$}} & 0 & 0 & 0 \\
    0 & \e^{iq_{lh}\mb{\tiny $\xi$}} & 0 & 0 \\
    0 & 0 & \e^{iq_{lh}\mb{\tiny $\xi$}} & 0 \\
    0 & 0 & 0 & \e^{iq_{hh}\mb{\tiny $\xi$}}
   \ea
   \right\Vert \,,\;\;
   (\mb{\large \bn{\varphi}}^{\mb{\tiny R}}_{\mb{\tiny +}})^{-1} =
   (\mb{\large \bn{\varphi}}^{\mb{\tiny L}}_{\mb{\tiny +}})^{*}
   \,, \;\; \mb{and} \;\;
   (\mb{\large \bn{\varphi}}^{\mb{\tiny L}}_{\mb{\small -}})^{-1} =
  \mb{\large \bn{\varphi}}^{\mb{\tiny L}}_{\mb{\tiny +}}\;\;.
$$
$$
 \bn{\alpha}-\bn{\beta}\,\bn{\delta}^{-1}\,\bn{\gamma} =
   \left\Vert
   \ba{cccc}
    \lambda_{11} & \lambda_{12} & \lambda_{13} & \lambda_{14} \\
    \lambda_{21} & \lambda_{22} & \lambda_{23} & \lambda_{24} \\
    \lambda_{31} & \lambda_{32} & \lambda_{33} & \lambda_{34} \\
    \lambda_{41} & \lambda_{42} & \lambda_{43} & \lambda_{44} \\
   \ea
   \right\Vert\,,\;\;
 -\bn{\delta}^{-1}\,\bn{\gamma} =
   \left\Vert
   \ba{cccc}
    \rho_{11} & \rho_{12} & \rho_{13} & \rho_{14} \\
    \rho_{21} & \rho_{22} & \rho_{23} & \rho_{24} \\
    \rho_{31} & \rho_{32} & \rho_{33} & \rho_{34} \\
    \rho_{41} & \rho_{42} & \rho_{43} & \rho_{44} \\
   \ea
   \right\Vert\,.
$$
\ni If we obtain $\bn{t}$ and $\bn{r}$, as defined in (\ref{coef-Disp3-4}), we get as consequence that $T_{ji}$ and $R_{ji}$ of  cases (i) and (iii) are equal, \textit{i.e.}:
\begin{eqnarray}
 \label{coef-Disp5b}
 \left.{\hspace{1mm}}
  \begin{tabular}{cc}
   &  \underline{Case(i)}  \\
   \\
   & $ \Bigl|\mb{\Large {$\lambda$}}_{ji}\Bigr|^{2}$  \\
   & $ \forall \; $  {\mb {\small $i,j$}}  \\
   \\
  \end{tabular}
 \right\}
  =  T_{ji}  =
 \left\{
   \begin{tabular}{cccc}
        & \underline{Case(iii) } & \\
        $ |\lambda_{ji}|^{2};$  $\;\;\forall$ {\mb {\small $i = j\;\;$}} or
          $\,$ {\mb {\small $i+j = 5$}} & \\
                                 &
           $ = \Bigl|\mb{\Large {$\lambda$}}_{ji}\Bigr|^{2}; $
           $  \;\; \forall \,$ {\mb {\small $i,j$}}\,,   \\
        $ |\e^{i(q_{n}-q_{m})\mb{\tiny $\xi$}}|^{2}|\lambda_{ji}|^{2};\;\;$ else & \\
        \\
  \end{tabular}
  \right.
\end{eqnarray}
\ni and then
\begin{eqnarray}
 \label{coef-Disp6b}
 \left.{\hspace{1mm}}
  \begin{tabular}{cc}
   &  \underline{Case(i)}  \\
   \\
   & $ \Bigl|\mb{\Large {$\rho$}}_{ji}\Bigr|^{2}$  \\
   & $ \forall \; $  {\mb {\small $i,j$}}  \\
   \\
  \end{tabular}
 \right\}
  =  R_{ji}  =
 \left\{
   \begin{tabular}{cccc}
        & \underline{Case(iii) }& \\
        $ |\e^{i2(q_{n})\mb{\tiny $\xi$}}|^{2}
        |\rho_{ji}|^{2};$  $\;\;\forall$ {\mb {\small $i = j\;\;$}} or
          $\,$ {\mb {\small $i+j = 5$}} & \\
                                 &
           $ = \Bigl|\mb{\Large {$\rho$}}_{ji}\Bigr|^{2}; $
           $  \;\; \forall \,$ {\mb {\small $i,j$}}\,.   \\
        $ |\e^{i(q_{n}+q_{m})\mb{\tiny $\xi$}}|^{2}|\lambda_{ji}|^{2};\;\;$ else & \\
        \\
   \end{tabular}
 \right. \nn  \\
\end{eqnarray}

The solved example, clearly shows that the matrices of transmission  and reflection amplitudes of cases (i) and (iii), are equal up to a phase factor, that do not matter in the values of the corresponding coefficients, as can be seen directly on (\ref{coef-Disp5b}) and (\ref{coef-Disp6b}). Nevertheless, the difference between these two cases, must be seen in the revision of the FC law in matrix form \cite{Diago05} [see for instance (\ref{sim(r-t)a2})], because this expression contains directly the matrices $\bn{t}$ and $\bn{r}$.

\subsection{Probabilities Flux Conservation}
 \label{Cap:SM::Tun-Amplit:::Conserva-Flux}

\hspace*{6mm} Before we analyzed that select an arbitrary basis of LI functions lead to the pseudo-unitary property. Here we analyzed qualitatively which restrictions imposes the pseudo-unitary property of the scattering operator on the other elements in matrix form for the KL model.
The unitary condition on $\bn{S}\,\bn{S}\t = \bn{S}\t\,\bn{S}=\bn{I}_{\mb{\tiny 2N}}$, applied to a group of important problems implies in details that:
\begin{equation}
 \label{CC9}
  \left.{\hspace{1mm}}
   \begin{tabular}{ccc}
    & (a) & $ R  \leq 1 $ \\
    &  & $ T \leq 1 $  \\
    & (b)& $R$ y $T$ are anti-resonant
   \end{tabular}
 \right\} \,.
\end{equation}

When one is dealing with the scattering of holes, one intuitively be placed in front of the continuity of the probability current density of the $k$-channel satisfies:
\begin{equation}
 \label{CC10}
  \sum^{N}_{i}T_{ki} + \sum^{N}_{i}R_{ki} = 1 \,,
\end{equation}
\ni but when one carefully analyze this problem that this relation is non-always true. Let us take the situation proposed in \cite{Sanchez95}, to start the analysis. We will suppose incidence only from the left, this means that $f = h = 0$ on Figure \ref{Fig:Potdis2}(a). From (\ref{cuasi-uni2}) one can obtain directly
\begin{eqnarray}
 \label{sim(r-t)3}
 j_{\mb{\tiny H}}|t_{\mb{\tiny HH}}|^{2} + j_{\mb{\tiny L}}|t_{\mb{\tiny LH}}|^{2}+
 j_{\mb{\tiny H}}|r_{\mb{\tiny HH}}|^{2} + j_{\mb{\tiny L}}|r_{\mb{\tiny LH}}|^{2}
 & = & j_{\mb{\tiny H}} \\
 \label{sim(r-t)4}
 j_{\mb{\tiny L}}|t_{\mb{\tiny LL}}|^{2} + j_{\mb{\tiny H}}|t_{\mb{\tiny HL}}|^{2}+
 j_{\mb{\tiny L}}|r_{\mb{\tiny LL}}|^{2} + j_{\mb{\tiny H}}|r_{\mb{\tiny HL}}|^{2}
 & = & j_{\mb{\tiny L}}\,.
\end{eqnarray}

\paragraph{\textsf{The problem of one incident quasi-particle}$\,$:} If one considers only one heavy hole propagating from $-\infty$, then one must use
(\ref{sim(r-t)3}), that when divided by $j_{\mb{\tiny H}}$, lead us to $T_{k \mb{\tiny H}} \leq 1, \; R_{k \mb{\tiny H}} \leq 1$ ant to the anti-resonant character of both magnitudes. This indicates that it is satisfied (\ref{CC9}), then a problem like this reproduces the conditions of incident particle/emergent without mixing. As can be seen, under the imposed considerations, the conductance in one channel is reduced to its transmission coefficient, then one has:
\begin{equation}
 \label{G1}
  G^{(1)}_{k} = T_{ki}\Bigl|_{i=k} \; \leq \; 1 \Bigr.\,.
\end{equation}
\paragraph{\textsf{The problem of two incident cuasi-particles}$\,$:}  Let`s consider that from $-\infty$ are synchronized incident to the scatterer one heavy hole and one light hole. To simplify we will take the holes softly interacting -we suppose small values of $k_{x}$ y $k_{y}$ at regions $R$ and $L$- this supposition is enough to our propose and then to have a problem mathematically well definid you need to take (\ref{sim(r-t)3}) and (\ref{sim(r-t)4}), and after some transformations we obtain:
\begin{equation}
 \label{CC12}
 T_{\mb{\tiny HH}} + T_{\mb{\tiny LH}} + T_{\mb{\tiny LL}} + T_{\mb{\tiny HL}} +
 R_{\mb{\tiny HH}} + R_{\mb{\tiny LH}} + R_{\mb{\tiny LL}} + R_{\mb{\tiny HL}}
 = 1 + \frac{j_{\mb{\tiny L}}}{j_{\mb{\tiny H}}}\,.
\end{equation}
\ni If one wants again a relation of the type (\ref{CC10}), it is necessary to normalize (\ref{CC12}) conveniently. To do so we multiply both members of (CC12) by $\eta = \frac{j_{\mb{\tiny H}}}{j_{\mb{\tiny L}} + j_{\mb{\tiny H}}}$, and then we have:
\begin{equation}
 \label{CC14}
  \eta \left\{\sum^{\mb{\tiny L,H}}_{k} T_{k \mb{\tiny H}} +
  \sum^{\mb{\tiny L,H}}_{k} T_{k \mb{\tiny L}} +
  \sum^{\mb{\tiny L,H}}_{k} R_{k \mb{\tiny H}} + \sum^{\mb{\tiny L,H}}_{k} R_{k \mb{\tiny L}}
  \right\}
   = 1 \,.
\end{equation}
\ni In this proper sense we underline the concordance between $\,\eta\;$ and what appears in Wessel´s report  \textit{et al} \cite{Wess89}, where the authors show in the transmission coefficient of channel $n$ that
$$
 D_{n}= \frac{\langle \bn{f}_{n}^{(t)}|j_{z}|\bn{f}_{n}^{(t)}\rangle}
         {\langle \bn{f}_{1}^{(i)}|j_{z}|\bn{f}_{1}^{(i)}\rangle}
      =  \mathrm{N}|t_{n}|^{2} \,,
$$
\ni where $\mathrm{N}$ is a normalization with no explicit definition in \cite{Wess89}. A similar idea was demand by the authors of \cite{Chao91}, who starting from a relation similar to (\ref{CC12}) obtain (\ref{CC14}), redefining the coefficients of transmission and reflection  in a way such that its sum is normalized to unit.

\begin{figure}[ht!]
 \begin{center}
 \includegraphics[width=0.5\linewidth,height=0.4\linewidth]{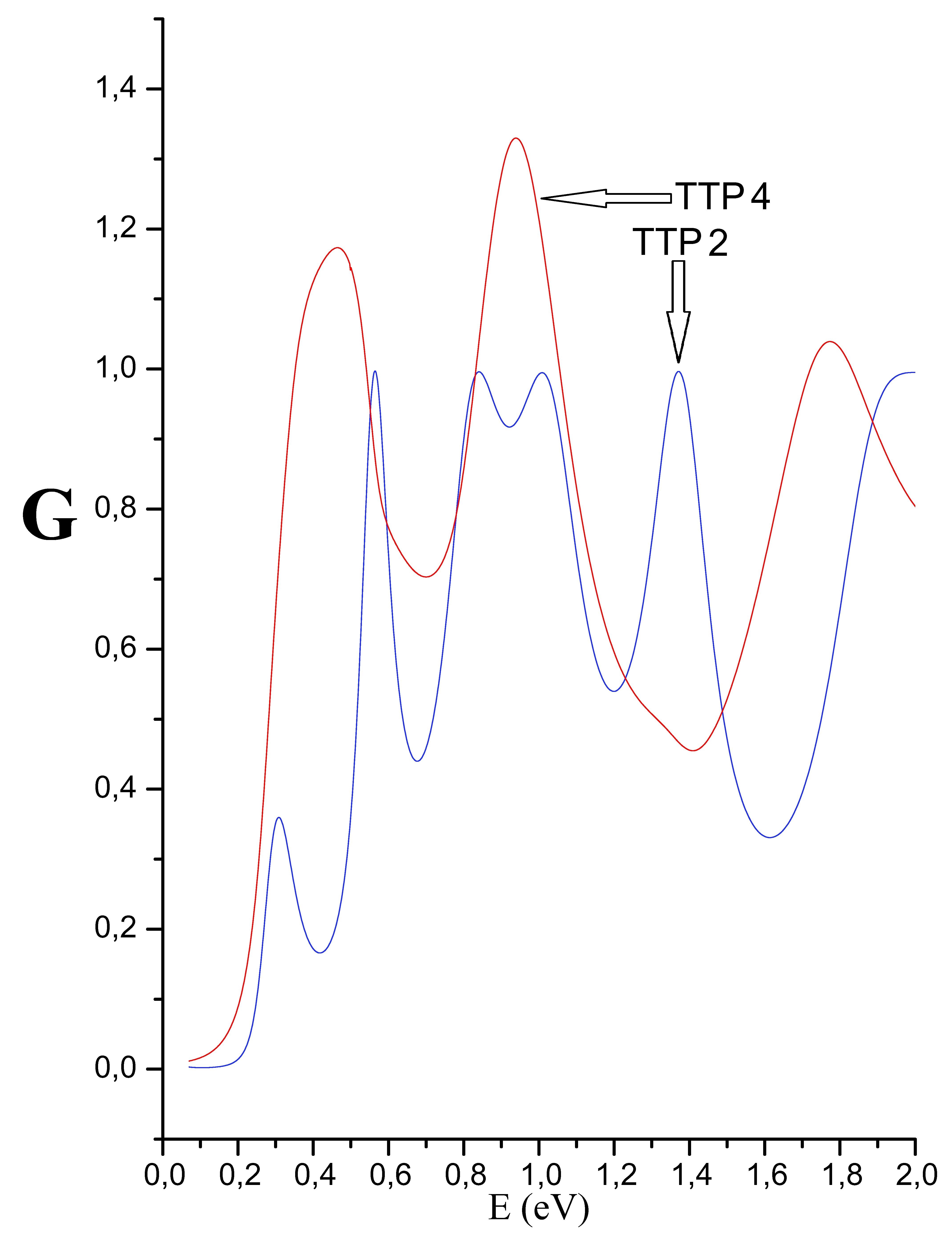}\\
 \caption{\label{Fig:Gbar} (Color online). Conductance of $hh$ and $lh$ throughout output channels $2$ and $4$ [see Fig.\ref{Fig:BVprofile}]. TTP stands for the Total Transmission Probability (\ref{Gk-def}). The TTP is shown as function of the incident energy for $hh_{\pm 3/2}$ and $lh_{\pm 3/2}$, impinging simultaneously on an $AlAs$ scattering barrier of $10\,\AA$ thick, embedded in $GaAs$ layers. Among the $4$ accesible channels [see Fig.\ref{Fig:BVprofile}], we display: the output channel $4$ (TT4), correspondent to $hh_{\mb{\tiny -3/2}}$ (red line) and the output channel $2$ (TTP2), correspondent to $lh_{\mb{\tiny -1/2}}$ (blue line).}
 \end{center}
\end{figure}

From (\ref{CC12}) it can be evaluated the \emph{\underline{one channel conductance}}. This term could be more appropriated than the one \emph{Total Probability of Transmission by one channel} (TTP), usually used in literature \cite{Sanchez95,Pereyra00}, because it reflects the better the information that gives the problem of several particles mixed incident and also it prevents confusions in the numerical evaluation as will be seen later. Conductance $G_{k}$ over the $k$-th channel is defined by 
 \begin{equation}
  \label{Gk-def}
   G_{k} = \sum^{\mb{\tiny N}}_{i} T_{ki}\,,
 \end{equation}
 \ni whose physical meaning is the collective transmission throughout the system emerging \emph{via} the $k$-th channel and quoted by the TTP (\ref{coef-Disp5b}). The last will be sampled for the output channel $H$ under incidence of $H$ and $L$, that is to say:
\begin{equation}
 \label{G2a}
  G_{\mb{\tiny H}} = T_{\mb{\tiny HH}} + T_{\mb{\tiny HL}}
  = 1 + \frac{j_{\mb{\tiny L}}}{j_{\mb{\tiny H}}} -
  \left\{T_{\mb{\tiny LL}} + T_{\mb{\tiny LH}} + R_{\mb{\tiny HH}} + R_{\mb{\tiny LH}} +
  R_{\mb{\tiny LL}} + R_{\mb{\tiny HL}}
  \right\}\,.
\end{equation}
\ni If one use an arbitrary basis of LI functions it can be shown that it is possible to find values of $\; G_{\mb{\tiny H}} \;$ for which $\;G_{\mb{\tiny H}} > 1\,,$ in the whole incident energy range $\;(E_{i})\,.$ To obtain the contrary as a result, the difference between the second and third terms of the right hand side of  (\ref{G2a}) must be zero. To do so, it must fulfill that $t_{\mb{\tiny LL}} = r_{\mb{\tiny HH}} = r_{\mb{\tiny HL}} = 0 $, but this implies that the transmission channel of light hole to himself and the reflection channel of heavy hole are closed $\forall\;E_{i}$. Then the scattering system is opaque to the wave going from a light hole channel to the L channel and transparent to the wave moving to channel H from any channel.  This is not acceptable, it is enough to mention that in a barrier of $\,10\;\AA$ $\;$ de $AlAs$, for $E_{i}\,< 200 \;meV, \; R_{\mb{\tiny HH}} \simeq 1$ \cite{Sanchez95}. Then the initial supposition is true $\forall\;E_{i}$ of holes. Note that what was said does not exclude, eventually, $G_{\mb{\tiny H}} \leq 1$ for certain values of energy.

\subsubsection{\textsf{Estimation of Conductance}:}
 \label{Cap:SM::Tun-Amplit:::Conserva-Flux::::EstimaG}

\hspace*{6mm} You can estimate the maximum value of $G_{k}$. The idea under this calculation is to take into account the complementary contribution of crossed paths to the flux of a direct transition when you include different particles that make a non-zero mix. Each channel produces a contribution independent of the other, then this behavior is a point of the Principle of Superposition. The following is not a rigorous demonstration but a criteria for the evaluation of the results and it is valid for problems with sectional constant potential, with arbitrary basis of LI functions.
You can take into account that we will operate with \emph{superior bounds}  or \emph{maximum values} and not with the possible numerical graduation of physical magnitudes. Let us suppose we have the problem of \textsf{an incident particle}. In this case, the \textbf{upper bound} of the one channel conductance is $1$, as it is said in (\ref{G1}). Let us add immediately another charge carrier, so you have move to a \emph{two incident particles} with coupling. In this case we have changed to the \textsf{two incident particles} with coupling and in this case the \textbf{superior bound} of the one channel conductance, that was of $1$, now has an addition that is observed in  (\ref{G2a}). Let us think on Wave Superposition: the maximum value that a channel can transfer to himself is the whole incident energy- this will get $1$ for the probability of occurrence of this event. The next step could be the addition of the fraction that represents the crossed transition. The added carrier can add to the observation channel a maximum value of \mb{\Large{$\frac{1}{2}$}}, because the other \mb{\Large{$\frac{1}{2}$}} that ``rest to it" is the maximum value it can assigned to itself. This equipartition of energy can be considered as a manifestation of the phenomenon of interference due to constructive superposition and it is the key point of what we want to describe. Let us see the problem of \textsf{N incident particles} with coupling under the assumption that we are adding a fraction of the flux when we are adding particles in the analysis. Making an analysis by induction progressively, one can obtain:
\begin{eqnarray}
 \label{G3}
 G_{k}\Bigl|_{\mb{\tiny N}}\; \approx \; 1 + \frac{1}{N} +  \frac{1}{N} +
  \ldots + \frac{1}{N} \; \approx \; 1 + (N-1)\frac{1}{N} \;
  \approx \; \frac{2N-1}{N} \; < \; \limsup = 2 \Bigr.; \nn \\
  \hspace{10cm}  \forall  \, N \geq 2 \,. \nn \\
\end{eqnarray}

This analysis made lead us to conclude that conductance $G_{k}$ by the $k$-\textit{th} channel is strictly \textbf{less than} $\mathbf{2}$ and is independent of the number of particles of the incident flux.

\begin{figure}[ht!]
 \centering
 \subfigure[(Color online). Superlattice of $\{GaAs/AlAs/GaAs\}^{n}$; with cell dimension $(25-20-25) \AA$, $n = 24$ cells, $k_{x}=0$ and $k_{y}=0$]{\includegraphics[width=0.6\linewidth]{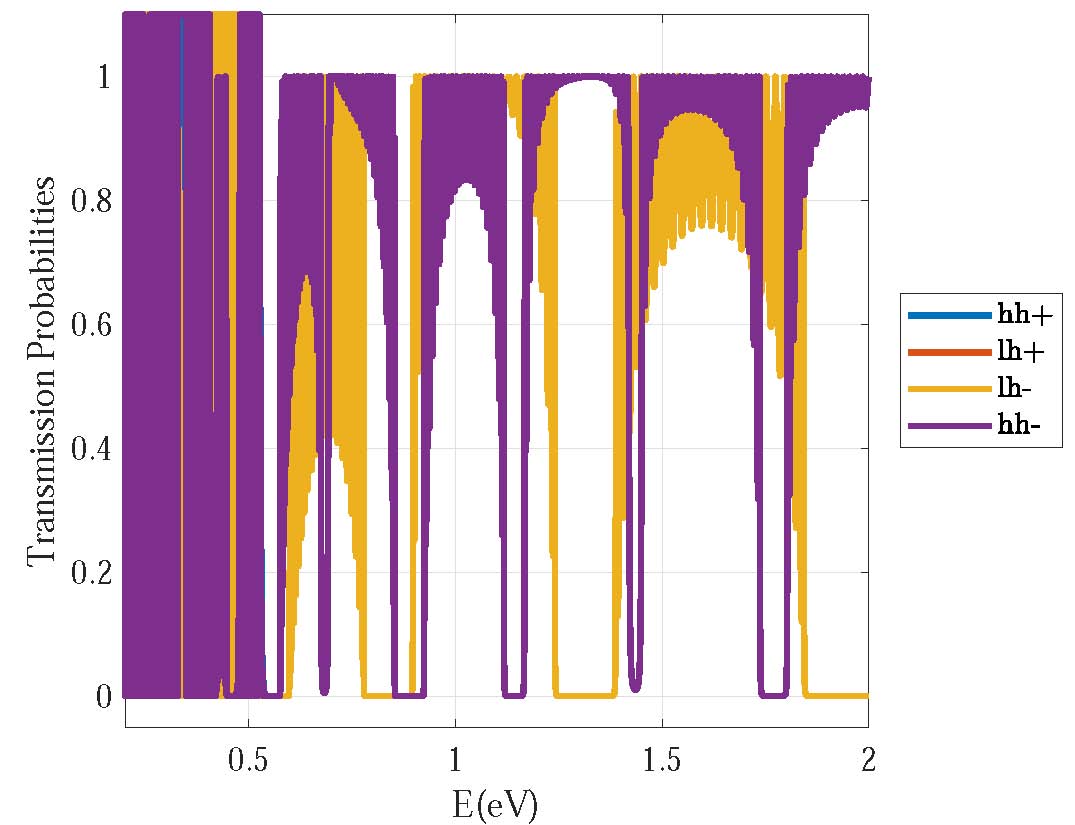}}\\
 \subfigure[(Color online). Condition number of the state-vector matrix $\bn{M}_{sv}(z,z_{0})$ (see \ref{Sec:AppB}) and the amplitude-matrix $\bn{t}$ for $\{GaAs/AlAs/GaAs\}^{n}$, $n=24$, are shown in the upper frame. Lower frame displays the same for the block $\bn{\delta}$ of $\bn{M}_{sv}(z,z_{0})$.] {\hspace{-17mm}\includegraphics[width=0.5\linewidth]{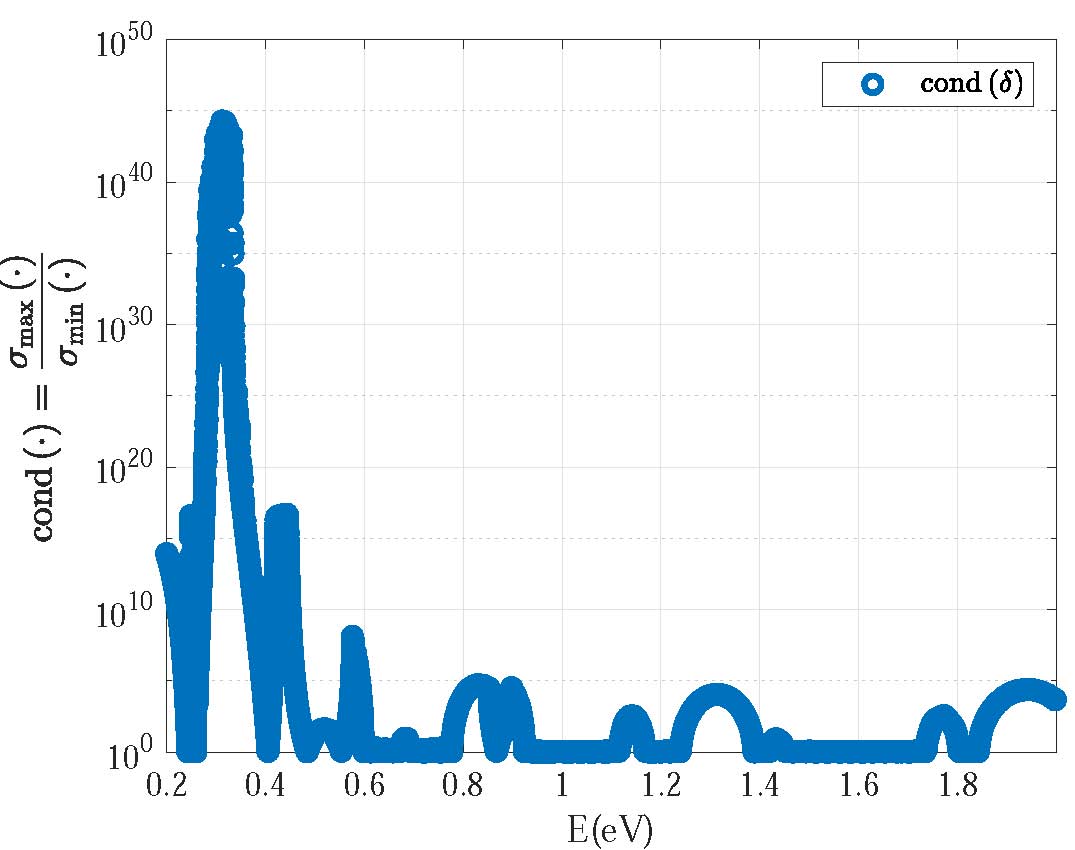}}\\
    \caption{\label{Fig:Nomix24} Transmission Probabilities of $hh_{\pm 3/2}$ and $lh_{\pm 1/2}$, for all the channels of the system, as function of the incident energy. The hole incidence upon a layered heterostructure is assumed as simultaneous. In panels (a)/(b), we have taken $V_{b} = 0.498$ eV.}
\end{figure}

For the conductance $G$ of the system one reaches:\cite{Landauer94,Diago19}
\begin{defn}
 \label{Gsis-def}
 \begin{equation}
    G = Tr\Bigl(\bn{t}\,\bn{t}\t\Bigr) = \sum^{{\mb {\tiny N}}}_{k} G_{k}\,,
 \end{equation}
\end{defn}
\ni and the formulation is similar.

When we add a new carrier, the conductance of the system now have added fractions that can be taken into account by introducing in (\ref{Gsis-def}) ``the sum" of the upper bounds of conductance in each channel. We considered that the introduced carrier can only add to the inicial channel is \mb{\Large{$\frac{1}{2}$}}, meanwhile the other \mb{\Large{$\frac{1}{2}$}} is the maximum value it can transfer to itself and all together is the contribution of this charge carrier to the conductance of the system. We suppose there are not sources and drains of charge carriers. Using complete induction one obtains:
\begin{eqnarray}
 \label{G4}
  G \Bigl|_{\mb{\tiny N}} \; \approx \; 1 + \frac{N-1}{N} +
  \ldots + \frac{N-1}{N} \; \approx \;
  1 + N\left(\frac{N-1}{N}\right) = 1 + (N-1)
  \; \leq \; \sup = N  \Bigr.\,. \nn \\
  \hspace{11cm}  \forall  \, N \geq 1 \, \nonumber \\
\end{eqnarray}

\begin{figure}[ht!]
 \centering
 \subfigure[(Color online). Superlattice of $\{GaAs/AlAs/GaAs\}^{n}$; with cell dimension $(20-20-20) \AA$, $n = 8$ cells, $k_{x}=0.01\AA^{-1}$ and $k_{y}=0$]{\includegraphics[width=0.6\linewidth]{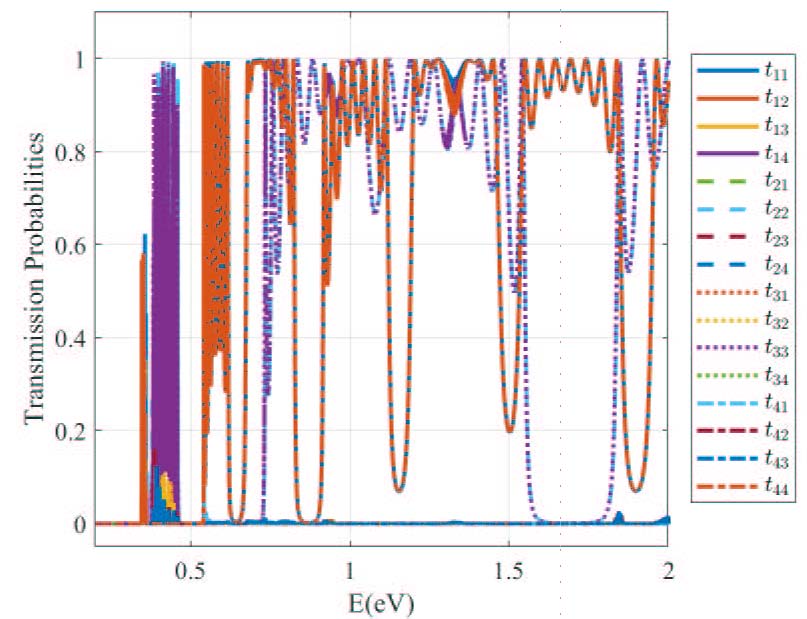}}\\
 \subfigure[(Color online). Condition number of the state-vector matrix $\bn{M}_{sv}(z,z_{0})$ (see \ref{Sec:AppB}) for $\{GaAs/AlAs/GaAs\}^{n}$, $n=8$]{\hspace{-17mm}\includegraphics[width=0.5\linewidth]{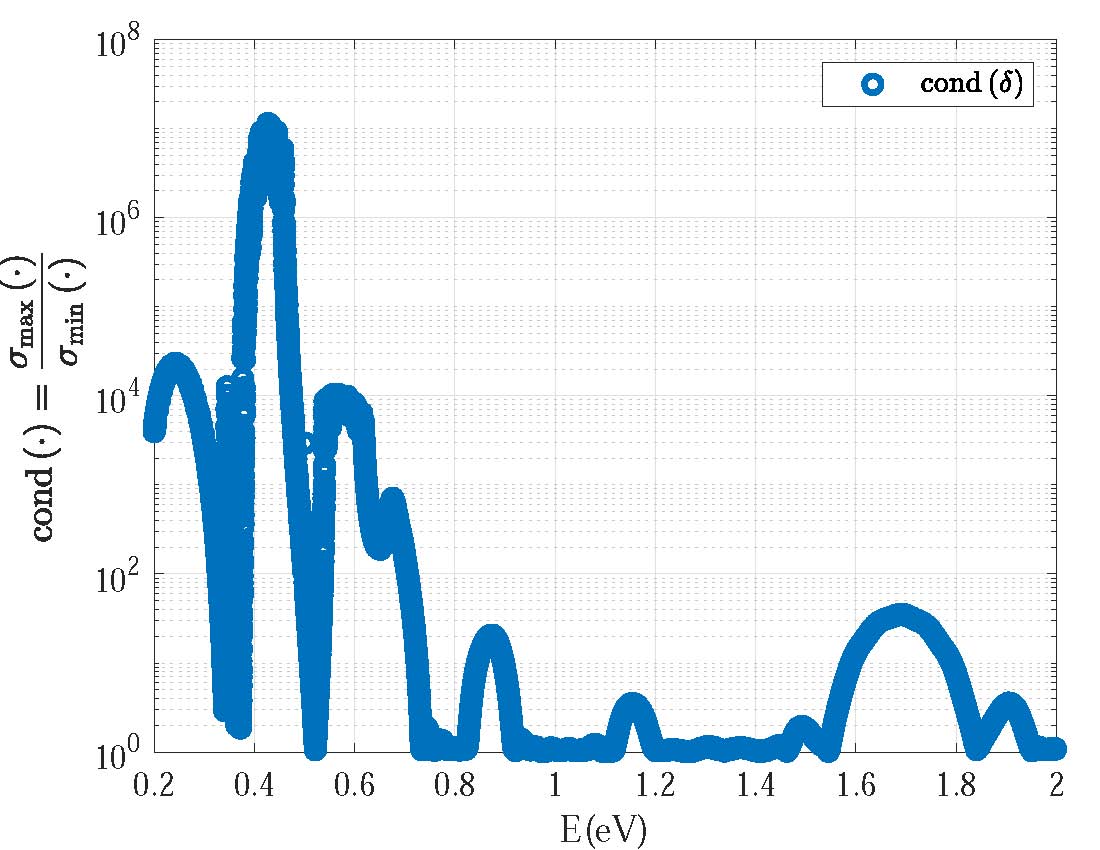}}\\
    \caption{\label{Fig:Mix8} Transmission Probabilities of $hh_{\pm 3/2}$ and $lh_{\pm 1/2}$, for all the channels of the system, as function of the incident energy. The hole incidence upon a layered heterostructure is assumed as simultaneous. In panels (a)/(b), we have taken $V_{b} = 0.498$ eV.}
\end{figure}

Conductance of $hh$ and $lh$ throughout output channels $2$ and $4$ [see Fig.\ref{Fig:BVprofile}]. TTP stands for the Total Transmission Probability (\ref{Gk-def}). The TTP is shown as function of the incident energy for $hh_{\pm 3/2}$ and $lh_{\pm 3/2}$, impinging simultaneously on an $AlAs$ scattering barrier of $10\,\AA$ thick, embedded in $GaAs$ layers. Among the $4$ accesible channels [see Fig.\ref{Fig:BVprofile}], we display: the output channel $4$ (TT4), correspondent to $hh_{\mb{\tiny -3/2}}$ (red line) and the output channel $2$ (TTP2), correspondent to $lh_{\mb{\tiny -1/2}}$ (blue line).

Figure \ref{Fig:Gbar} shows the scattering process of $hh$ and $lh$, through a single $AlAs$ QB of $\,10\;\AA$ thick. Here, we do not impose the basis to fulfills (\ref{Orto-QEP(gen)}). The phenomenology is comparable to that shown in Reference \cite{Sanchez95}, for a $GaAs$ QW of $50\,\AA\,$ width, embedded in $Al_{\,0.3}\,Ga_{\,0.7}As$ [see Fig.\ref{Fig:Potdis2}(b)]. In that reference, the conductance throughout the channel L, becomes $G_{\mb{\tiny L}}\approx 1.2$ for $E = 35\; meV$, and obviously do not fulfill the flux conservation principle (\ref{CC10}). In that case, the incoming quasi-particle is solely a heavy hole H (means that for the incoming L, the amplitud becomes strictly zero)\footnote{The right-hand symbol stands for the input channel, meanwhile the other is the output one}. The result of Reference \cite{Sanchez95} and ours [vea la Fig.\ref{Fig:Gbar}], are both obtained for a single cell ($n = 1$)[see Fig.\ref{Fig:BVprofile}]. We plot TTP curves of Fig.\ref{Fig:Gbar} quoting (\ref{Gk-def}) as a function of $hh_{\pm 3/2}$ and $lh_{\pm 3/2}$ incident energy. It is assumed a simultaneous incidence upon a single scattering QB of $AlAs$ ($10\,\AA$ thick), embedded into $GaAs$ layers. For the channel $4$, which describes the tunneling of a $hh_{-3/2}$ (red solid line), it turns that $G_{hh-3/2} \approx 1.2$ for $E = 0.4\,$eV and $G_{hh-3/2} \approx 1.34$ for $E= 0.9\,$eV. Notice the agreement of these results with the estimation predicted by (\ref{G3}). However, is straightforward the violation of unitarity flux requirement (\ref{Unitar}) for holes, and correspondingly they take apart from the statistical rule (\ref{CC10})\footnote{The reader should keep in mind that $hh$($lh$), represent heavy(light) hole, respectively.}. It is convenient to stress the origin of the prompt breaks of a general conservation principle, since we have taken an arbitrary basis of LI functions, that do not satisfy the definition (\ref{Orto-QEP(gen)}). To make this fact evident, let us next re-examine the vector-space of solutions, following at this time the procedure to build a completely orthonormalized basis as described in definitions (\ref{Def:Ortnonor-Full}) and (\ref{Orto-QEP(gen)}).

 Figure \ref{Fig:Nomix24} and Figure \ref{Fig:Mix8} show an analog situation to that of Fig.\ref{Fig:Gbar}, but with a clue difference; now we expand the envelope function (\ref{eq2}) for describing a system's state, on certain $(N\times 1)$ super-spinors $\mb{\bm $\Gamma$}_j$, that have been derived as \emph{eigen}-solutions of the QEP (\ref{QEP1}) satisfying as well, the complete orthonormalization conditions (\ref{Orto-QEP(1)}) and (\ref{Orto-QEP(1)}). We will discuss a superlattice (SL) of $\{GaAs/AlAs/GaAs\}^{n}$, with $n = 24(8)$ cells [Fig.\ref{Fig:Nomix24}(\ref{Fig:Mix8})]. We have taken $k_{x} = 0(0.01)\AA^{-1}$, $k_{y} = 0$ and $V_{b} = 0.498$ eV, for the Fig. \ref{Fig:Nomix24}(\ref{Fig:Mix8}), respectively.

 Figure \ref{Fig:Nomix24}(a), shows transmission probabilities for direct paths [\textit{i.e.}, when $i = j$, see Fig.\ref{Fig:BVprofile}], completely superposed. This picture fulfills for channels $hh_{\pm 3/2}$ and the same for those of $lh_{\pm 1/2}$. A basic coincidence displayed in the present panel, derives from the interplay of $hh$ and $lh$ quasi-particles with the $\{GaAs/AlAs/GaAs\}^{24}$ scattering system, whose SL's potential do not seem powerful enough to take apart $hh_{+3/2}(lh_{+1/2})$ from $hh_{-3/2}(lh_{-1/2})$, respectively \cite{Diago06}. Given the absence of $hh-lh$ mixing, shortly $k_{x} = k_{y} = 0\AA^{-1}$, it is straightforward the lack of crossed transitions [\textit{i.e.}, when $i \neq j$, see Fig.\ref{Fig:BVprofile}], which remain forbidden by the flux conservation principle (\ref{CC10}). We remark that no violation of this last general law was detected --no matter how long the SL one takes--, and it is not difficult to explain, since completely orthonormalized basis (\ref{Def:Ortnonor-Full}) accounts for the observed results.

 Figure \ref{Fig:Mix8}(a), displays the same as Fig.\ref{Fig:Nomix24}(a), but for the SL $\{GaAs/AlAs/GaAs\}^{8}$, in the presence of strong subband $hh-lh$ mixing, which means that $k_{x} = 0.01\AA^{-1}$, although $k_{y}$ remains zeroed. Hence the large entry for $k_{x}$, the presence of finite values for transmission probabilities throughout crossed paths, whose contributions turns into a competitor with those of the direct paths, becomes expected. It is simple to argue that, because the in-plane quasi-momentum, is widely-accepted as a trigger of $hh-lh$ mixing. Thus, the allowed transverse movement (\emph{via} $k_{x} = 0.01\AA^{-1}$), now induces transitions trough crossed paths \cite{Diago06}. Worthwhile to remark, the preservation of unitarity (\ref{Unitar}) for the outgoing flux, as can be straightforwardly observed in panel Figure \ref{Fig:Mix8}(a), despite the strong $hh-lh$ mixing regime. We have taken a completely orthonormalized basis (\ref{Def:Ortnonor-Full}), leading our results to a good agreement with (\ref{CC10}). Importantly, there is another difference with calculations reported in Reference \cite{Sanchez95}: none of the incoming amplitudes for propagating $hh$($lh$) modes, have been assumed as zero \emph{a priori}. Instead, they were calculated by solving the QEP (\ref{QEP1}). This procedure of a synchronous incidence of $hh-lh$, was amply discussed and successfully applied elsewhere for a $III-V$ semiconducting double-barrier resonant tunneling and a perfect-interface SL \cite{Diago06}.

When quoting scattering amplitudes in the framework of the SM theory, one must be specially careful with the equations (\ref{coef-Disp3-4}), for two reasons. Firstly, the TM formalism has well-known intrinsic numerical instabilities, that could yield mathematical-artifacts. Secondly, the inversion-matrix operations, might lead to heavily distorted results. In both cases, one can overcome the difficulties, by using the filters defined in the proposition (\ref{Propo:Filtros}). In the present calculations, we have avoided the first problem, by taking a proper partition of the single cell \textit{z}-coordinate interval \cite{Diago06}, \emph{via} (\ref{Propo:Filtros})(b). Meanwhile, the second challenge was faced, by monitoring the condition number of the state-vector matrix $\bn{M}_{sv}(z,z_{0})$ [see \ref{Sec:AppB}]. The condition number of a norm $2$ squared matrix $\bn{M}$, is
\begin{equation}
 \label{NumC}
   \mb{cond}(\bn{M}) = \frac{\sigma_{max}(\bn{M})}{\sigma_{min}(\bn{M})},
\end{equation}
\noindent where $\sigma_{max,min}(\bn{M})$, stands for the singular maximum(minimum) of the matrix $(\bn{M})$, respectively, which are positive-defined numbers. This way, (\ref{NumC}), evaluates somehow the numerical artifacts, that could possible rise from the presence of $(\bn{\delta})^{-1}$ in the expression for the transmission amplitudes (\ref{coef-Disp3-4}). As a bonus, (\ref{NumC}) measures the accuracy when dealing with a matrix inversion operation and/or the algebraic system's solutions.

Therefore, a major aim of Figure \ref{Fig:Nomix24}(b) and Figure \ref{Fig:Mix8}(b), is to provide a reliability index on the scattering coefficients' data. However, worthwhile remarking that the information extracted from (\ref{NumC}), is not enough for a complete characterization of data confidence: It is the flux conservation principle (\ref{CC10}), the one that does so. The larger the $\mb{cond}(\bn{M}_{sv}(z,z_{0}))$, the lest reliable transmission probabilities. The last, could be explained based on the lost of significant numbers ($\approx 10$ or more), due to accumulative machine's accuracy errors. Thereby our results here, have at least a $6$-digit significant-number accuracy, taking into account the double-precision for real-number representation, we have assumed. We have verified two main trends during tunneling amplitudes calculations, within the framework of the chosen physical parameters, namely: (i) the $\mb{cond}(\bn{M}_{sv}(z,z_{0}))$, rises with the number of SL's layers; and (ii) when $\mb{cond}(\bn{M}_{sv}(z,z_{0})) \gg 10^{10}$, the transmission probabilities do not remains $\leq 1$, thus being far from unitary condition (\ref{CC10}). Top[bottom] panel of Fig.\ref{Fig:Nomix24}(b), plots $\mb{cond}(\bn{M}_{sv},\bn{t}/[(\bn{\delta})^{-1}])$, respectively. In the interval of $E \in [0.5, 0.6]$ eV, despite $\mb{cond}(\bn{M}_{sv})\gg 10^{10}$, it is observed that $\mb{cond}(\bn{t})< 10^{10}$, which is a consequence of a dominant contribution of $\mb{cond}(\bn{\delta})< 10^{10}$ [see bottom panel]. For the rest of the energy interval, $\mb{cond}(\bn{M}_{sv},\bn{t}) < 10^{2}$ [see top panel (blue x)/(red circles), respectively]. Figure \ref{Fig:Mix8}(b), shows that $\mb{cond}(\bn{\delta})< 10^{10}$ for $E \in [0.5, 0.6]$ eV, meanwhile $\mb{cond}(\bn{\delta})< 10^{5}$ elsewhere. Then, the correctness of transmission probabilities, in the rank of interest for the incoming $E$, for both $\{GaAs/AlAs/GaAs\}^{(8,24)}$ is guaranteed.

A widely discussed problem in quantum physics, is the meaning of an eigenvalue's problem solutions. Within the framework of the propagating modes approximation [see reference \cite{Diago06} and references therein], the eigenvalues $\lambda$ of $\bn{M}_{sv}(z,z_{0})$ have been calculated \footnote{The reader should consider these characteristic values as no longer related to $hh$ or $lh$.}. In this case, the numerical simulation focuses an attempt to determine the evasive threshold energy ($E_{\tiny{TH}}$) for quantum tunneling of $hh$ and $lh$, throughout semiconducting scattering systems. We have exercised different samples of the $\{GaAs/AlAs/GaAs\}^{n}$ SL, and observed that $\lambda(GaAs)\in\Re$; $\forall E$, while $\lambda(AlAs) \in \Im$ for $ E < 0.5$ eV; and $\lambda(AlAs) \in \Re$ for $ E \geq 0.5$ eV. Besides, for $n = 24$, in the interval $E < 0.5$ eV, we have detected that $\mb{cond}(\bn{\delta}) \gg 10^{20}$, with nonsense accompanying tunneling probabilities. With these preliminary results, we foretell the interplay for $\bn{M}_{sv}(z,z_{0})$ eigenvalues $\lambda(GaAs/AlAs)$, together with the large values of $\mb{cond}(\bn{\delta})$, as complementary tools for a more accurate definition of $E_{\tiny{TH}}$ for tunnelling channels in a scattering experiment, a novel mathematical procedure yet to be refined.

\section{Concluding remarks}
Once the choice of a completely orthonormalized basis has been settled on, and provided we used it consequently, no FC numerical inconsistencies should arise, when dealing with quantum transport calculation in the framework of the EFA, within the MMST. Thus, rather arbitrary conditions to the basis-set and/or to the output scattering coefficients to preserve FC, should not be necessary. The symmetry requirements on the TM objects, the so called filters, represent  paramount complements to the FC and unitarity condition on the SM, whose advantages have been demonstrated. We foretell that the present general theoretical modelling, is valid for different kind of multiband-multicomponent physical systems of mixed charge-spin carriers, within the EFA, with minor transformations if any.

\section{Appendixes}
\begin{appendix}
\section{\hspace{15mm} Kohn-L\"{u}ttinger model Parameters}
 \label{Sec:AppA}
\hspace{4mm}The following parameters correspond to the KL Hamiltonian \cite{Broido85,Diago06}
\beann
  P = \frac{\hbar^{2}}{2m_{0}} \gamma_{1} (\kappa_{\mb{\tiny T}}^{2} + k_{z}^{2}) & \mb{;} &
  Q = \frac{\hbar^{2}}{2m_{0}} \gamma_{2} (\kappa_{\mb{\tiny T}}^{2} - 2k_{z}^{2}) \\
  R = \frac{\hbar^{2}\sqrt{3}}{2m_{0}} (\mu k_{+}^{2} - \gamma k_{-}^{2}) &
  \mb{;} &
  S = \sqrt{3} \;\; \frac{\hbar^{2}}{2m_{0}} \;\; \gamma_{3} k_{-} k_{z}\\
  T = - \frac{\hbar^{2}}{2m_{0}} \;\; \beta k_{-} & \mb{;} &
  T^{\prime} = -\frac{2}{\sqrt{3}} \;\;
                      \frac{\hbar^{2}}{2m_{0}} \;\; \beta k_{z}\\
  k_{\pm} = k_{x} \pm \imath k_{y} & \mb{;} &
  \kappa_{\mb{\tiny T}}^{2} = k_{x}^{2} + k_{y}^{2} \\
  \gamma = \frac{1}{2} (\gamma_{2}+\gamma_{3}) & \mb{;} &
  \mu = \frac{1}{2} (\gamma_{3}-\gamma_{2})\\
  A_{1} = \frac{\hbar^{2}}{2m_{0}} (\gamma_{1}+\gamma_{2}) & \mb{;} &
  A_{2} = \frac{\hbar^{2}}{2m_{0}} (\gamma_{1}-\gamma_{2}) \\
  B_{1} = \frac{\hbar^{2}}{2m_{0}} (\gamma_{1}+2\gamma_{2}) & \mb{;} &
  B_{2} = \frac{\hbar^{2}}{2m_{0}} (\gamma_{1}-2\gamma_{2}) \\
  C_{xy} = \sqrt{3}\;\frac{\hbar^{2}}{2m_{0}}\;\sqrt{\gamma_{2}^{2}
  (k_{x}^{2}-k_{y}^{2})^{2}+4\gamma_{3}^{2}k_{x}^{2}k_{y}^{2}} & \mb{;} &
  D_{xy} = \sqrt{3}\;\frac{\hbar^{2}}{m_{0}}\;\gamma_{3}\;\kappa_{\mb{\tiny T}} \\
  \mathcal{A}_{1} = \gamma_{1} + \gamma_{2} & \mb{;} &
  \mathcal{A}_{2} = \gamma_{1} - \gamma_{2}\\
  \mathcal{B}_{1} = \gamma_{1} + 2\gamma_{2} & \mb{;} &
  \mathcal{B}_{2} = \gamma_{1} - 2\gamma_{2}\\
  q_{i} = k_{i} a_{0} \;\;\; & \mb{;} & \;\;\;i=x,y,z\\
  q_{\mb{\tiny T}}^{2} = q_{x}^{2}+q_{y}^{2} & \mb{;} &
  q = \lambda a_{0}\\
  t_{xy} = \frac{C_{xy}}{Ry} & \mb{;} &
  S_{xy} = \frac{D_{xy}}{Ry} \\
  {\cal E}=V(z)-E,\;\;\; \tilde{\cal E} = \frac{{\cal E}}{Ry}\;\;\;& \mb{;} & \;\;\;V_{0}=0 \\
  \beta_{\mb{\tiny T}} = \frac{1}{2} \left[(\mathcal{A}_{1}\mathcal{B}_{1} +
    \mathcal{A}_{2}\mathcal{B}_{2})q_{\mb{\tiny T}}^{2}
    +(\mathcal{B}_{1} + \mathcal{B}_{2})\tilde{\cal E}-S_{xy}^{2}\right] & \mb{;} &
  \alpha_{\mb{\tiny T}}= \mathcal{B}_{1} \mathcal{B}_{2} \\
  \delta_{\mb{\tiny T}} = \mathcal{A}_{1}\mathcal{A}_{2}q_{\mb{\tiny T}}^{4} + (\mathcal{A}_{1} & + &
                          \mathcal{A}_{2})\tilde{\cal E}q_{\mb{\tiny T}}^{2}
    +\tilde{\cal E}^{2}-t_{xy}^{2}
\eeann

\textbf{Parameters from the Hamiltonians} $\hat{\bn{\cal H}}_{u}$ \textbf{and} $\hat{\bn{\cal H}}_{l}$
 \label{Sec:AppA::Hul}
\be
 \left.{\hspace{1mm}} \begin{tabular}{ccc}
  $g_{11}$ & = & $ \mathcal{A}_{2}q_{\mb{\tiny T}}^{2} + \mathcal{B}_{1}q_{1}^{2} + \tilde{\cal E} $ \\
  $g_{13}$ & = & $ \mathcal{A}_{2}q_{\mb{\tiny T}}^{2} + \mathcal{B}_{1}q_{3}^{2} + \tilde{\cal E} $ \\
  $g_{22}$ & = & $ -(t_{xy} + \imath S_{xy}q_{2}) $ \\
  $g_{24}$ & = & $ -(t_{xy} + \imath S_{xy}q_{4})$ \\
  $g_{31}$ & = & $ \mathcal{A}_{1}q_{\mb{\tiny T}}^{2} + \mathcal{B}_{2}q_{1}^{2} + \tilde{\cal E} $ \\
  $g_{33}$ & = & $ \mathcal{A}_{1}q_{\mb{\tiny T}}^{2} + \mathcal{B}_{2}q_{3}^{2} + \tilde{\cal E} $ \\
  $g_{42}$ & = & $ -(t_{xy} + \imath S_{xy}q_{2}) $ \\
  $g_{44}$ & = & $ -(t_{xy} + \imath S_{xy}q_{4}) $
 \end{tabular}\right\},
\ee

\ni Due to the peculiarities of the Hamiltonians, it can be deduced that:
\be
 \left.{\hspace{1mm}} \begin{tabular}{ccc}
  $q_{2}$ & = & $-q_{1}$ \\
  $q_{4}$ & = & $-q_{3}$ \\
  $g_{11}$ & = & $ g_{12}$ \\
  $g_{13}$ & = & $ g_{14}$ \\
  $g_{22}$ & = & $ g_{21}^{*}$ \\
  $g_{24}$ & = & $ g_{23}^{*}$ \\
  $g_{31}$ & = & $ g_{32}$ \\
  $g_{33}$ & = & $ g_{34}$ \\
  $g_{42}$ & = & $ g_{41}^{*}$ \\
  $g_{44}$ & = & $ g_{43}^{*}$
 \end{tabular}\right\},
\ee
\ni where
\beann
 \left.{\hspace{1mm}} \begin{tabular}{ccc}
  $g_{1j},g_{3j}$ & is real \\
  $g_{2j},g_{4j}$ & is complex
 \end{tabular}\right\}.
\eeann
\ni The L\"{u}ttinger parameters: $\gamma_{1},\gamma_{2},\gamma_{3}$ characterize each layer of the structure.\\

\textbf{Matrix elements of the FTM} $\bn{M}_{u}(z,z_{o})$
 \label{Sec:AppA::Mu}

This matrix corresponds to the system of differential equations of the sub-space \textit{up} \cite{Diago06}, described by the Hamiltonian of this sub-space and its matrix elements are given by:
\begin{eqnarray*}
  \mb{\Large {(}}M_{u}(\xi,\xi_{o})\mb{\Large {)}}_{ij} = \frac{1}{\Delta_{\mb{\tiny T}}}\;
   \left[
    \ell^{\,(1)}_{ij}\cos(\tilde{q_{\mb{\tiny \textit{lh}}}})+\ell^{\,(2)}_{ij}
    \cos(\tilde{q_{\mb{\tiny \textit{hh}}}})
    \ell^{\,(3)}_{ij} sin(\tilde{q_{\mb{\tiny \textit{lh}}}})+\ell^{\,(4)}_{ij}
    sin(\tilde{q_{\mb{\tiny \textit{hh}}}})
   \right]\,,
\end{eqnarray*}

\begin{table}
 \caption{\label{TabAppA} Matrix elements of $\bn{M}_{u}(z,z_{o})$ that correspond to the sub-space \textit{up} of KL. This matrix belongs to the system of differential equations of the sub-space ($2 \times 2$) \cite{Diago06}}

 \begin{center}
  \begin{tabular}{||c|cccc||} \hline\hline
    Coefficients &              &              &              &      \\
    of the matrix elem. &$\ell^{(1)}_{ij}$&$\ell^{(2)}_{ij}$&$\ell^{(3)}_{ij}$
                      & $\ell^{(4)}_{ij}$\\ \hline
    $M_{11}$&
        $\mathcal{B}_{2}q_{1}\alpha_{1}\Omega_{3}$&
        $-\mathcal{B}_{2}q_{1}\alpha_{2}\Omega_{1}$&
        $\alpha_{1}\alpha_{2}t_{xy}S_{xy}$&$-\alpha_{1}\alpha_{2}
         \frac{q_{1}}{q_{3}}t_{xy}S_{xy}$\\
    $M_{12}$&
        $-\mathcal{B}_{2}q_{1}\alpha_{1}\alpha_{2}t_{xy}$&
        $\mathcal{B}_{2}q_{1}\alpha_{1}\alpha_{2}t_{xy}$&
        $-S_{xy}\Theta_{2}$&
        $-\frac{q_{1}}{q_{3}}S_{xy}\Theta_{2}$ \\
    $M_{13}$&
        $0$&
        $0$&
        $\alpha_{1}\Theta_{1}$&
        $-\frac{q_{1}}{q_{3}}\alpha_{2}\Theta_{1}$ \\
    $M_{14}$&
        $-\mathcal{B}_{2}q_{1}\alpha_{1}\alpha_{2}S_{xy}$&
        $\mathcal{B}_{2}q_{1}\alpha_{1}\alpha_{2}S_{xy}$&
        $-\mathcal{B}_{2}\alpha_{1}\alpha_{2}t_{xy}$&
        $\frac{q_{1}}{q_{3}}b_{2}\alpha_{1}\alpha_{2}t_{xy}$ \\ \hline
    $M_{21}$&
        $q_{1}t_{xy}\Theta_{3}$&
        $-q_{1}t_{xy}\Theta_{1}$&
        $S_{xy}\Theta_{4}$&
        $-\frac{q_{1}}{q_{3}}S_{xy}\Theta_{4}$ \\
    $M_{22}$&
        $-\alpha_{2}q_{1}\Theta_{1}$&
        $\alpha_{1}q_{1}\Theta_{1}$&
        $-S_{xy}t_{xy}\alpha_{1}\alpha_{2}$&
        $\frac{q_{1}}{q_{3}}S_{xy}t_{xy}\alpha_{1}\alpha_{2}$ \\
    $M_{23}$&
        $-q_{1}S_{xy}\Theta_{1}$&
        $q_{1}S_{xy}\Theta_{1}$&
        $t_{xy}\Theta_{1}$&
        $-\frac{q_{1}}{q_{3}}t_{xy}\Theta_{1}$ \\
    $M_{24}$&
        $0$&
        $0$&
        $-\mathcal{B}_{2}\alpha_{2}\Omega_{1}$&
        $\mathcal{B}_{2}\alpha_{1}\Omega_{3}\frac{q_{1}}{q_{3}}$ \\ \hline
    $M_{31}$&
        $q_{1}\alpha_{1}\alpha_{2}S_{xy}t_{xy}$&
        $-q_{1}\alpha_{1}\alpha_{2}S_{xy}t_{xy}$&
        $-\mathcal{B}_{2}q_{1}^{2}\alpha_{1}\Omega_{3}$&
        $\mathcal{B}_{2}q_{1}q_{3}\alpha_{2}\Omega_{1}$ \\
    $M_{32}$&
        $-q_{1}\alpha_{1}\alpha_{2}S_{xy}\Theta_{2}$&
        $q_{1}\alpha_{1}\alpha_{2}S_{xy}\Theta_{2}$&
        $\mathcal{B}_{2}q_{1}^{2}\alpha_{1}\alpha_{2}t_{xy}$&
        $-\mathcal{B}_{2}q_{1}q_{3}\alpha_{1}\alpha_{2}t_{xy}$ \\
    $M_{33}$&
        $\alpha_{1}q_{1}\Theta_{1}$&
        $-\alpha_{2}q_{1}\Theta_{1}$&
        $0$&
        $0$ \\
    $M_{34}$&
        $-\mathcal{B}_{2}q_{1}\alpha_{1}\alpha_{2}t_{xy}$&
        $\mathcal{B}_{2}q_{1}\alpha_{1}\alpha_{2}t_{xy}$&
        $\mathcal{B}_{2}q_{1}^{2}\alpha_{1}\alpha_{2}S_{xy}$&
        $-\mathcal{B}_{2}q_{1}q_{3}\alpha_{1}\alpha_{2}S_{xy}$ \\ \hline
    $M_{41}$&
        $q_{1}S_{xy}\Theta_{4}$&
        $-q_{1}S_{xy}\Theta_{4}$&
        $-q_{1}^{2}t_{xy}\Theta_{3}$&
        $q_{1}q_{3}t_{xy}\Theta_{1}$ \\
    $M_{42}$&
        $-q_{1}\alpha_{1}\alpha_{2}S_{xy}t_{xy}$&
        $q_{1}\alpha_{1}\alpha_{2}S_{xy}t_{xy}$&
        $\alpha_{1}q_{1}^{2}\Theta_{1}$&
        $-q_{1}q_{3}\alpha_{1}\Theta_{1}$ \\
    $M_{43}$&
        $q_{1}t_{xy}\Theta_{1}$&
        $-q_{1}t_{xy}\Theta_{1}$&
        $\alpha_{2}q_{1}^{2}\Theta_{1}$&
        $-\alpha_{1}q_{1}q_{3}\Theta_{1}$ \\
    $M_{44}$&
        $-\mathcal{B}_{2}q_{1}\alpha_{2}\Omega_{1}$&
        $\mathcal{B}_{2}q_{1}\alpha_{1}\Omega_{3}$&
        $0$&
        $0$ \\ \hline\hline
   \end{tabular}
 \end{center}
\end{table}

Some parameters used in the above table are:

\beann
    \alpha_{1}  =  \mathcal{B}_{2}q_{1}^{2} + \Theta_{2} & \mb{;} &
    \alpha_{2}  =  \mathcal{B}_{2}q_{3}^{2} + \Theta_{2} \\
    \Delta_{\mb{\tiny T}} & = & \mathcal{B}_{2}q_{1}(q_{1}^{2} - q_{3}^{2})\Theta_{1} \\
    \Theta_{1}  =  \mathcal{B}_{2}t_{xy}^{2} - \Theta_{2}S_{xy}^{2} & \mb{;} &
    \Theta_{2}  =  \mathcal{A}_{1}q_{\mb{\tiny T}}^{2} + \tilde{\cal E} \\
    \Theta_{3}  =  \mathcal{B}_{2}\Omega_{3} - \alpha_{2}S_{xy}^{2} & \mb{;} &
    \Theta_{4}  =  \mathcal{B}_{2}q_{1}^{2}\Omega_{3} + \alpha_{2}t_{xy}^{2} \\
    \Omega_{1} = t_{xy}^{2}+q_{1}^{2}S_{xy}^{2} & \mb{;} &
    \Omega_{3} = t_{xy}^{2}+q_{3}^{2}S_{xy}^{2}
\eeann
To obtain the TM of the first kind, in the general case, the following expression can be applied \cite{RPA04}:
\be
\label{FTM}
    \bn{M}_{fd}(\xi_{1},\xi_{o}) = \bn{N}(\xi_{1}) \cdot \bn{N}(\xi_{o})^{-1},
\ee
\noindent where $\bn{N}(\xi)$ is a $(N \times N)$ matrix of the linearly independent solutions of the system of equations (\ref{Eqmaestra}), and their derivatives. For illustration, if $N = 2$, it can be cast as:
\be
 \bn{N}(\xi) = \left\Vert \begin{array}{cccc}
                h_{11}\e^{iq_{1}\xi} & h_{12}\e^{iq_{2}\xi} & h_{13}\e^{iq_{3}\xi} &
                h_{14}\e^{iq_{4}\xi}\\
                h_{21}\e^{iq_{1}\xi} & h_{22}\e^{iq_{2}\xi} & h_{23}\e^{iq_{3}\xi} &
              h_{24}\e^{iq_{4}\xi}\\ iq_{1}h_{11}\e^{iq_{1}\xi} & iq_{2}h_{12}\e^{iq_{2}\xi} &
           iq_{3}h_{13}\e^{iq_{3}\xi} & iq_{4}h_{14}\e^{iq_{4}\xi}\\ iq_{1}h_{21}\e^{iq_{1}\xi}
           & iq_{2}h_{22}\e^{iq_{2}\xi} & iq_{3}h_{23}\e^{iq_{3}\xi} &
           iq_{4}h_{24}\e^{iq_{4}\xi}
\end{array}\right\Vert .
\ee
\end{appendix} 
\section{\hspace{15mm} Applicability bounds of the multi-component MSA model}
 \label{Sec:AppB}

Among the limiting points of the model there are: (i) The system considered must be seen in the approximation of flat band (semi-empirical band parameters must be in the approximation of flat band (semi-empirical band parameters must be sectionally constant)).  With an electric external field, the electrodes must be modeled as plane bands. (ii) Those coming from bounds of the Hamiltonian \mbox{\boldmath $k \cdot p$} from which we start, in the vicinity of high symmetry points of the Brillouin Zone. This restriction bounds the energy of the incident flux to some electron-volts and the values of $\kappa_{\mb{\tiny T}}$ to a small fraction of the Brillouin Zone (approximately the $25$ per cent). (iii) Those coming from the known numerical instabilities of the FTM for layers of several decades of \AA. But this disadvantage seems to be eliminated under certain conditions that will be commented lately.

One limitation of the model, comes out from the restrictions of working in the neighborhood of Brillouin Zone high-symmetry points. It is then worthy to extend the MSA analysis, to regions much more away from the above mentioned singularities. A simple way to solve this limitation, is to change the starting Hamiltonian, to other that explicitly consider a major number of bands. Another alternative, could be to consider a bigger number of cells, that yields  coherent-resonance states of the superlattice, which are non-localized. Thus, the magnitudes of interest must depend less of the starting Hamiltonian and more on the potential profile of the structure \cite{Diago06,Diago17}.\\

\textbf{Numerical Instabilities of the Transfer Matrix}
\label{Sec:AppB::TM}

\hspace{4mm}To elude the instabilities of the formalism TM found in structures of more than some decennial of \AA \cite{Wess89,Chao91} you have several algorithms \cite{Rokhlin02,PernasRPA15}. In our procedure we have taken one layer of superlattice and divide it in sub-regions at which the matrix ${\bn
M}_{fd}(z,z_{o})$ satisfied the general properties \cite{RPA04} and then it is possible to make the calculation successfully. For example let us suppose a layer \textit{A} whose length is $z_{2} - z_{1}$. We can divide this portion in $m$ parts, each one of length $\Delta_{z}$ at which the matrix satisfies:
\begin{equation}
 M_{fd}(z_{2},z_{1}) = M_{fd}(z_{2},z'_{m-1})\cdots M_{fd}(z'_{1},z_{1})
 =[M_{fd}(z_{1}+\Delta_{z},z_{1})]^{m}\,,
\end{equation}
\ni where $m=(z_{2}-z_{1})/\Delta_{z}$. What follows is the usual procedure of matching of the corresponding matrices and in one simple cell we obtain
\begin{equation}
 \bn{M}_{fd}(z_{3},z_{\mb{\tiny L}}) = \bn{M}_3(z_{3},z_2)\bn{C}_{2}\bn{M}_{2}(z_{2},z_{1})
 \bn{C}_{1}\bn{M}_{1}(z_1,z_{\mb{\tiny L}}),
\end{equation}
\ni meanwhile for the periodic heterostructure of $n$-cells we have:
\begin{equation}
 \label{ec:AppB::TM:::SL}
 \bn{M}_{fd}(z_{\mb{\tiny R}},z_{\mb{\tiny L}}) =
 \{\bn{M}_{fd}(z_{3},z_{\mb{\tiny L}})\}^{n}.
\end{equation}
\ni Matrices $\bn{M}_{1,2,3}$ correspond to \textit{Layer \textbf{L} / Layer \textbf{M} / Layer \textbf{R}}, respectively  [See Fig.\ref{Fig:BVprofile}]. Notice that for $n = 1$, layers \textit{R} and \textit{B} are coincident. Matrices $\bn{C}_{1,2}$, are the continuity matrices at the points where the potential and the band parameters jump from a group of values corresponding to a layer to the group for the next layer. This procedure has been verified in a superlattice of $(GaAs/AlAs)^n$ with $n=11$ (this is equivalent to a length of $660$ \AA). And one can see that in a great range of energy and for $\kappa_{\mb{\tiny T}}$, the following symmetry demands are satisfied:
\begin{eqnarray}
\nonumber
 \Re \{\det\{\bn{M}_{fd}(z_{3},z_{\mb{\tiny L}})\}^{n}\}=1, \\
 \Im \{\det\{\bn{M}_{fd}(z_{3},z_{\mb{\tiny L}})\}^{n}\}=0, \\
 \nonumber \\
 \label{ec:AppB::FC}
 \nonumber
  \Re \{\bn{M}_{sv}(z_{\mb{\tiny R}},z_{\mb{\tiny L}})\t \bn{\Sigma}_{z}
  \bn{M}_{sv}(z_{\mb{\tiny R}},z_{\mb{\tiny L}})-
  \bn{\Sigma}_{z}\}= \bn{O}_{8}, \\
  \Im \{\bn{M}_{sv}(z_{\mb{\tiny R}},z_{\mb{\tiny L}})\t \bn{\Sigma}_{z}
  \bn{M}_{sv}(z_{\mb{\tiny R}},z_{\mb{\tiny L}})-
  \bn{\Sigma}_{z}\}= \bn{O}_{8},
\end{eqnarray}
\ni corresponding to the unity of the determinant and flux conservation respectively. In these expressions $\bn{O}_{8}$ is the $(8 \times 8)$ null matrix. For an orthonormal basis of linearly independent solutions formed by the eigenvalues and eigenfunctions of the Hamiltonian, $\bn{\Sigma}_{z}$ is the extension matrix of $\bn{\sigma}_{z}$ which is the $(8 \times 8)$ Pauli matrix. The specific form of this matrix in another basis is found in appendix  (\ref{Sec:AppD}). It could be convenient, to avoid the dispersion of errors due to truncate the numeric simulation and to optimize the computer work, to diagonalize this matrix.        $$
\bn{M}_{fd}(z_{\mb{\tiny R}},z_{\mb{\tiny L}}) =
 \{\bn{M}_{fd}(z_{3},z_{\mb{\tiny L}})\}^{n}=
\bn{T}^{-1}\bn{J}_{or}^{n}\bn{T},
$$
although this is not always recommended because this could deteriorate the potentialities of the formalism TM \cite{RPA04,PernasRPA15}. The diagonal matrix $\bn{J}_{or}$ is the Jordan matrix $\bn{M}_{fd}(z_{\mb{\tiny R}},z_{\mb{\tiny L}})$.

There is a crucial relation between $\bn{M}_{sv}(z_{\textsc r},z_{\textsc l})$ and $\bn{M}_{fd}(z_{\textsc r},z_{\textsc l})$ , which reads
\be
 \label{ec:AppB::Mfd-Msv}
 \bn{M}_{sv}(z_{\textsc r},z_{\textsc l}) =  \bn{\cal N}^{-1}
 \bn{M}_{fd}(z_{\textsc r},z_{\textsc l}) \bn{\cal N},
\ee
\noindent with $\bn {\cal N}$ a transformation matrix depending upon the specific $N$-component Hamiltonian \cite{Diago06}.The transformation matrix
$\bn{\mathcal{N}}$, is of the form
\be
 \label{Ncal}
  \bn{\mathcal{N}}= \left[ \ba{cccc}
  \bn{g}_{1} & \bn{g}_{2} & \;\;...\;\; & \bn{g}_{2N} \\
  d_{1}\bn{g}_{1} & d_{2}\bn{g}_{2} & \;\;...\;\; & d_{2N}\bn{g}_{2N}
  \ea \right],
\ee and can be obtained when each LI solution is written as a $(N\times 1)$ super-spinor, with no coordinate dependence (represented here by $\bn{g}_{j}$), times a plane wave. By $d_{j}$ we denote the coefficient of $z$ in the exponent of the plane waves.


\section{\hspace{15mm} Reconstruction of the space $(4 \times 4)$ of the KL model}
 \label{Sec:AppC}

To obtain the matrix $\bn{M}_{fd}(z,z_{o})$ in each layer, that is, matrix $\bn{M}_{1,2,3,...,n}$ and the continuity matrix $\bn{C}_{1,2,...,n-1}$. These last matrices allow the matching where the potential and the band parameters jump from a group of values in one layer to another group in the next layer. First thing to do is to look for inside the KL model $(4 \times 4)$ the corresponding matrices of the sub-spaces $(2 \times 2)$ \textit{up (u)} and \textit{low (l)}. It is convenient to build first the TM of the first type in the subspace \textit{up} and then generate the TM of the other subspace applying the relation:
\begin{equation}
 \label{ec:AppB::TRI4Mu}
 \bn{M}_{u,\;l}(z,z_{o}) = \bn{\Gamma}_{x} \bn{M}^*_{l,\;u}(z,z_{o}) \bn{\Gamma}_{x}\,,
\end{equation}
\ni where $\bn{\Gamma}_{x}=\bn{I}_{2}\bigotimes\bn{\sigma}_{x}$. To obtain the FTM in the original space of the KL model $(4 \times 4)$ we apply the following transformation:
\begin{equation}
 \bn{M}_{fd}(z,z_0)=\bn{\cal U}^{\dag}\bn{Z}
 \left[ \ba{cc}
  \bn{M}_u(z,z_0) &    \bn{O}_4  \\
    \bn{O}_4    &  \bn{M}_l(z,z_0)
 \ea \right] \bn{Z}\bn{\cal U}\,.
\end{equation}
\ni The orthogonal transformation $\bn{Z}$ provides the appropriate order in the vectors $(8 \times 1)$ formed by the wave functions $\bn{F}_{u,l}(z)$ and its derivatives and has the form:
\begin{eqnarray*}
 \bn{Z} = \left\vert \ba{cccc}
    \bn{I}_2 & \bn{O}_2 & \bn{O}_2 & \bn{O}_2 \\
    \bn{O}_2& \bn{O}_2 & \bn{I}_2 & \bn{O}_2 \\
    \bn{O}_2 & \bn{I}_2 & \bn{O}_2 & \bn{O}_2 \\
    \bn{O}_2 & \bn{O}_2 & \bn{O}_2 & \bn{I}_2
    \ea \right\vert.
\end{eqnarray*}
\ni Here we have used $\bn{\cal U}$ to represent the generalization of the unitary transformation $\bn{U}_{b}$ of Broido and Sham~\cite{Broido85} and it is written as:
\begin{eqnarray*}
 {\bn{\cal U}} =
       \left[ \ba{cc}
       \bn{U}_{b} & \bn{O}_4 \\
       \bn{O}_4 & \bn{U}_{b}
     \ea \right]\,.
\end{eqnarray*}


\textbf{The Continuity Matrix}
\label{Sec:AppC::CM}

\hspace*{6mm} The continuity matrices in the sub-spaces are given by:
\begin{equation}
 \bn{C}_{u,l}(z) =
 \left[
  \ba{cc}
   \bn{I}_{2} & \bn{O}_{2} \\
   -\frac{\imath}{2}\left(\bn{A}^{u,l}_{+}\right)^{-1}
   \left[\bn{B}^{u,l}_{+} - \bn{B}^{u,l}_{\mb{\Large {-}}} \right]
   & \left(\bn{A}^{u,l}_{+}\right)\left(\bn{A}^{u,l}_{\mb{\Large {-}}} \right)
  \ea
 \right]\,.
\end{equation}
\ni The signs $+/-$ mean that the band parameters are evaluated on the \emph{right/left} of the matching plane. Here $\bn{A}^{u,l}$ and $\bn{B}^{u,l}$ are matrices who appear as coefficients in the equation of motion belonging to the $(2 \times 2)$ subspace of the KL problem, for a homogenous layer. To obtain the continuity matrix corresponding to the space $(4 \times 4)$ of the KL model we perform the unitary inverse transformation to that of Broido and Sham
\begin{eqnarray}
 \bn{C}(z) = \bn{\cal U}^{\dag}\bn{Z}
 \left[
  \ba{cc}
    \bn{C}_{u}(z) &    \bn{O}_{4}  \\
    \bn{O}_{4}    &  \bn{C}_{l}(z)
  \ea
 \right] \bn{Z}\,\bn{\cal U}\,.
\end{eqnarray}

In the KL model and probably in others of similar type it is usual to work in reduced spaces to analyze some spectral and transport properties that do not change with the reduction of the original dimension of the space. In our case, to study the spectrum and transport phenomena in holes with an applied electric field we start from the reduced spaces and go to the bigger space later.  Although we will not demonstrate it here, we have shown that the form of the continuity matrix $\bn{C}(z)\;(4 \times 4)$ is invariant to the order in which the operations for the matching are performed.

In studying the transmission of holes without electric field is useful to express the Continuity Matrix  in terms of the matrices of the QEP associated to the equation (\ref{Eqmaestra}), and then:
\begin{equation}
 \bn{C}(z) = \left( \ba{cc}
         -\bn{I}_{4} &  \bn{O}_{4} \\
          \frac{\imath}{2}(\mathbb{C})_{+} & (\mathbb{M})_{+}
         \ea \right)^{-1}\,
         \left( \ba{cc}
         \bn{I}_{4} &  \bn{O}_{4} \\
          \frac{\imath}{2}(\mathbb{C})_{\mb{\Large {-}}} &
          (\mathbb{M})_{\mb{\Large {-}}}
         \ea \right)\,,
\end{equation}
\ni where
\begin{eqnarray*}
 \label{M-QEP}
\mathbb{M}=
 \left(
  \begin{array}{cccc}
   B_{2} & 0 & 0 & 0 \\
   0 & B_{1} & 0 & 0 \\
   0 & 0 & B_{1} & 0 \\
   0 & 0 & 0 & B_{2}
  \end{array}
 \right)\;\;\;\; y \;\;\;\;
 \label{C-QEP}
 \mathbb{C} =
 \left(
  \begin{array}{cccc}
   0 & 0 & \widetilde{H}_{13} & 0 \\
   0 & 0 & 0 & -\widetilde{H}_{13} \\
   \widetilde{H}_{13}^{*} & 0 & 0 & 0 \\
   0 & -\widetilde{H}_{13}^{*} & 0 & 0
  \end{array}
 \right)\,,
\end{eqnarray*}
\ni where $\widetilde{H}_{13}$ is basically the same ${H}_{13}$ we had defined at appendix \ref{Sec:AppA}, but eliminating $\hat{k}_{z}$ because the eigenvalue $q_{j}$ of the QEP directly appears in (\ref{QEP1}) and (\ref{QEP2}). To complete this presentation we add:
\begin{eqnarray*}
 \label{K-QEP}
 \mathbb{K} =
 \left(
  \begin{array}{cccc}
   A_{1}\kappa_{\mb{\tiny T}}^{2} + V(z) - E & H_{12} & 0 & 0 \\
   H_{12}^{*} & A_{2}\kappa_{\mb{\tiny T}}^{2} + V(z) - E & 0 & 0 \\
   0 & 0 & A_{2}\kappa_{\mb{\tiny T}}^{2} + V(z) - E & H_{12} \\
   0 & 0 & H_{12}^{*} & A_{1}\kappa_{\mb{\tiny T}}^{2} + V(z) - E
  \end{array}
 \right)\,.
\end{eqnarray*}
\ni These expressions are valid for a layer modeled as plane wave.


\section{\hspace{15mm} Auxiliary Matrices}
 \label{Sec:AppD}

\textbf{Matrix Formalisms}

Considering the case in which the interesting region is only one and we do not add any symbol to the magnitudes to identify the domain to which they refer \cite{RPA04}. We define in this case:
\begin{eqnarray}
 \label{Supervec-ffl}
\bn{\Omega}(z) & = & \begin{array}{||c||}
                       \bn{F}(z) \\
                       \bn{P}(z) \cdot \bn{F}(z) + \bn{B}(z) \cdot \bn{F}\p(z)
                     \end{array} \;,\\
                   &    &  \nonumber \\
\bn{\Omega}_j(z) & = & \begin{array}{||c||}
                         \bn{F}_j(z) \\
                         \bn{P}(z) \cdot \bn{F}_j(z) + \bn{B}(z) \cdot \bn{F}_j\p(z)
                       \end{array} \;,\\
                   &    & \nonumber \\
 \label{Q-def}
\bn{Q}(z) & = & \begin{array}{||ccc||}
                 \bn{\Omega}_1(z) \bn{\Omega}_2(z) & \cdots & \bn{\Omega}_{2N}(z)
                \end{array} \;,\\
 \label{R-def}
\bn{R}(z) & = & \begin{array}{||cc||}
                  \bn{I}_{\mb{\tiny N}} & \bn{O}_{\mb{\tiny N}} \\
                  \bn{P}(z)  & \bn{B}(z)
                \end{array} \;.
\end{eqnarray}
Then, the following relations are fulfilled:
\begin{eqnarray}
 \label{Omega-Psi-01}
  \bn{\Omega}(z) & = & \bn{R}(z) \; \cdot \; \bn{\Psi}(z) \, ,\\
 \label{Omega-Psi-02}
  \bn{\Omega}_j(z) & = & \bn{R}(z) \; \cdot \; \bn{\Psi}_j(z) \,,\\
 \label{Omega-Psi-03}
  \bn{Q}(z) & = & \bn{R}(z) \cdot \; \bn{N}(z) \,.
\end{eqnarray}

Let us suppose that all information about the intermedia region ${\rm M}$ [see Figure \ref{Fig:PotDis1}] is given by the TM  $\bn{T}(z,z_{0})$ (o en la $\bn{M}_{fd}(z,z_{0})$) that is defined and it is known $\forall \; z, z_{0} \in {\rm M}$. Then:
\begin{eqnarray}
 \label{Supervec-ffl2}
\bn{\Omega}(z) & = &
         \left\{
         \begin{array}{lcc}
          \bn{Q}({\rm L}:z) \; \cdot \; \bn{a}({\rm L}) & \hspace{2cm} & \mbox{{\normalsize $z\leq z_{\mb{\tiny L}}$}}\\
          \bn{T}(z,z_{{\mb{\tiny L}}}) \; \cdot \; \bn{\Omega}(z_{{\mb{\tiny L}}}) & & \mbox{{\normalsize
          $z_{\mb{\tiny L}}\leq z\leq z_{\mb{\tiny R}}$}}\\
          \bn{Q}({\rm R}:z) \; \cdot \; \bn{a}({\rm R}) &  & \mbox{{\normalsize $z\geq z_{\mb{\tiny R}}$}}
         \end{array}
         \right.\,.
\end{eqnarray}
We are supposing also that $\bn{\Omega}$ is continuous in $z_{\mb{\tiny L}}$ and $z_{\mb{\tiny R}}$.\\

\textbf{Matrices to define the Charge Conservation Law}
 \label{Sec:AppD::CC}

The $(8 \times 8)\;$ matrix $\bn{X}$ given by:
\be
 \bn{X}=-\imath\bn{Q}\t\bn{\Sigma}_{y}\bn{Q}=\left[ \ba{cc}
  \bn{X}_{11} &  \bn{X}_{11} \\
   \bn{X}_{21} & \bn{X}_{22}
 \ea \right],
\ee
\ni where
$$
 \bn{\Sigma}_{y}= \left[ \ba{cc}
  \bn{O}_{4} &  -\imath\bn{I}_{4} \\
   \imath\bn{I}_{4} & \bn{O}_{4}
 \ea \right],
$$
\ni is the generalized Pauli matrix $\bn{\sigma}_{y}$.  Matrix $\bn{Q}(z)$ satisfies
\be
 \bn{Q}(z)=\bn{R}(z)\bn{N}(z).
\ee

\ni The matrix $\bn{N}(z)$ is defined in the usual way from the linearly independent solutions and its derivatives \cite{RPA04}, meanwhile matrix $\bn{Q}(z)$ has components of second order formed as a linear combination of the wave functions $\bn{F}(z)$ and its derivatives. In the $(4 \times 4)$ space of KL, matrix $\bn{R}(z)$ is defined as:
\be
  \bn{R}(z) =
 \left\vert \ba{cc}
   \bn{I}_4 & \bn{O}_4 \\
     \,\, &  \,\,    \\
   2\bn{U}\t \left[\ba{cc}
            \bn{B}^u & \bn{O}_2 \\
            \bn{O}_2 & \bn{B}^l
          \ea \right]\bn{U} &  \bn{U}\t
          \left[\ba{cc} \bn{A}^u & \bn{O}_2 \\
            \bn{O}_2 & \bn{A}^l
          \ea \right]\bn{U}
 \ea \right\vert.
\ee


\textbf{Matrices to define the Time-Reversal Invariance}

The time reversal operator $\hat{T}$ becomes
\be
 \hat{T} = \bn{K}\hat{C} =
 \left[\begin{array}{cc}
   \bn{O}_{2} & \bn{\sigma}_{x} \\
   -\bn{\sigma}_{x} & \bn{O}_2
  \end{array}
 \right]\hat{C}.
\ee
\noindent being $\hat{C}$, the complex-conjugation operator.


\textbf{Transformation matrices: Discrete symmetries}
 \label{Sec:AppD::Tran}

\hspace{4mm} From the transformation of matrices $\bn{M}_{fd}(z,z_{0})$ and $\bn{M}_{sv}(z,z_{0})$, it is simple to see that the requirement of flux conservation over matrix $\bn{M}_{fd}(z,z_{0})$ \cite{Diago05} implies, for an arbitrary basis, that:
\be
 \bn{\Sigma}_{z}=\left(
 \bn{\mathcal{N}}^{\;\dag }\right)^{-1}\bn{J}_{fd} \bn{\mathcal{N}}.
\ee
\ni Nevertheless, it is important to emphasize that if one chooses a non orthogonal basis of linearly independent solutions -which is the case very often for the KL model- than we have:
\[
\bn{J}_{fd}= \left[
\begin{array}{ccc}
 \bn{B}^{\dag }(z)-\bn{B}(z) & \;\; & \bn{A}(z) \\
 \bn{A}(z) & \;\; & \bn{O}_4
\end{array}
\right].
\]

For the requirements of invariance under time reversion and spatial inversion  of the TM $\bn{M}_{sv}(z,z_{0})$ \cite{Diago05}, the matrices $\bn{\Sigma}_{x}$ and $\bn{S}_{sv}$, were used respectively and they write like:
\bea
 \bn{\Sigma}_{x} & = & \bn{\mathcal{N}}^{-1}
 \bn{\Sigma}^{-1}\bn{\mathcal{N}}^{*} \\
 \bn{S}_{sv} & = & \bn{\mathcal{N}}\bn{S}
 \bn{\mathcal{N}}^{-1}.
\eea

\textbf{Transformation Matrix: Differential equations}
 \label{Sec:AppD::EcuDif}

\hspace{4mm}For the $up\,$ sub-space of the $(2\times 2)$ KL model one obtains in atomic units:
$$
 \mathcal{P}_{\psi}(\mb{\small $\xi$}) =
 \bn{A}_{\mb{\tiny D}}^{-1}\mathcal{P}_{\psi}(z)\bn{A}_{\mb{\tiny D}}
  =
  \left\Vert
   \begin{array}{cccc}
     0 & 0 & 1 & 0 \\
     0 & 0 & 0 & 1 \\
     w_{31}(\mb{\small $\xi$}) & w_{32} & 0 & w_{34} \\
     w_{41} & w_{42}(\mb{\small $\xi$}) & w_{43} & 0
   \end{array}
  \right\Vert\,,\; {\mb con}\;\;
  \bn{A}_{\mb{\tiny D}} =
  \left\Vert
    \begin{array}{cc}
       a^{\mb{\tiny $-1/2$}}_{o}\bn{I}_{2} & \bn{0}_{2} \\
       \bn{0}_{2} & a^{\mb{\tiny $-3/2$}}_{o}\bn{I}_{2}
     \end{array}
    \right\Vert\,.
$$
\ni We have defined
$$
  w_{31}(\mb{\small $\xi$})
    =  \frac{(\gamma_{1} + \gamma_{2})q^{2}_{\mb{\tiny T}} +
       {\cal V}_{\mb{\tiny E}}(\mb{\small $\xi$})}{2\gamma_{2} - \gamma_{1}}\,,\;\;\;
  w_{32}
    =
   -\frac{t_{xy}}{2\gamma_{2} - \gamma_{1}}\,,\;\;\;
   w_{34}
    =
   -\frac{\sqrt{3}\gamma_{3}q_{\mb{\tiny T}}}{2\gamma_{2} - \gamma_{1}}\,,
$$
$$
   w_{41}
    =
   -\frac{t_{xy}}{2\gamma_{2} + \gamma_{1}}\,,\;\;\;
   w_{42}(\mb{\small $\xi$})
    =
   -\frac{(\gamma_{1} - \gamma_{2})q^{2}_{\mb{\tiny T}} +
    {\cal V}_{\mb{\tiny E}}(\mb{\small $\xi$})}{2\gamma_{2} + \gamma_{1}}\,,\;\;\;
   w_{43}
    =  \frac{\sqrt{3}\gamma_{3}q_{\mb{\tiny T}}}{2\gamma_{2} + \gamma_{1}}\,;
$$
\ni and the non-dimensional potential was taken as ${\cal V}_{\mb{\tiny E}}(\mb{\small
$\xi$}) = \frac{V(\mb{\tiny $\xi$})}{Ry} - \frac{E}{Ry}\,.$




\section{References}
\begin{thebibliography}{45}
\bibitem{Schneider85} H. Schneider, H. T. Granh, K. v. Klitzing  and K. Ploog, \textit{Phys. Rev. B}. (b) \textbf{40}, 10040 (1985).
\bibitem{Wess89} R. Wessel and M. Altarelli, Phys. Rev. B {\bf 74}, 045308 (2006).
\bibitem{Erdogan93} M. U. Erdo\v{g}an, K. W. Kim, M. A. Stroscio, \textit{App. Phys. Lett.} \textbf{62}, 1423 (1993).
\bibitem{Kumar97} D. A. Broido and L. J. Sham, \textit{Phys. Rev. B} \textbf{58}, 167 (1985).
\bibitem{Broido85} D. A. Broido and L. J. Sham, \textit{Phys. Rev. B} \textbf{58}, 167 (1985).
\bibitem{Morifuji95} M. Morifuji and C. Hamaguchi, \textit{Phys. Rev. B} \textbf{52}, 14131 (1995).
\bibitem{Sanchez95} A. D. S\'anchez and C. R. Proetto, \textit{J. Phys.: Condens. Matter} \textbf{7}, 2059 (1995).
\bibitem{Diago02} L. Diago-Cisneros, P. Pereyra, R. P\'{e}rez-\'{A}lvarez, and H. Rodr\'{\i}guez-Coppola, \textit{Phys. Stat. Sol.} (b) \textbf{232}, 125 (2002).
\bibitem{Diago06} L. Diago-Cisneros, H. Rodr\'{\i}guez-Coppola, R. P\'erez-\'Alvarez and P. Pereyra, \textit{Phys. Rev. B} \textbf{74}, 045308 (2006).
\bibitem{Klimeck01} G. Klimeck, R. C. Bowen, and T. B. Boykin, \textit{Supperlatt. Mic.} \textbf{29}, 188 (2001).
\bibitem{RPA04} Rolando P\'erez-\'Alvarez and Federico Garc\'{\i}a-Moliner, {\it ``Transfer Matrix, Green Function and related techniques:
       Tools for the study of multilayer heterostructures"} (Universitat Jaume I, Castell\'{o}n de la Plana, Espa\~{n}a, 2004).
\bibitem{FGM92} F. Garc\'{\i}a-Moliner and V. R. Velasco, {\it Theory of Single and Multiple Interfaces} (World Scientific, Singapore, 1992).
\bibitem{Malik99} A. M. Malik, M. J. Godfrey, and P. Dawson, \textit{Phys. Rev. B} \textbf{59}, 2861 (1999).
\bibitem{RPA01} R. P\'erez-\'Alvarez, C. Trallero-Herrero, and F. Garc\'{\i}a-Moliner, \textit{Eur. J. Phys.} \textbf{22}, 275 (2001).
\bibitem{Diago05} L. Diago-Cisneros,  H. Rodr\'{\i}guez-Coppola, R. P\'{e}rez-\'{A}lvarez and P. Pereyra, \textit{Phys. Scr.}. (b) \textbf{71}, 582 (2005).
\bibitem{Borland61} R. E. Borland, and , \textit{Proc. Roy. Soc. (London)}. (b) \textbf{84}, 926 (1961).
\bibitem{Mello88} P. A. Mello, P. Pereyra and N. Kumar, \textit{Ann. Phys.}. (b) \textbf{181}, 290 (1988).
\bibitem{Pereyra98} P. Pereyra, \textit{J. Phys. A: Math. Gen.} \textbf{31}, 4521 (1998).
\bibitem{Tisseur03} D. S. MacKey, N. MacKey, and F. Tisseur, {\it Structured Tools for Structured Matrices}, Numerical Analysis Report No. 419 (Manchester Centre for Computational Mathematics, England, 2003).
\bibitem{Tisseur01} F. Tisseur and K. Meerbergen, \textit{SIAM Review} \textbf{43}, 235 (2001).
\bibitem{Diago11} A. Mendoza-\'{A}lvarez, J. J. Flores-Godoy, G. Fern\'{a}ndez-Anaya and L. Diago-Cisneros, \textit{Phys. Scr.} \textbf{84}, 055702 (2011).
\bibitem{Diago19} E. Nieva-P\'erez, E. A. Mendoza-\'Alvarez, L. Diago-Cisneros, C. A. Duque, J. J. Flores-Godoy and G. Fern\'andez-Anaya, \textit{Phys. Scr.} \textbf{94}, 035205 (2019).
\bibitem{Davydov65} A. S. Davydov, \textit{Quantum Mechanics} (Instituto del Libro, La Habana, 1965).
\bibitem{Landauer94} R. Landauer and Th. Martin, textit{Rev. Mod. Phys.} \textbf{66}, 217 (1994).
\bibitem{Pereyra00} P. Pereyra, \textit{Phys. Rev. Lett.} \textbf{84}, 1772 (2000).
\bibitem{Chao91} C. Y. Chao and S. L. Chuang, \textit{Phys. Rev. B} \textbf{43}, 7027 (1991).
\bibitem{Rokhlin02} S. I. Rokhlin and L. Wang, \textit{J. Acoustic. Soc. Am} \textbf{12}, 822 (2002).
 \bibitem{PernasRPA15} R. P\'erez-\'Alvarez, R. Pernas-Salom\'{o}n and V. R. Velasco, \textit{SIAM J. Appl. Math.} \textbf{75}, 1403 (2015).
\end{thebibliography}
\end{document}